\documentclass[10pt]{article}
\usepackage[cp1250]{inputenc}
\usepackage{amsmath} 
\usepackage{amssymb} 
\usepackage{a4wide} 
\usepackage[left=3cm]{geometry} 
\usepackage{braket}
\usepackage{slashed}
\usepackage{bbm}
\usepackage{yfonts}
\usepackage{simplewick}
\usepackage{hyperref}
\usepackage{tikz}
\usetikzlibrary{matrix,arrows}
\usetikzlibrary{decorations.markings}

\allowdisplaybreaks

\DeclareMathOperator{\Tr}{Tr}

\DeclareMathOperator{\End}{End}

\title{$\mathcal{W}_{\infty}$ in the quadratic basis}
\author{Tomáš Procházka}

\begin{document}

\bibliographystyle{hieeetr}

\vskip 2.1cm

\centerline{\large \bf Exploring $\mathcal{W}_{\infty}$ in the quadratic basis}
\vspace*{8.0ex}

\centerline{\large \rm Tom\'{a}\v{s} Proch\'{a}zka \footnote{Email: {\tt prochazkat at fzu.cz}}}

\vspace*{8.0ex}

\centerline{\large \it Institute of Physics AS CR, Na Slovance 2, Prague 8, Czech Republic}
\vspace*{2.0ex}

\vspace*{6.0ex}

\centerline{\bf Abstract}
\bigskip

We study the operator product expansions in the chiral algebra $\mathcal{W}_{\infty}$, first using the associativity conditions in the basis of primary generating fields and second using a different basis coming from the free field representation in which the OPE takes a simpler quadratic form. The results in the quadratic basis can be compactly written using certain bilocal combinations of the generating fields and we conjecture a closed-form formula for the complete OPE in this basis. Next we show that the commutation relations as well as correlation functions can be easily computed using properties of these bilocal fields. In the last part of this paper we verify the consistency with results derived previously by studying minimal models of $\mathcal{W}_{\infty}$ and comparing them to known reductions of $\mathcal{W}_{\infty}$ to $\mathcal{W}_N$. The results we obtain illustrate nicely the role of triality symmetry in the representation theory of $\mathcal{W}_{\infty}$.

 \vfill \eject

\tableofcontents

\setcounter{footnote}{0}

\newpage

\section{Introduction}

The study of the conformal field theories with extended higher spin symmetries is problem which is almost as old as conformal field theory itself. First example of extended symmetry algebra, $\mathcal{W}_3$ was constructed by Zamolodchikov in \cite{Zamolodchikov:1985wn}. This algebra was obtained by extending the Virasoro algebra by addition of spin $3$ field. In \cite{Fateev:1987zh, Lukyanov:1987xg, Lukyanov:1990tf} the authors extended the construction of $\mathcal{W}_3$ and used the free fields to construct whole family $\mathcal{W}_N$ of algebras with $N=3, 4, \ldots$. Each of these algebras contains the Virasoro algebra together with additional generating local fields of spin $3, 4, \ldots, N$. These algebras exist for arbitrary values of central charge $c$. One property that distinguishes these extended algebras from the more simple algebras such as the Virasoro algebra or affine Lie algebras is their nonlinearity. Unless we largely extend the set of fields that we are considering, they can no longer be seen as Lie algebras.

A second source of (classical) $\mathcal{W}$ algebras came from the theory of integrable partial differential equations. After \cite{Magri:1977gn, Gervais:1982nw, Gervais:1985fc, Bakas:1988mq} found the connection between the Virasoro algebra and Korteweg-de Vries equation, the Gelfand-Dickey algebra of pseudodifferential operators \cite{Gelfand:1995qu, Adler:1979ib, Mathieu:1988pm, Bakas:1989um, Dickey:1991xa} was introduced to study the generalized KdV hierarchies. Later the geometric picture was given as the Hamiltonian reduction of the coadjoint orbits of loop groups by Drinfeld and Sokolov \cite{Drinfeld:1984qv}.

There are now various ways known of systematically producing the quantum chiral algebras with higher spin generators \cite{Bouwknegt:1992wg}. For example, they appear naturally as subalgebras of the universal enveloping algebras of affine Lie algebras - the so-called Casimir algebras \cite{Bais:1987dc}, as well as in various GKO coset constructions \cite{Goddard:1984vk, Goddard:1986ee, Bais:1987zk}. There are also many direct constructions in the spirit of the original Zamolodchikov's construction \cite{Blumenhagen:1990jv, Blumenhagen:1991ia, Bowcock:1990ku, Kausch:1990bn} and there is also a quantum version of the Drinfeld-Sokolov reduction.

The idea of the quantum Drinfeld-Sokolov reduction is that instead of quantizing the classical $\mathcal{W}$-algebra, realized as the symplectic quotient, it is easier to quantize the manifold before taking the quotient and to implement the constraints at the quantum level. In our context one can start with affine Lie algebra and use the BRST procedure to obtain the quantum W-algebra \cite{Bershadsky:1989mf, Diaz:1990mg, Feigin:1990pn, FigueroaO'Farrill:1990dz, Kawai:1991qt}. Overview of this construction with many references to the literature are given in \cite{Bouwknegt:1992wg}. For the family $\mathcal{W}_N$ that we are mainly interested in, this procedure reduces to simpler procedure of quantizing the free field representation of $\mathcal{W}_N$ (which is also called the Miura transformation) as done in \cite{Fateev:1987zh, Lukyanov:1987xg, Lukyanov:1990tf}.

The Drinfeld-Sokolov reduction has recently found an interesting physical application in the context of higher spin theories \cite{Henneaux:2010xg, Campoleoni:2010zq, Campoleoni:2011hg, Gaberdiel:2011wb}. There is an interesting class of extensions of cosmological Einstein gravity in three spacetime dimensions. Einstein theory of gravity in three dimensions can be formulated \cite{Achucarro:1987vz, Witten:1988hc} as a Chern-Simons theory with gauge group $SL(2)$ \footnote{The precise real form depends on the signature of the metric and sign of the cosmological constant.}. We can extend the gauge group from $SL(2)$ to any simple Lie group. Depending on the choice of the Lie group and embedding of $SL(2)$ inside of it, we get a theory that describes cosmological Einstein gravity with higher spin fields. Because these theories live in three spacetime dimensions, there are no locally propagating degrees of freedom corresponding to graviton or higher spin fields. But specializing to theories with AdS spacetime as a solution, one can study the degrees of freedom associated to the conformal boundary of the AdS spacetime. One particularly interesting property to study is the so-called asymptotic symmetry algebra. In the context of the cosmological Einstein gravity this was originally studied in \cite{Brown:1986nw}. What authors found was that choosing properly the asymptotic boundary conditions at the conformal boundary of AdS one finds the Virasoro algebra as symmetry algebra acting in the space of solutions of the theory (one could also consider this as a spectrum generating algebra of the theory). In \cite{Henneaux:2010xg, Campoleoni:2010zq, Campoleoni:2011hg, Gaberdiel:2011wb} the asymptotic symmetry algebra was computed in the case of higher spin theories. It was found that the asymptotic Virasoro algebra is extended to classical $\mathcal{W}$-algebra. The precise $\mathcal{W}$-algebra that one obtains in this way depends on the choice of the Chern-Simons gauge group of the bulk theory as well as the embedding of $SL(2)$ subgroup. Mathematically, as was explained in \cite{Campoleoni:2011hg}, one is effectively performing the Drinfeld-Sokolov reduction and the particular embedding of $SL(2)$ determines the constraint that is part of the input for the Drinfeld-Sokolov reduction.

Another interesting $\mathcal{W}$-algebra very closely related to $\mathcal{W}_N$ series is their formal limit $\mathcal{W}_{\infty}$. The first two quantum versions of this algebra were studied in \cite{Pope:1989ew, Pope:1989sr, Pope:1990kc}. From today's point of view, the authors were studying $\mathcal{W}_{\infty}$ for special value of parameters where the algebra linearizes. Although already at that time there were signs of connection between $\mathcal{W}_{\infty}$ and $\mathcal{W}_N$ \cite{Lu:1991pe}, the understanding that $\mathcal{W}_{\infty}$ is actually a two-parametric family algebras came only recently \cite{Gaberdiel:2011wb, Gaberdiel:2011zw, Gaberdiel:2012ku}. The connection between one-parametric algebras $\mathcal{W}_N$ and two-parametric family $\mathcal{W}_{\infty}$ is analogous to the construction of the higher spin algebra $\mathfrak{hs}(\lambda)$ in Vasiliev theory \cite{Vasiliev:1999ba} and its universal relation to $\mathfrak{sl}(N)$ algebras. If we understand $\mathcal{W}_{\infty}$, we can truncate it to $\mathcal{W}_N$ for any $N$. In the other direction, in suitable basis the structure constants of $\mathcal{W}_N$ are rational functions of $N$ and $\mathcal{W}_{\infty}$ is an `analytic continuation'.

Shifting the point of view from $\mathcal{W}_N$ to $\mathcal{W}_{\infty}$, in their study of the holographic duality \cite{Gaberdiel:2010ar, Gaberdiel:2010pz} Gaberdiel and Gopakumar made a surprising discovery of a triality symmetry \cite{Gaberdiel:2012ku}. The mapping between the rank parameter $N$ and the structure constants of $\mathcal{W}_{\infty}$ is $3:1$, so if we continue $N$ to complex numbers, there are generically $3$ different values of $N$ which give an isomorphic algebra. This extends the picture of level-rank duality discovered in coset models related to $\mathcal{W}_N$ earlier \cite{Altschuler:1990th, Kuniba:1990zh}. The triality symmetry has many consequences for the representation theory of $\mathcal{W}_{\infty}$ which are still to be understood.

Although $\mathcal{W}_N$ algebras have been studied for a long time and many ways of constructing them are available, we still do not have an explicit form for the structure constants. The operator product expansions in $\mathcal{W}_N$ are only known for small values of $N$. There are known results for the classical limit $\mathcal{W}_{\infty}$ in the quadratic basis of generating fields \cite{FigueroaO'Farrill:1992cv, Khesin:1994ey} and in primary basis \cite{Campoleoni:2011hg}. The quantum case is considered in \cite{Lukyanov:1988} where the quadraticity of OPE in basis of fields coming from Miura transformation is proved.

The mail goal of this work was to try to fill in this gap by finding a closed-form formula for $\mathcal{W}_{1+\infty}$ \footnote{The algebra $\mathcal{W}_{1+\infty}$ is very closely related $\mathcal{W}_{\infty}$ as will be explained later.}operator product expansions in the quadratic basis of generating fields. Although our computation is based on solving the associativity conditions, the resulting operator product expansions are most easily written using bilocal fields similar to \cite{Lukyanov:1988}. We show that these bilocal fields are very useful objects for finding the commutation relations of modes or for computing the correlation functions in $\mathcal{W}_{1+\infty}$. We also make many independent verifications of consistency of the conjectured formulas, the most non-trivial being the correct results on $\mathcal{W}_N$ minimal models and the preservation of triality symmetry which is not at all manifest in the basis of fields we are using.

\subsection{Overview of this paper}
In the first part of this paper we focus on the construction of $\mathcal{W}_{\infty}$ in the primary basis as was done in \cite{Gaberdiel:2012ku}. The procedure is quite straightforward: one starts with the stress-energy tensor whose modes satisfy the Virasoro algebra and extends the algebra by adding one additional independent primary field $W_n(z)$ of every integer dimension $n \geq 3$. These fields generate all other fields using the operations of taking normal ordered products and derivatives. To fully specify the chiral algebra, we need to determine the singular part of the operator product expansion of $W_j(z) W_k(w)$ in terms of fields of lower total dimension. For this one uses the conditions of the associativity of operator products. One starts with the general ansatz for given field content and solves the algebraic equations that must be satisfied in order for the operator product algebra to be associative. We carry out the analysis up to total spin $j+k \leq 10$ which is slightly further than what was done in \cite{Gaberdiel:2012ku}. One of the original motivations for extending the analysis further was a hope that the structure constants in the primary basis could follow some pattern which could be understood. But it turns out that the number of primary fields grows too much and there is not even any nice canonical way know to enumerate the primary fields, so at this point it seems that this approach is not so useful. Nevertheless, knowing some structure constants in the primary basis is useful to comparison with different approaches. Furthermore, it is this primary basis where the nontrivial triality symmetry is easy to observe \cite{Gaberdiel:2012ku}. In fact, the primary generators can be chosen in such a way that everything (fields and structure constants) is manifestly invariant under the triality symmetry. This together with the determination of the number of parameters that $\mathcal{W}_{\infty}$ depends on (the central charge and an additional coupling constant) are the main results of the analysis in the primary basis. We also verify the expectation that for suitably chosen value of the $\mathcal{W}_\infty$ parameters, the operator product algebra can be consistently truncated to a quotient algebra which has Virasoro generator together with generators of spins $3, 4, \ldots, N$. In this sense, $\mathcal{W}_{\infty}$ contains information about all the $\mathcal{W}_N$. Since the structure constants are rational functions of parameters, we can freely move between $\mathcal{W}_N$ for various $N$ and $\mathcal{W}_{\infty}$.

In the next part, we use a different starting point, which is the free field representation of $\mathcal{W}_{N}$ as used in \cite{Fateev:1987zh}. From this construction, the Miura transformation, we obtain another generating set of fields $U_j(z)$ different from the primary fields $W_j(w)$. The very special property of these generating fields is that their operator product expansions only contain fields that are at most quadratic normal ordered products of $U_j(w)$. This gives a hope that using this basis of the algebra, we might actually be able to determine all the structure constants. The computation using the free fields can lead to quite compliated combinatorics \footnote{See however \cite{Lukyanov:1988} for important results in this direction}. So we proceed similarly to what we did in the first part and use the associativity conditions of the operator product algebra to compute the singular parts of the operator product expansions. It turns out that the results can be organized in very simple form (\ref{opequadinv}) which captures all the information that is contained in the singular part of OPE. Furthermore, we conjecture a closed-form formula for all the structure constants of $\mathcal{W}_{1+\infty}$. Since the way of writing the operator product expansion differs from the usual way of describing the algebra, we make many consistency checks. We show how to compute all the correlation functions and discuss transformation from $U_j(w)$ fields to quasiprimary and to primary basis. In this way we can make comparison between results of the first and second parts of the article. Another ingredient that is easy to introduce using the free field representation is the coproduct. Apart from providing us with many non-trivial consistency conditions for the conjectured structure constants, it also shows up later in the properties of $\mathcal{W}_{\infty}$ minimal models.

In the last part of this article we study to some extent the representation theory of $\mathcal{W}_{1+\infty}$. In particulal, we look at discrete values of parameters of $\mathcal{W}_{1+\infty}$ for which the vacuum representation is maximally degenerate. This is the analogue in $\mathcal{W}_{1+\infty}$ of the Virasoro or $\mathcal{W}_N$ minimal models. We expect to find all the $\mathcal{W}_N$ minimal model but in fact we find also some new minimal models which do not come from $\mathcal{W}_N$. The fact that the resulting set of special values of parameters is triality invariant is another nontrivial check of the results found in the previous parts of the paper.

\section{$\mathcal{W}_{\infty}$ algebra in the primary basis}
\label{secprimary}

In this section we will study the algebra $\mathcal{W}_{\infty}$ in the basis of the Virasoro primary fields. To get the $\mathcal{W}_{\infty}$ algebra, we will extend the Virasoro algebra generated by stress-energy tensor $T(z)$ \footnote{We will always work with the holomorphic part of the symmetry algebra of the field theory. All the generating fields that we will consider have spin equal to the scaling dimension, so we will use these terms interchangeably. For example, $T(z)$ is the $T_{zz}$ component of the full stress-energy tensor and is a holomorphic function because of the energy-momentum conservation. All the fields that we will consider will transform well under the scaling and with respect to translations.} by higher spin primary generators $W_s(z)$ of spin $3, 4, \ldots$. The algebraic structure will be fixed by imposing the conditions of associativity of the operator algebra.

We start by reviewing the construction and algebraic properties of the normal ordering and mode expansions in the radial quantization. Although this material is quite standard \cite{DiFrancesco:1997nk}, many of the results of this article rely heavily on computations performed within this formalism. In the next part, we will discuss the primary content of $\mathcal{W}_\infty$ algebra which is the first ingredient when making an ansatz for the OPE of $\mathcal{W}_{\infty}$ generators. With this input, we can use the Mathematica package \texttt{OPEdefs} to find the associativity conditions expressed as algebraic relations between the parameters of the ansatz. The ansatz that we consider determines the OPE of $W_j(z) W_k(w)$ up to $j + k \leq 10$. The associativity conditions are solved up to sum of spins $12$. Our result matches what was found in \cite{Gaberdiel:2012ku} - we find a two-parametric family of solutions to the associativity constraints. Furthermore we make a comparison to parametrization used when studying $\mathcal{W}_{N}$ algebras and discuss the triality symmetry found in \cite{Gaberdiel:2012ku}. In the last part of this section we compare the universal properties of $\mathcal{W}_{\infty}$ with respect to the family of $\mathcal{W}_N$ algebras to the similar universality of $\mathfrak{hs}(\lambda)$ and $\mathfrak{sl}(N)$.

\subsection{OPE and Jacobi identities}
We will be interested in the behaviour of the local operators as they approach each other which is captured by the operator product expansion. In this section we will summarize the main properties of these expansions. Further details can be found in \cite{Bais:1987dc, DiFrancesco:1997nk, Thielemans:1991uw, Thielemans:1994er}.

The general form of OPE of two local operators $A(z)$ and $B(w)$ that we will consider is
\begin{equation}
\label{opebracket}
A(z) B(w) = \sum_{k=-\infty}^{h_A+h_B} \frac{\left\{AB\right\}_k(w)}{(z-w)^k}
\end{equation}
All local operators will have a definite non-negative integral scaling dimension; the scaling dimension of operator $A(z)$ will be denoted by $h_A$. The upper bound in the summation over $k$ follows from the fact that there will be no fields with negative scaling dimension. Note that this expression is not symmetric in $A$ and $B$, since the right-hand side is expanded at $w$. Choosing a different expansion point would modify $\left\{AB\right\}_n$ by a combination of derivatives of $\left\{AB\right\}_k$ with $k>n$.

We can write a contour integral representation of $\left\{AB\right\}_k$ which will be useful in the following:
\begin{equation}
\left\{AB\right\}_k(w) = \oint_w \frac{dz}{2\pi i} A(z) B(w) (z-w)^{k-1}.
\end{equation}
In particular, the zeroth order operator
\begin{equation}
\label{normalorder}
(AB)(w) \equiv \left\{AB\right\}_0(w) = \oint_w \frac{dz}{2\pi i} \frac{A(z) B(w)}{z-w}
\end{equation}
is called the \emph{normal ordered product} of $A$ and $B$ \cite{Bais:1987dc}. By \emph{contraction} of two operators we will mean the singular part of the operator product expansion,
\begin{equation}
{\contraction{}{A}{(z)}{B}{}{A}{(z)}{B}}(w) \equiv \sum_{k=1}^{h_A+h_B} \frac{\left\{AB\right\}_k(w)}{(z-w)^k}
\end{equation}
and we can write
\begin{equation}
\label{normalorder2}
(AB)(w) = \lim_{z \to w} \Big[ A(z) B(w) - {\contraction{}{A}{(z)}{B}{}{A}{(z)}{B}}(w) \Big].
\end{equation}

\paragraph{Derivatives}
By taking a derivative of (\ref{opebracket}) we can easily derive the following equations
\begin{eqnarray}
\label{opeleibniz}
\partial \left\{A B\right\}_k & = & \left\{A^\prime B\right\}_k + \left\{A B^\prime\right\}_k \\
\label{opederivative1}
\left\{A^\prime B\right\}_{k+1} & = & -k \left\{AB\right\}_k
\end{eqnarray}
The first of them expresses the Leibniz formula for derivative of product of operators, while the second one can be used to relate the regular terms of OPE to normal ordered product,
\begin{equation}
\left\{AB\right\}_{-k}(w) = \frac{1}{k!} (A^{(k)} B)(w) \quad \quad \text{for} \; k \geq 0.
\end{equation}
In fact, by taking a formal Taylor series we arrive at the operator
\begin{equation}
\label{nobilocal}
(A(z)B(w)) \equiv \sum_{k=0}^{\infty} \frac{(z-w)^k}{k!} (A^{(k)}B)(w) = \sum_{k \leq 0} \frac{\left\{AB\right\}_k(w)}{(z-w)^k} = A(z)B(w) - {\contraction{}{A}{(z)}{B}{}{A}{(z)}{B}}(w)
\end{equation}
which just represents all the regular terms of the OPE. Inserted in any correlation function, this bilocal operator has no singularities as $z \to w$. Later we will encounter other operators with similar properties.

\paragraph{Symmetry of normal ordering}
As noted earlier, although the operator product expansion is symmetric under exchange of $A(z)$ and $B(w)$, the normal ordered product that we introduced above is not, since we are expanding at $w$. By Taylor expanding at $z$ instead, we can easily see that
\begin{equation}
\label{opeasymmetry}
\left\{BA\right\}_k(w) = \sum_{l \geq 0} \frac{(-1)^{k+l}}{l!} \partial^{l}\left\{AB\right\}_{k+l}(w)
\end{equation}
One could consider the OPE with midpoint expansion, which would lead to (anti)symmetric version of the normal ordered product but other properties like the associativity or the correspondence between states and local operators would become more complicated. Furthermore, we will see later that there is another way to produce regular bilocal operators in $\mathcal{W}_{1+\infty}$ which will be more useful.

\paragraph{Wick theorem and associativity}
The generalized Wick theorem
\begin{equation}
\contraction{}{A}{(z)}{(BC)}{A(z)(BC)}(w) = \oint_{w} \frac{dx}{2\pi i} \frac{1}{x-w} \left( \contraction{}{A}{(z)}{B}{A(z) B}(x) C(w) + \contraction{}{A}{(z) B(x)}{C}{A(z) B(x) C}(w) \right)
\end{equation}
expresses the fact that the singularities in OPE of $A(z)$ with $(BC)(w)$ come only from singularities of either $A(z)$ and $B(w)$ or $A(z)$ and $C(w)$ (see \cite{Bais:1987dc} and the discussion in appendix 6.B. of \cite{DiFrancesco:1997nk}). This formula is important for manipulations with operator products.

We can translate the associativity conditions of OPE to the language of normal ordered products. We start by using the contour integral representation of the normal ordered product to write
\begin{equation}
\left\{A\left\{BC\right\}_k\right\}_l = \oint_w \frac{dz}{2\pi i} A(z) (z-w)^{l-1} \left( \oint_w \frac{dx}{2\pi i} B(x) C(w) (x-w)^{k-1} \right)
\end{equation}
where the contour of $x$-integration is inside of the $z$-integration contour. Now we make the usual contour deformation by pulling the $x$-contour outside of $z$-contour and adding a contour encircling $z$:
\begin{equation}
\label{contourpic}
\centering
\begin{tikzpicture}
\draw[very thick,
  decoration={markings, mark=at position 0.25 with {\arrow{>}}, mark=at position 0.625 with {\filldraw[black] circle (1pt);}},
	postaction={decorate}
	]
	(0,0) circle (0.75);
\draw[very thick,
  decoration={markings, mark=at position 0.125 with {\arrow{>}}, mark=at position 0.865 with {\filldraw[black] circle (1pt);}},
	postaction={decorate}
	]
	(0,0) circle (1.5);
\draw[very thick,
  decoration={markings, mark=at position 0.25 with {\arrow{>}}, mark=at position 0.625 with {\filldraw[black] circle (1pt);}},
	postaction={decorate}
	]
	(4,0) circle (0.75);
\draw[very thick,
  decoration={markings, mark=at position 0.125 with {\arrow{>}}, mark=at position 0.865 with {\filldraw[black] circle (1pt);}},
	postaction={decorate}
	]
	(4,0) circle (1.5);
\draw[very thick,
  decoration={markings, mark=at position 0.375 with {\arrow{>}}, mark=at position 1 with {\filldraw[black] circle (1pt);}},
	postaction={decorate}
	]
	(7.7,0) circle (1.2);
\draw[very thick,
  decoration={markings, mark=at position 0.125 with {\arrow{>}}, mark=at position 0.865 with {\filldraw[black] circle (1pt);}},
	postaction={decorate}
	]
	(8.9,0) circle (0.8);
\draw[font=\bf] (2,0) node {=};
\draw[font=\bf] (6,0) node {+};
\filldraw [black] (0,0) circle (1pt);
\draw[font=\bf] (0.15,-0.2) node {w};
\filldraw [black] (4,0) circle (1pt);
\draw[font=\bf] (4.15,-0.2) node {w};
\filldraw [black] (7.7,0) circle (1pt);
\draw[font=\bf] (7.55,0.20) node {w};
\draw[font=\bf] (-0.6,-0.75) node{x};
\draw[font=\bf] (1.2,-1.2) node{z};
\draw[font=\bf] (3.4,-0.75) node{z};
\draw[font=\bf] (5.2,-1.2) node{x};
\draw[font=\bf] (9.05,-0.2) node{x};
\draw[font=\bf] (9.6,-0.75) node{z};
\end{tikzpicture}
\end{equation}
This gives us equation
\begin{equation}
\big\{A\left\{BC\right\}_j\big\}_k - \big\{B\left\{AC\right\}_k\big\}_j = \sum_{l>0} {k-1 \choose l-1} \left\{\left\{AB\right\}_l C\right\}_{j+k-l}
\end{equation}
These relations taken with $j$ and $k$ positive are the associativity conditions which will lead to non-trivial algebraic equations satisfied by structure constants of $\mathcal{W}_\infty$.

\paragraph{Mode expansions}
Given a local operator $A(z)$ with scaling dimension $h$ we can expand it in terms of mode operators around the origin of the complex plane,
\begin{equation}
A(z) = \sum_m z^{-m-h} A_m.
\end{equation}
The inverse relation is
\begin{equation}
\label{modeoperator}
A_m = \oint_0 \frac{dz}{2\pi i} A(z) z^{m+h-1}
\end{equation}
By the operator-state correspondence we want to associate a state to an operator by writing
\begin{equation}
\label{operatorstate}
\ket{A} \equiv \lim_{z \to 0} A(z) \ket{0}
\end{equation}
For this expression to be regular, we need to have
\begin{equation}
\label{vacmode}
A_m \ket{0} = 0 \quad \quad \text{for} \; m>-h
\end{equation}
For example, we must have
\begin{equation}
L_{-1} \ket{0} = 0
\end{equation}
where $L_{-1}$ is the mode operator associated to dimension $2$ operator. If these conditions are satisfied, we can write
\begin{equation}
\ket{A} = \lim_{z \to 0} A(z) \ket{0} = A_{-h_A} \ket{0}.
\end{equation}
The mode operators of derivative $\partial A(z)$ are proportional to mode operators of $A(z)$ itself,
\begin{equation}
(\partial A)_m = -\big(h_A+m\big) A_m
\end{equation}
and similarly for higher derivatives. The state corresponding to the derivative of an operator is thus
\begin{equation}
\ket{\partial A} = (\partial A)_{-h_A-1} \ket{0} = A_{-h_A-1} \ket{0}.
\end{equation}
In this way the negative modes of operator $A(z)$ acting on vacuum produce either zero or states which correspond to derivatives of $A(z)$. Using the same contour deformation argument as above we can relate the mode expansion of $(AB)(w)$ to that of $A(w)$ and $B(w)$,
\begin{equation}
\label{normordmodes}
(AB)_n = \sum_{k \leq -h_A} A_k B_{n-k} + \sum_{k>-h_A} B_{n-k} A_k
\end{equation}
which is quite similar to the normal ordering of free fields, but note the asymmetry between $A$ and $B$ in this formula as discussed above. Acting on vacuum, we see that the operator mapped to normal ordered product of fields is
\begin{equation}
\ket{(AB)} = (AB)_{-h_A-h_B} \ket{0} = A_{-h_A} B_{-h_B} \ket{0}
\end{equation}
and similarly
\begin{equation}
\ket{(A(BC))} = A_{-h_A} (BC)_{-h_B-h_C} \ket{0} = A_{-h_A} B_{-h_B} C_{-h_C} \ket{0}.
\end{equation}
In this way, the right-nested normal ordered products of operators correspond to states obtained by successive application of the mode operators. This property is one of advantages of the asymmetric normal ordering prescription (\ref{normalorder}). Finally, we want to state the formula for commutation relations between modes of two local operators. It is again a direct consequence of the contour deformation argument above:
\begin{eqnarray}
\label{modecommutator}
\left[ A_m, B_n \right] & \equiv & \oint_0 \frac{dw}{2\pi i} \oint_{|z|>|w|} \frac{dz}{2\pi i} z^{m+h_A-1} w^{n+h_B-1} A(z) B(w) \\
\nonumber
& & - \oint_0 \frac{dz}{2\pi i} \oint_{|w|>|z|} \frac{dw}{2\pi i} z^{m+h_A-1} w^{n+h_B-1} A(z) B(w) \\
\nonumber
& = & \oint_0 \frac{dw}{2\pi i} \oint_w \frac{dz}{2\pi i} z^{m+h_A-1} w^{n+h_B-1} A(z) B(w)
\end{eqnarray}
By definition, the operator ordering on the left hand side corresponds in the radial quantization to radial ordering of local fields on the right hand side. This formula tells us how to extract the commutator from the OPE of $A(z)$ and $B(w)$.

\paragraph{Virasoro algebra and primaries}
In the presence of stress-energy tensor $T(z)$ the Virasoro algebra imposes further constraints on the form of OPE of primary fields. In fact, for primary fields $A(z)$ and $B(w)$ we can write the OPE as
\begin{equation}
\label{primaryope}
A(z) B(w) \sim \sum_{j} C_{AB}^j \sum_Y \frac{\beta_Y(h_A, h_B, h_j, c) \mathcal{O}_j^Y(w)}{(z-w)^{h_A+h_B-h_j-|Y|}}
\end{equation}
Here the sum over all local fields factorizes into the sum over primaries indexed by $j$ and for each primary $j$ a sum over Virasoro descendants of $\mathcal{O}_j$ indexed by Young diagrams $Y$. The coefficients $\beta$ are universal which means that they depend on the primary fields only through their dimensions and do not depend on any other details of the chiral algebra (apart from the Virasoro central charge $c$). There are explicit expressions known for them involving the inverse of the Shapovalov form for the Virasoro algebra, but they (unlike for instance the conformal blocks) depend on the choice of basis of the Virasoro descendants.

Using the form (\ref{primaryope}) of OPE of primary fields, we can `improve' the normal ordered product of two primaries to make it primary. The way to proceed is a simple variation of (\ref{normalorder2}). Instead of subtracting just the singular terms, this time we subtract with each primary field appearing in the singular part also all of its descendants appearing in singular and regular parts of the OPE. This leads to a primary projection of the normal-ordered product
\begin{equation}
\label{normalprimary}
[AB](w) = (AB)(w) - \sum_{j: h_j < h_A+h_B} C_{AB}^j \sum_{|Y|=h_A+h_B-h_j} \beta_Y(h_A, h_B, h_j, c) \mathcal{O}_j^{Y}(w)
\end{equation}
The sum over $j$ is restricted to those operators that appear in the singular part of $A(z) B(w)$ OPE (we don't want to subtract the primary operator appearing in $(AB)(w)$). The resulting operator is a primary operator constructed out of stress-energy tensor and primary operators of lower dimension (the examples will be given later). There is a straightforward generalization of this construction which extracts the primary fields appearing deeper in the regular part of $A(z)B(w)$ OPE.

\subsection{Bootstrap and $\mathcal{W}_\infty$ in primary basis}
Having reviewed the basic properties of operator product expansions in the previous section, we are now ready to apply them to compute the OPE of $\mathcal{W}_\infty$. Since the computations are quite involved, we will use the Mathematica package \texttt{OPEdefs} by Kris Thielemans. For short overview of \texttt{OPEdefs} see \cite{Thielemans:1991uw}, in \cite{Thielemans:1994er} much more detailed description is given together with details of the implementation. The usefullness of Thielemans' package cannot be overestimated.

\subsubsection{Field content of $\mathcal{W}_\infty$}
Our starting point in this section will be the definition of $\mathcal{W}_\infty$ as the algebra obtained by extending the Virasoro algebra by primary fields $W_s(z)$ of spin $s=3, 4, 5, \ldots$ \footnote{We will usually denote the Virasoro generator by $T(z)$, but sometimes it will be convenient to consider it as a generator of spin $2$, so sometimes we will call it $W_2(z)$.}. All other fields of $\mathcal{W}_\infty$ will be obtained by taking derivatives and normal ordered products of these fields. As we will see, there is a two-parametric family of these algebras. For generic values of these parameters there will be no extra relations between the local operators apart from those that are required by the consistency of the algebra of local fields discussed in the previous section.

Another way of phrasing this is through the operator-state correspondence. To each local operator we can assign a state via (\ref{operatorstate}). All the states corresponding to all local operators of $\mathcal{W}_\infty$ can be produced by acting on the vacuum with product of mode operators corresponding to the generating fields. These states are not independent, but all the relations between them follow only by applying the commutation relations of $\mathcal{W}_\infty$. Using the generalization of the Poincar\'{e}-Birkhoff-Witt theorem\footnote{The generalization since we don't have a Lie algebra. The usual form of PBW theorem for linear algebras works because there is a natural grading on the space of states by number of mode operators. Everytime we commute two mode operators, we obtain a state which has a smaller number of mode operators acting on the vacuum. For $\mathcal{W}_\infty$ there is another grading which does not simply count the mode operators, but gives them a weight according to their spin. See also \cite{Watts:1990pd}.}, we can choose a basis of the local fields by choosing a specific ordering of mode indices, for instance
\begin{equation}
\cdots W_{3, -n_{3, 1}} W_{3, -n_{3, 2}} \cdots W_{3, -n_{3, k_3}} W_{2, -n_{2, 1}} \cdots W_{2, -n_{2, k_2}} \ket{0}
\end{equation}
with
\begin{equation}
n_{s, 1} \geq n_{s, 2} \geq \cdots \geq n_{s, k_s} \geq s.
\end{equation}
(see \cite{Gaberdiel:2010ar}). The genericity property discussed above is the fact that after ordering the modes in a chosen way, all these states are linearly independent for generic values of $\mathcal{W}_\infty$ parameters. As consequence of this genericity, we can immediatelly write down the character of the vacuum representation of $\mathcal{W}_\infty$ for generic value of parameters:
\begin{equation}
\label{vacvermachar}
\chi_{0}(c,q) = \Tr_{vac} q^{L_0 - \frac{c}{24}} = q^{-\frac{c}{24}} \prod_{s=2}^{\infty} \prod_{j=s}^{\infty} \frac{1}{1-q^j}
\end{equation}
For any values of parameters, this is the character of the vacuum Verma module of $\mathcal{W}_\infty$. Generically, it will be irreducible, but for special values of parameters it can develop null states. It is easier to consider these later when discussing the representation theory of $\mathcal{W}_\infty$ so we will for the moment consider only the generic case.

\paragraph{Counting primaries}
Apart from the generating set of primary fields of spin $3, 4, \ldots$ there will be many other composite primary fields. Using the character formula (\ref{vacvermachar}) and known properties of Virasoro algebra characters, we can easily write down the function that counts the Virasoro primaries. It is enough to take the full $\mathcal{W}_\infty$ vacuum character and decompose it into a sum of Virasoro Verma module characters. For generic values of the $\mathcal{W}_\infty$ parameters, these are all irreducible except for the vacuum Verma module, which has a singular vector at level 1. We have
\begin{equation}
\chi_0(c,q) = \chi_{V0}(c,q) + \sum_{h=1}^{\infty} P_h \; \chi_{V}(h,c,q)
\end{equation}
where $\chi_{V}(h,c,q)$ is the character of the Virasoro algebra Verma module with highest weight $h$,
\begin{equation}
\chi_{V}(h,c,q) = q^{h-\frac{c}{24}} \prod_{j=1}^{\infty} \frac{1}{1-q^j},
\end{equation}
$\chi_{V0}(c,q)$ is the character of the Virasoro algebra vacuum Verma module
\begin{equation}
\chi_{V0}(h,c,q) = \chi_{V}(0,c,q) - \chi_{V}(1,c,q) = q^{-\frac{c}{24}} \prod_{j=2}^{\infty} \frac{1}{1-q^j}
\end{equation}
and $P_h$ is the integer counting the number of Virasoro primeries of conformal dimension $h$. Defining a generating function for number of primaries
\begin{equation}
P(q) = \sum_{h=0}^{\infty} P_h q^h
\end{equation}
and using the previous formulas we find a compact expression for $P(q)$
\begin{equation}
\label{primarycounting}
P(q) = q + (1-q) \prod_{s=3}^{\infty} \prod_{j=s}^{\infty} \frac{1}{1-q^j} \simeq 1 + q^3 + q^4 + q^5 + 2q^6 + 2q^7 + 5q^8 + 6q^9 + 11q^{10} + 14q^{11} + 26q^{12} + \cdots
\end{equation}
We see that the first composite primary field has dimension $6$ and is obtained from two spin $3$ fields. The same happens for the spin $7$ field composed of spin $3$ and spin $4$ fields. At dimension $8$, however, there are $5$ primary fields. One of them is the `elementary' primary field $W_8(z)$. Two of them are composites of $[W_3 W_5]$ and $[W_4 W_4]$. The remaining two are schematically $[W_3 W_4]^{(1)}$ and $[W_3 W_3]^{(2)}$ \footnote{Here these brackets are used symbolically to denote the composite primary fields. The primary projection of the normal ordered product was defined explicitly around (\ref{normalprimary}) only for two primary fields. For composite primaries involving derivatives we need to extract terms deeper in the regular part. Examples are given in appendix \ref{appendixcomposite}}. We can summarize this discussion by the following table:
\begin{center}
\begin{tabular}{|c|c|c|}
\hline
spin & count & composition of the primary field \\
\hline
0 & 1 & $\mathbbm{1}$ \\
3 & 1 & $W_3$ \\
4 & 1 & $W_4$ \\
5 & 1 & $W_5$ \\
6 & 2 & $W_6, [W_3 W_3]$ \\
7 & 2 & $W_7, [W_3 W_4]$ \\
8 & 5 & $W_8, [W_3 W_5], [W_4 W_4], [W_3 W_4]^{(1)}, [W_3 W_3]^{(2)}$ \\
9 & 6 & $W_9, [W_3 W_6], [W_4 W_5], [W_3 W_5]^{(1)}, [W_3 W_4]^{(2)}, [W_3 W_3 W_3]$ \\
\hline
\end{tabular}
\end{center}
Note that to understand better what $W_s$ primaries the composite primaries are made of, one can use the more refined counting function
\begin{equation}
\prod_{s=2}^{\infty} \prod_{j=s}^{\infty} \frac{1}{1-w_s q^j}
\end{equation}
(with $w_2=1$). For example, at order $9$ in $q$ we find the coefficient
\begin{equation}
w_9 + w_3 w_6 + w_4 w_5 + w_3 w_5 + w_3 w_4 + w_3 w_3 w_3
\end{equation}
which is simply reflected by the last line of the table above.

\subsubsection{Bootstrap and OPE associativity}
Having understood what are the primary fields of our algebra, we are ready to use \texttt{OPEdefs} to find the constraints on $\mathcal{W}_\infty$ structure constants following from the associativity of the operator algebra. In fact, we will use the extension of \texttt{OPEdefs} called \texttt{OPEconf} which is useful for working with Virasoro algebra and Virasoro primary fields. One starts by writing the ansatz for the singular part of OPE of the primary fields. In our case, it can be symbolically written as
\begin{eqnarray}
\label{primaryansatz1}
\nonumber
W_3 W_3 & \sim & C_{33}^0 \mathbbm{1} + C_{33}^4 W_4 \\
\nonumber
W_3 W_4 & \sim & C_{34}^3 W_3 + C_{34}^5 W_5 \\
\nonumber
W_3 W_5 & \sim & C_{35}^4 W_4 + C_{35}^6 W_6 + C_{35}^{[33]} [W_3 W_3] \\
W_4 W_4 & \sim & C_{44}^0 \mathbbm{1} + C_{44}^4 W_4 + C_{44}^6 W_6 + C_{44}^{[33]} [W_3 W_3]
\end{eqnarray}
The structure constants $C_{jk}^l$ do not depend on the choice of the expansion point of the normal ordering prescription, since these differ by derivatives of operators, which are Virasoro descendants. So the structure constants have symmetry
\begin{equation}
\label{opepsymmetry}
C_{kj}^l = (-1)^{h_j+h_k-h_l} C_{jk}^l
\end{equation}
In general, it is known that there is a $\mathbbm{Z}_2$ symmetry of $\mathcal{W}_\infty$ flipping the sign of odd spin generators,
\begin{equation}
W_s \to (-1)^s W_s,
\end{equation}
but in (\ref{primaryansatz1}) it is a consequence of the symmetry of the OPE under the exchange of the two operators (\ref{opepsymmetry}).

There is an obvious freedom to rescale the $W_s$ generators but in fact we can also add to any new primary field, say $W_6$, any combination of composite primaries of the same spin, in this case just $[W_3 W_3]$. Doing this changes the coefficient of identity of $W_6$ with itself (the two-point function on a sphere) as well as with $[W_3 W_3]$. There is a natural condition that can be used to fix this shift symmetry which is to require $W_s$ to have zero two-point function with other primaries of the same dimension. Equivalently, we can ask $W_s$ to have zero coefficient of identity with all composite primaries of spin $s$.

Unfortunatelly, fixing this shift symmetry using the orthogonality of two-point function has certain disadvantages. First of all, computing directly the two-point function of primary operators requires us to compute OPE of $W_j(z) W_k(w)$ for higher $j+k$. Second, and more importantly, fixing the shift symmetry in this way does not considerably simplify the structure constants. For this reason, when performing the computations, we left the shift symmetry unfixed. But later, when comparing the results of the computations in the quadratic basis of $\mathcal{W}_{\infty}$, we will be using this two-point function orthogonality condition.

As for the normalization of $W_s$, the original convention dating back to \cite{Zamolodchikov:1985wn} is to have
\begin{equation}
W_s(z) W_s(w) \sim \frac{c/s}{(z-w)^{2s}} + \cdots
\end{equation}
This choice of normalization proportional to the central charge $c$ removes some singularities of the structure constants but leaves others. In fact, in the context of $\mathcal{W}_{\infty}$ this normalization does not seem to be very well motivated. For this reason, we will leave also the normalization constant $C_{ss}^0$ free during our computations.

Finally we are ready to compute the first implications of Jacobi identities. The first two lines of ansatz (\ref{primaryansatz1}) let us find the expression for primary operators $[W_3 W_3]$ and $[W_3 W_4]$ and we find
\begin{multline}
[W_3 W_3] = (W_3 W_3) + C_{33}^4 \left( - \frac{22(T W_4)}{3 (c+24)} -\frac{(5 c+76) W_4^{\prime\prime}}{36 (c+24)} \right) \\
+ C_{33}^0 \left( -\frac{6 (67c^2+178c-752) (T^{\prime\prime}T)}{c (2c-1)(5c+22)(7c+68)} -\frac{3(225c^2+1978c+776) (T^\prime T^\prime)}{2c(2c-1)(5c+22)(7c+68)} \right. \\
\left. - \frac{16(191c+22) (T(TT))}{c(2c-1)(5c+22)(7c+68)} -\frac{(c-8) (5c^2+60c+4) T^{(4)}}{2c(2c-1)(5c+22)(7c+68)} \right) \\
\end{multline}
and
\begin{multline}
[W_3 W_4] = (W_3 W_4) + C_{34}^5 \left( -\frac{94 (T W_5)}{11c+350} -\frac{(c+19) W_5''}{11+350} \right) + C_{34}^4 \left( -\frac{(5c+22) (313 c^2+5783 c+2964) (T^\prime W_3^\prime)}{36 (c+2) (c+23) (5 c-4) (7 c+114)} \right. \\
-\frac{(437c^3+9089 c^2+22454 c-76152) (T^{\prime\prime} W_3)}{12 (c+2)(c+23)(5c-4)(7c+114)} -\frac{(355c^3-329c^2-52214c-12072) (T W_3^{\prime\prime})}{18(c+2)(c+23)(5c-4)(7c+114)} \\
\left. -\frac{4(257c+83) (T(T W_3))}{(c+23)(5c-4)(7c+114)} -\frac{(25c^4-930c^3-17157c^2+115358c+26904) W_3^{(4)}}{432(c+2)(c+23)(5c-4)(7c+114)} \right).
\end{multline}
Although the coefficients look very complicated, they are all fixed by the Virasoro algebra. Every pair of parentheses contains all the descendants at given level, multiplied by the corresponding $\beta$ coefficients. Using \texttt{OPEconf} one only needs to call the function \texttt{OPEPPole} to arrive at this result. Now that we know all the fields appearing on the right hand side of (\ref{primaryansatz1}), we can find what the Jacobi identities imply for the structure constants appearing in (\ref{primaryansatz1}). The Jacobi identities for tripple $(W_3 W_3 W_3)$ do not give anything, but already $(W_3 W_3 W_4)$ give us five interesting relations. Note that we need to know OPE of $W_j(z)$ and $W_k(w)$ up to $j+k=8$ to be able to check this Jacobi identity. The resulting relations are
\begin{eqnarray}
\nonumber
C_{34}^3 & = & \frac{C_{33}^4 C_{44}^0}{C_{33}^0} \\
\nonumber
C_{44}^4 & = & \frac{3(c+3)}{(c+2)} \frac{C_{33}^4 C_{44}^0}{C_{33}^0} -\frac{288(c+10)}{c(5c+22)} \frac{C_{33}^0}{C_{33}^4} \\
C_{35}^4 & = & \frac{5 (c+7)(5c+22)}{(c+2)(7c+114)} \frac{(C_{33}^4)^2 C_{44}^0}{C_{33}^0 C_{34}^5} -\frac{60}{c} \frac{C_{33}^0}{C_{34}^5} \\
\nonumber
C_{44}^6 & = & \frac{4}{5} \frac{C_{34}^5 C_{35}^6}{C_{33}^4} \\
\nonumber
C_{44}^{[33]} & = & \frac{30(5c+22)}{(c+2)(7c+114)} \frac{C_{44}^0}{C_{33}^0} +\frac{4}{5} \frac{C_{34}^5 C_{35}^{[33]}}{C_{33}^4}
\end{eqnarray}
One can proceed further to compute the OPE of $W_j$ and $W_k$ with higher and higher $j+k$. We reached $j+k \leq 10$ and Jacobi identities were satisfied for $j+k+l \leq 12$. Apart from OPEs given in (\ref{primaryansatz1}) we have
\begin{eqnarray}
\label{primaryansatz2}
\nonumber
W_3 W_6 & \sim & C_{36}^3 W_3 + C_{36}^5 W_5 + C_{36}^7 W_7 + C_{36}^{[34]} [W_3 W_4] + C_{36}^{[34]^\prime} [W_3 W_4]^{(1)} \\
\nonumber
W_4 W_5 & \sim & C_{45}^3 W_3 + C_{45}^5 W_5 + C_{45}^7 W_7 + C_{45}^{[34]} [W_3 W_4] + C_{45}^{[34]^\prime} [W_3 W_4]^{(1)} \\
\nonumber
W_3 W_7 & \sim & C_{37}^4 W_4 + C_{37}^6 W_6 + C_{37}^{[33]} [W_3 W_3] + C_{37}^8 W_8 + C_{37}^{[35]} [W_3 W_5] \\
& & + C_{37}^{[44]} [W_4 W_4] + C_{37}^{[33]^{\prime\prime}} [W_3 W_3]^{(2)} + W_{37}^{[35]^\prime} [W_3 W_5]^{(1)} \\
\nonumber
W_4 W_6 & \sim & C_{46}^4 W_4 + C_{46}^6 W_6 + C_{46}^{[33]} [W_3 W_3] + C_{46}^8 W_8 + C_{46}^{[35]} [W_3 W_5] \\
\nonumber
& & + C_{46}^{[44]} [W_4 W_4] + C_{46}^{[33]^{\prime\prime}} [W_3 W_3]^{(2)} + W_{46}^{[35]^\prime} [W_3 W_5]^{(1)} \\
\nonumber
W_5 W_5 & \sim & C_{55}^0 \mathbbm{1} + C_{55}^4 W_4 + C_{37}^6 W_6 + C_{55}^{[33]} [W_3 W_3] + C_{55}^8 W_8 + C_{55}^{[35]} [W_3 W_5] \\
\nonumber
& & + C_{55}^{[44]} [W_4 W_4] + C_{55}^{[33]^{\prime\prime}} [W_3 W_3]^{(2)} + W_{55}^{[35]^\prime} [W_3 W_5]^{(1)}
\end{eqnarray}
Although this looks complicated, the application of Jacobi identities proceeds exactly in the same way as before. Rather than giving all the relation found in this way, let us count the number of free structure constants in the ansatz and number of relations found from the Jacobi identities. In (\ref{primaryansatz1}) and (\ref{primaryansatz2}) we have $46$ unknown coefficients. Solving Jacobi to order $12$ determines $34$ of them. But exactly $11$ of them correspond to field redefinitions and normalization. These can be chosen to be for example
\begin{equation}
C_{33}^0, C_{44}^0, C_{34}^5, C_{35}^6, C_{36}^7, C_{37}^8, C_{35}^{[33]}, C_{36}^{[34]}, C_{37}^{[35]}, C_{37}^{[44]}, C_{37}^{[33]^{\prime\prime}}
\end{equation}
So in total there remains $46-34-11 = 1$ combination of coefficients that is independent of field redefinitions and rescalings. One possible choice is to take
\begin{equation}
\label{opexdef}
x^2 = \frac{(C_{33}^4)^2 C_{44}^0}{(C_{33}^0)^2}.
\end{equation}
Together with the central charge $c$ which appears already in the Virasoro subalgebra, we find two-parametric family of algebras, as was first found in \cite{Gaberdiel:2012ku}. For completeness, the values of the structure constants that we found are given in appendix \ref{appendixprimary}.

\subsubsection{Connection to $\mathcal{W}_N$}

One knows from the coset construction or from the free field representation, that for each integer $N \geq 3$ there exists a one-parametric family of algebras extending the Virasoro algebra by generating primary fields of dimension $3, 4, \ldots, N$. For each fixed $N$ the additional parameter is just the central charge. For example, $\mathcal{W}_3$ of Zamolodchikov \cite{Zamolodchikov:1985wn} is the simplest algebra of this family.

It is clear from the construction of $\mathcal{W}_\infty$ above that by putting all fields $W_s$ with $s>N$ to zero we arrive at ansatz for $\mathcal{W}_N$ algebra. So there should be a discrete number of values for $x^2$ such that when $x^2$ takes one of these values, the $\mathcal{W}_\infty$ can be truncated to $\mathcal{W}_N$. To determine these values, Gaberdiel and Gopakumar in \cite{Gaberdiel:2012ku} used the representation theory of $\mathcal{W}_N$ and found
\begin{equation}
\label{opexton}
x^2 = \frac{144(c+2)(N-3)(c(N+3)+2(4N+3)(N-1))}{c(5c+22)(N-2)(c(N+2)+(3N+2)(N-1))}.
\end{equation}
Another possibility is to use the free field representation of $\mathcal{W}_\infty$ of the second part of this article to compute $x^2$. For  lower values of $N$ the formula (\ref{opexton}) can be verified also by comparing directly to $\mathcal{W}_N$. Either way, there is an important fact that Gaberdiel and Gopakumar noticed. Since all the structure constants of $\mathcal{W}_\infty$ that were computed up to dimension $12$ of Jacobi identities are algebraic function of $c$, $x^2$ and normalization-dependent coefficients (see formulas in appendix \ref{appendixprimary}), choosing a different value of parameter $N$ in (\ref{opexton}) gives exactly the same $\mathcal{W}_\infty$ as long as $c$ and $x^2$ are the same. Since equation for $N$ in terms of $c$ and $x^2$ is a cubic equation for $N$, there are generically three different values of $N$ at fixed $c$ which correspond to identical algebra. The equations for three solutions of (\ref{opexton}) as an equation for $N$ are quite complicated, but one can instead try to compute $c$ and $x^2$ in terms of these three roots. Let us denote the three roots by $\lambda_1, \lambda_2$ and $\lambda_3$. What we find is the set of equations
\begin{eqnarray}
\label{lambdaquadratic}
0 & = & \lambda_1 \lambda_2 + \lambda_1 \lambda_3 + \lambda_2 \lambda_3 \\
c & = & (\lambda_1 - 1) (\lambda_2 - 1) (\lambda_3 - 1) \\
x^2 & = & \frac{144(c+2)(\lambda_1-3)(\lambda_2-3)(\lambda_3-3)}{c(5c+22)(\lambda_1-2)(\lambda_2-2)(\lambda_3-2)}
\end{eqnarray}
These equations are manifestly symmetric in three roots. Also all the zeros and poles of $x^2$ in this parametrization have a representation-theoretic meaning. For example the poles at $\lambda = 2$ are consequences of the possibility of truncation of $\mathcal{W}_{\infty}$ to Virasoro subalgebra for $\lambda = 2$. Zeros at $\lambda = 3$ similarly represent the possible truncation of $\mathcal{W}_{\infty}$ to $\mathcal{W}_3$. The value of central charge $c = -22/5$ is the Lee-Yang singularity minimal model that is at the same time minimal model of Virasoro algebra and $\mathcal{W}_3$. More discussion about $\mathcal{W}_{\infty}$ will follow later.

\subsubsection{Universality}

Since the relationship between $\mathcal{W}_{\infty}(\lambda)$ and $\mathcal{W}_N$ is important, let's compare it to the construction of $\mathfrak{hs}(\lambda)$ which is analogous but simpler. \footnote{There are various connections between $\mathcal{W}_{\infty}$ and $\mathfrak{hs}(\lambda)$. For example, the Drinfeld-Sokolov procedure applied to $\mathfrak{hs}(\lambda)$ produces $\mathcal{W}_{\infty}$ and in the other direction, in the limit of large central charge the vacuum-preserving subalgebra of $\mathcal{W}_{\infty}$ contains $\mathfrak{hs}(\lambda)$ \cite{Bowcock:1991zk, Perlmutter:2012ds}} Let us start with the Lie algebra $\mathfrak{sl}(2)$,
\begin{eqnarray}
\left[ J_0, J_{\pm} \right] & = & \pm J_{\pm} \\
\left[ J_+, J_- \right] & = & 2J_0
\end{eqnarray}
and its $N$-dimensional irreducible representation. This naturally gives us an embedding of $\mathfrak{sl}(2) \subset \End(N)$ under which the space $\End(N)$ of $N \times N$ matrices decomposes as
\begin{equation}
\End(N) \simeq 1 + 3 + 5 + \cdots + (2N-1).
\end{equation}
Here we denote the irreducible representations of $\mathfrak{sl}(2)$ by their dimensions. This means that there exists a basis $T^l_m$ ($l=0,1,\ldots,N-1$ and $m=-l,\ldots,l$) of $\End(N)$ with the subspace spanned by $T^l_m$, $m=-l,\ldots,l$ being a $(2l+1)$-dimensional irreducible representation of $\mathfrak{sl}(2)$ and $m$ is the eigenvalue under $J_0$. If choose the normalization of $T^l_m$ suitably, we can write the associative product of $\End(N)$ in the form \footnote{The structure constants $F$ are in fact zero unless $m_1+m_2=m$ because of the $U(1)$ symmetry generated by $L_0$ but this is not important.}
\begin{equation}
T^{l_1}_{m_1} \star T^{l_2}_{m_2} = \sum_{l=0}^{l_1+l_2} \sum_{m=-l}^l F^{l_1 l_2 l}_{m_1 m_2 m}(N) T^l_m
\end{equation}
The explicit expressions for the structure constants in terms of $3j$ and $6j$ symbols are given in \cite{Pope:1989sr, Pope:1990kc}. What is important is that as $N$ varies, the structure constants $F$ are rational functions of $N$. This allows us to consider the universal associative algebra $\mathcal{A}(\lambda)$ defined on the vector space of $T^l_m$, $l = 0, 1, 2, \ldots$; $m = -l, \ldots, l$ with the same structure constants but no longer restricting $\lambda$ to be a positive integer,
\begin{equation}
\label{hsstarproduct}
T^{l_1}_{m_1} \star T^{l_2}_{m_2} = \sum_{l=0}^{l_1+l_2} \sum_{m=-l}^l F^{l_1 l_2 l}_{m_1 m_2 m}(\lambda) T^l_m
\end{equation}
To get back from $\mathcal{A}(\lambda)$ (which as a vector space has an infinite dimension) to finite-dimensional $\End(N)$, one realizes that for $\lambda$ equal to positive integer $N$, there is an ideal in associative algebra $\mathcal{A}(\lambda)$ generated by $T^l_m$ with $l \geq N$ and we get back $\End(N)$ if we factor this ideal out.

There is a simple one-line construction of $\mathcal{A}(\lambda)$ as a quotient of the universal enveloping algebra of $\mathfrak{sl}(2)$ by the Casimir ideal,
\begin{equation}
\mathcal{A}(\lambda) = \frac{\mathcal{U}(\mathfrak{sl}(2))}{\mathcal{I}(J_0^2 + \frac{1}{2} J_+ J_- + \frac{1}{2} J_- J_+ - \frac{\lambda^2-1}{4})}
\end{equation}
It is quite easy to see that as vector space, the right hand side decomposes into direct sum of all odd-dimensional irreducible representations of $\mathfrak{sl}(2)$, each with multiplicity $1$. Furthermore, taking $N$-dimensional irreducible representation of $\mathfrak{sl}(2)$ gives us map from the universal enveloping algebra to $\End(N)$ and if $\lambda = \pm N$ this map is compatible with the quotient, so the kernel of this map is exactly the ideal that we need to quotient out to get from $\mathcal{A}(N)$ to $\End(N)$. Note that both ways of constructing the algebra give us a preferred $\mathfrak{sl}(2)$ subalgebra. This $\mathfrak{sl}(2)$ and the preferred basis of $\mathcal{A}(\infty)$ (or equivalently the decomposition of $\mathcal{A}(\lambda)$ into irreducible representations of $\mathfrak{sl}(2)$) is what distinguishes $\mathcal{A}(\lambda)$ from other variants of $\mathfrak{gl}(\infty)$.

Geometrically, this construction is the quantum analogue of the construction of the 2-sphere $S^2$ embedded in $\mathbbm{R}^3$ via
\begin{equation}
X^2 + Y^2 + Z^2 = R^2.
\end{equation}
Defining the usual $\mathfrak{su}(2)$ generators $X,Y$ and $Z$ by
\begin{eqnarray}
J_+ & = & (X + iY) \\
J_- & = & (X - iY) \\
J_0 & = & Z
\end{eqnarray}
the Casimir constraint becomes
\begin{equation}
X^2 + Y^2 + Z^2 \sim \frac{\lambda^2-1}{4}
\end{equation}
which gives us relation between the radius of the sphere and the dimension of the quantized Hilbert space (which is the usual connection between the number of states and the symplectic volume of the quantized manifold). In the large radius limit $\lambda \to \infty$ one should reduce to the classical case and in fact one can check that with proper identification of $T^l_m$ with spherical harmonics $Y^l_m$ the limit of the structure constants in (\ref{hsstarproduct}) has the leading term corresponding to multiplication of the spherical harmonics and the subleading term corresponding to their Poisson brackets induced from the natural symplectic form on $S^2$ which is the rotationaly-invariant volume form.

Usually one defines the higher spin algebra $\mathfrak{hs}(\lambda)$ to be the Lie algebra associated to the associative algebra $\mathcal{A}(\lambda)$ with the center (the zero-dimensional representation $T^0_0$) quotiented out. Eliminating this one-dimensional center is the same thing that is done for $N$ integer when going from $\mathfrak{gl}(N)$ to simple Lie algebra $\mathfrak{sl}(N)$.

The situation in $\mathcal{W}_{\infty}$ is entirely analogous. There are two infinitely generated algebras, $\mathcal{W}_{\infty}$ and $\mathcal{W}_{1+\infty}$. The first one is the extension of the Virasoro algebra by primary generators of spin $3, 4, \ldots$, while in the second algebra we also include the spin $1$ generator. The spin $1$ part of $\mathcal{W}_{1+\infty}$ can be factored out, so $\mathcal{W}_{1+\infty}$ is in fact a product of $\mathcal{W}_{\infty}$ with $\hat{\mathfrak{u}}(1)$ algebra. If we choose $N$ in (\ref{opexton}) to be an integer $\geq 2$, the algebra $\mathcal{W}_{\infty}$ develops a large ideal generated by $W_j(z)$ with $j > N$ and factoring this ideal out, we get back to $\mathcal{W}_N$. The structure constants of $\mathcal{W}_N$ family of algebras are (with suitable normalization of the fields) again rational functions of $c$ and $N$ and $\mathcal{W}_{\infty}$ can be obtained by `analytic continuation'. In this sense, working with $\mathcal{W}_{\infty}$ is equivalent to working with all $\mathcal{W}_N$ at the same time. But the triality symmetry typically connects $\mathcal{W}_N$ to $\mathcal{W}_{\infty}$ with rational or negative values of $\lambda$ (just like the special case of the triality, the level-rank duality \cite{Altschuler:1990th, Kuniba:1990zh}) so working with the full $\mathcal{W}_{\infty}$ reveals more symmetry than if we only work with individual $\mathcal{W}_N$.

\section{$\mathcal{W}_{1+\infty}$ in the quadratic basis}

In the previous section we studied the $\mathcal{W}_\infty$ in the basis of the Virasoro primary fields. The main advantage of that approach was that it was quite easy at lower spins to see that there is in fact a two-parametric family of algebras and furthermore that the connection between `physical' parameters, the central charge $c$ and the rank parameter $N$, is $3:1$ - the triality symmetry.

Looking at the structure constants in OPE of lower spin fields, the systematic understanding of these as rational functions of parameters seems to be very difficult. The main problem is that we are decomposing a big algebra under a small Virasoro subalgebra. The number of possible primaries is growing roughly as the number of plane partitions (more precisely it is given by the counting function (\ref{primarycounting})). But more important than the large number of coefficients is the fact that there is no easy and systematic way to enumerate them which would make the structure constants `nice' and canonical.

Furthermore, the operators appearing in the OPE of primary operators $W_j(z) W_k(w)$ have unbounded nonlinearity with respect to basic fields $W_l(w)$. In general, there is no reason for the chiral algebra to be linear. The reason why simple algebras like Virasoro algebra of affine Lie algebras are linear is that in those cases one is considering only low spins and small field content. Quite often algebras of higher spins have certain restriction on spin interactions which in our case takes this form: the singular part of OPE of $X_1(z)$ and $X_2(w)$ with spins $s_1$ and $s_2$ only contains spins up to
\begin{equation}
s \leq s_1 + s_2 - 1
\end{equation}
which can be even stronger if there is a $\mathbbm{Z}_2$ symmetry. So for algebras with generators of low spin the composite fields have too high spin to appear in the singular part of the OPE. But already some of the superconformal algebras with more supersymmetries \cite{Bershadsky:1986ms, Knizhnik:1986wc} are nonlinear.

Luckily, in $\mathcal{W}_\infty$ there exists a different choice of the generating fields which are not primary but whose OPE has only quadratic nonlinearity. This will significantly reduce the number of possible terms on the right hand side of the OPE as well as possible field redefinitions. The simplest way to obtain these generating fields is through the free field representation of $\mathcal{W}_N$ algebras. This will be our starting point of this section. Using this free field representation, we will determine OPE of generating fields $U_j(z) U_k(w)$ with $j = 1, 2$. Basically, by doing this we will fix how the fields $U_k(w)$ transform under the $\hat{\mathfrak{u}}(1)$ and Virasoro subalgebras of $\mathcal{W}_{1+\infty}$. Having fixed this, all other OPE can be determined using the associativity conditions of OPE and quadraticity of this basis.

Having determined the OPE $U_j(z) U_k(w)$ up to $j+k \leq 15$ we can analyze the structure constants. There is still a certain ambiguity between the various normal ordering prescriptions. Instead of the usual normal ordering prescription (\ref{normalorder}) we will implicitly introduce a new set of bilocal fields (\ref{opefullform}). These fields will absorb all the terms with derivatives of the usual operator product expansion. Their introduction will allow us to specify the $\mathcal{W}_{1+\infty}$ in different form than is usually done (\ref{opequadinv}) and this new form will let us quickly derive the commutation relations of the modes of the $U$-fields. Also the correlation functions of $U$-operators on a sphere are recursively computable.

Starting from the Miura representation of $U$-fields in terms of the free fields, we will find a coproduct in $\mathcal{W}_{1+\infty}$, which will represent $\mathcal{W}_{1+\infty}$ with parameters $(\alpha_0,N_1+N_2)$ in the product of $\mathcal{W}_{1+\infty}$ with parameters $(\alpha_0,N_1)$ and $(\alpha_0,N_2)$. There are many non-trivial identities for structure constants $C_{jk}^{lm}(\alpha_0,N)$ that must hold in order for this coproduct to be consistent. We will find the linear combinations of $U$-fields and their derivatives which transform as quasiprimary fields with respect to the stress-energy tensor $T_{1+\infty}$ and describe how to construct the primary fields. Using these primary fields we can connect to the primary basis computation of the previous section and verify consistency of these computations. Finally, we discuss how the triality symmetry which is manifest in the primary basis implies the existence of other two $U$-bases. The transformation between primary and quadratic $U$-bases is non-linear and the same is true for the triality action exchanging various $U$-bases.

\subsection{Free field representation}
The starting point of this section is the free field representation of $\mathcal{W}_{1+N}$ in terms of $N$ free bosons - the Miura transformation \cite{Bouwknegt:1992wg, Fateev:1987zh, Lukyanov:1987xg}. We define the $\hat{\mathfrak{u}}(1)$ currents
\begin{equation}
J_j(z) = i \partial \phi_j(z)
\end{equation}
with OPE
\begin{equation}
J_j(z) J_k(w) \sim \frac{\delta_{jk}}{(z-w)^2}
\end{equation}
and an operator $R(z)$
\begin{equation}
\label{miurar}
\boxed{R(z) = \; : \prod_{j=1}^{N} \Big( \alpha_0 \partial + J_j(z) \Big) : = \sum_{k=0}^N U_k(z) (\alpha_0 \partial)^{N-k}}
\end{equation}
The double dots denote the normal ordering of free fields and $\alpha_0$ is a parameter which we will soon relate to the central charge of the algebra. Classically, this expression would correspond to a factorization of $N$-th order differential operator as a product of differential operators of the first order. For our purposes, it represents the generators of $\mathcal{W}_{1+N}$, $U_j(z)$, in terms of the free fields $J_k(z)$. It is not obvious from this definition that the chiral algebra generated by $U_j(z)$ closes, i.e. that all singular terms of OPE of $U_j(z)$ can be expressed only in terms of derivatives and normal ordered products of $U_j$, but this has been proved in \cite{Fateev:1987zh, Lukyanov:1987xg, Lukyanov1988}.

For an illustration, we can list the first few $U_j(z)$ explicitly:
\begin{eqnarray}
U_0 & = & \mathbbm{1} \\
U_1 & = & \sum_{j=1}^N J_j \\
U_2 & = & \sum_{j<k} :J_j J_k : + \alpha_0 \sum_{j=1}^N (j-1) J_j^\prime \\
U_3 & = & \sum_{j<k<l} :J_j J_k J_l: + \alpha_0 \sum_{j<k} (j-1) :J_j^\prime J_k: \\
\nonumber
& & + \alpha_0 \sum_{j<k} (k-2) :J_j J_k^\prime: + \frac{\alpha_0^2}{2} \sum_{j=1}^N (j-1)(j-2) J_j^{\prime\prime} \\
U_4 & = & \sum_{j<k<l<m} :J_j J_k J_l J_m: + \frac{\alpha_0^3}{6} \sum_j (j-1)(j-2)(j-3) J_j^{\prime\prime\prime} \\
\nonumber
& & + \alpha_0 \sum_{j<k<l} (j-1) :J_j^\prime J_k J_l: + (k-2) J_j J_k^\prime J_l: + (l-3) :J_j J_k J_l^\prime: \\
\nonumber
& & + \frac{\alpha_0^2}{2} \sum_{j<k} (j-1)(j-2) :J_j^{\prime\prime} J_k: + 2(j-1)(k-3) :J_j^\prime J_k^\prime: + (k-2)(k-3) :J_j J_k^{\prime\prime}
\end{eqnarray}

\subsection{$\hat{\mathfrak{u}}(1)$ and Virasoro subalgebras}
\label{secu1virasoro}

Let us now focus on the subalgebra generated by $U_1(w)$ and $U_2(w)$. Using the Wick theorem for free fields, we can compute
\begin{eqnarray}
U_1(z) U_1(w) & \sim & \frac{N}{(z-w)^2} \\
U_1(z) U_2(w) & \sim & \frac{N(N-1)\alpha_0}{(z-w)^3} + \frac{(N-1) U_1(w)}{(z-w)^2} \\
\label{opeu2u2}
U_2(z) U_2(w) & \sim & \frac{N(N-1)(1-2(2N-1)\alpha_0^2)}{2(z-w)^4} + \frac{-2U_2(w)}{(z-w)^2} + \frac{(N-1)(U_1 U_1)(w)}{(z-w)^2} \\
\nonumber
& & + \frac{\alpha_0 N(N-1) U_1^\prime(w)}{(z-w)^2} - \frac{U_2^\prime(w)}{z-w} + \frac{(N-1)(U_1^\prime U_1)(w)}{z-w} + \frac{\alpha_0 N(N-1) U_1^{\prime\prime}(w)}{2(z-w)}
\end{eqnarray}

\paragraph{$\hat{\mathfrak{u}}(1)$ subalgebra}
Since there is a unique dimension $1$ field in $\mathcal{W}_{1+\infty}$, $U_1(z)$, there are no fields redefinitions possible apart from the rescaling. The OPE of $U_1(z)$ with itself is that of $\hat{\mathfrak{u}}(1)$ current algebra. Normalizing $U_1(z)$ so that it has a $N$-independent OPE,
\begin{equation}
J(z) = -\frac{1}{\sqrt{N}} U_1(z)
\end{equation}
we see that
\begin{equation}
J(z) J(w) \sim \frac{1}{(z-w)^2}
\end{equation}

\paragraph{Virasoro subalgebras}
At dimension $2$ we have a three dimensional space of local operators. We can look for linear combinations of these which satisfy OPE of the Virasoro algebra. There are two one-parametric families and one discrete solution: one family is the Sugawara stress-energy tensor corresponding to $J(z)$ with background charge modification,
\begin{equation}
T_1(z) \equiv \frac{1}{2} (JJ)(z) + \beta J^\prime(z)
\end{equation}
which satisfies the Virasoro algebra with central charge
\begin{equation}
c_1(\beta) = 1 - 12 \beta^2 N.
\end{equation}
The discrete solution is the stress-energy tensor of $\mathcal{W}_\infty$ which has zero OPE with $J(z)$
\begin{equation}
T_{\infty}(z) = -U_2(z) + \frac{(N-1)\alpha_0}{2} U_1^\prime(z) + \frac{N-1}{2N} (U_1 U_1)(z)
\end{equation}
and satisfies Virasoro algebra with central charge
\begin{equation}
c = (N-1)\big(1-N(N+1)\alpha_0^2\big).
\end{equation}
The sum of these stress-energy tensors gives us the other one-parametric family of Virasoro algebras with central charge which is the sum of their central charges. In this family there is a natural stress-energy tensor from point of view of $\mathcal{W}_{1+\infty}$,
\begin{equation}
\label{t1inf}
T_{1+\infty}(z) = -U_2(z) + \frac{(N-1)\alpha_0}{2} U_1^\prime(z) + \frac{1}{2} (U_1 U_1)(z)
\end{equation}
The central charge of this stress-energy tensor is
\begin{equation}
c_{1+\infty} = c + c_1(0) = c+1 = N\big(1-(N-1)(N+1)\alpha_0^2\big).
\end{equation}
Everytime we discuss the scaling dimensions of the various fields, we mean the `engineering' dimensions, which are $j$ for $U_j(z)$, dimension $1$ for each derivative and which are additive under normal products.\footnote{It is this counting of dimensions that gives us the nice formula for the vacuum character of $\mathcal{W}_{1+\infty}$ as MacMahon function \cite{Gaberdiel:2010ar}.} The quadratic pole of $T_{1+\infty}(z)$ with any field of definite scaling dimension $A(w)$ measures exactly this scaling dimension (this is not true for instance for $T_\infty$ which assings dimension $0$ to $U_1(w)$ and with respect to which the other fields $U_j(w)$ do not even have definite scaling dimensions).

In the following (especially when discussing the triality), we will need to compare properties of $\mathcal{W}_{1+\infty}$ computed in primary and quadratic bases. Since we are using different parametrization of $\mathcal{W}_{\infty}$ in both cases, we will fix the following convention: the parameter $N$ of both primary basis computation and quadratic basis computation will be the same and equal to the first root $\lambda_1$. The central charge $c$ in the primary basis computation must match the central charge of stress-energy tensor which commutes with $U_1(z)$ (since restricting to fields commuting with $U_1(z)$ takes us from $\mathcal{W}_{1+\infty}$ to $\mathcal{W}_{\infty}$). This means that
\begin{equation}
\label{ctoalpha}
c = (\lambda_1-1)(\lambda_2-1)(\lambda_3-1) = (N-1)(1-N(N+1)\alpha_0^2)
\end{equation}
so that
\begin{equation}
\label{alphatolambda}
\alpha_0^2 = -\frac{(\lambda_2 + \lambda_3)^2}{\lambda_2 \lambda_3} = -\frac{\lambda_2 \lambda_3}{\lambda_1^2}
\end{equation}
We also see that the combination
\begin{equation}
N^3 \alpha_0^2 = -\lambda_1 \lambda_2 \lambda_3
\end{equation}
is triality invariant.

\subsubsection{OPE of higher spin fields with $U_1(z)$ and $U_2(z)$}
Having found the $\hat{\mathfrak{u}}(1)$ and Virasoro subalgebras of $\mathcal{W}_{1+\infty}$, we can now use the free field Wick theorem to compute OPE of $U_1(z)$ and $U_2(z)$ with $U_k(w)$. We have
\begin{eqnarray}
U_1(z) U_1(w) & \sim & \frac{N}{(z-w)^2} \\
U_1(z) U_2(w) & \sim & \frac{N(N-1)\alpha_0}{(z-w)^3} + \frac{(N-1) U_1(w)}{(z-w)^2} \\
U_1(z) U_3(w) & \sim & \frac{N(N-1)(N-2)\alpha_0^2}{(z-w)^4} + \frac{(N-1)(N-2)\alpha_0 U_1(w)}{(z-w)^3} + \frac{(N-2) U_2(w)}{(z-w)^2} \\
U_1(z) U_4(w) & \sim & \frac{N(N-1)(N-2)(N-3)\alpha_0^3}{(z-w)^5} + \frac{(N-1)(N-2)(N-3)\alpha_0^2 U_1(w)}{(z-w)^4} \\
\nonumber
& & + \frac{(N-2)(N-3)\alpha_0 U_2(w)}{(z-w)^2} + \frac{(N-3) U_3(w)}{(z-w)^2}
\end{eqnarray}
and in general
\begin{equation}
\label{opeu1uk}
U_1(z) U_k(w) \sim \sum_{l=0}^{k-1} \frac{(N-l)!\alpha_0^{k-1-l}}{(N-k)!} \frac{U_l(w)}{(z-w)^{k+1-l}}.
\end{equation}
One can verify that this form of OPE satisfies the $(U_1 U_1 U_j)$ Jacobi identity as it should. Computation of OPE of $U_2(z) U_k(w)$ using combinatorics of free field Wick contractions is more complicated. One can simplify things slightly using the Newton identities from the theory of symmetric polynomials. It is simpler to compute first OPE of $T_{1+\infty}(z)$ with $U_k(w)$. Using free fields, we find
\begin{equation}
T_{1+\infty}(z) = \frac{1}{2} \sum_{j=1}^N :J_j(z) J_j(z): + \frac{\alpha_0}{2} \sum_{j=1}^N (N+1-2j) J_j^\prime(z)
\end{equation}
and
{\small \begin{eqnarray}
T_{1+\infty}(z) U_1(w) & \sim & \frac{U_1(w)}{(z-w)^2} + \frac{U_1^\prime(w)}{z-w} \\
T_{1+\infty}(z) U_2(w) & \sim & \frac{\frac{1}{2}(N-1)N(N+1)\alpha_0^2}{(z-w)^4} + \frac{(N-1)\alpha_0 U_1(w)}{(z-w)^3} + \frac{2U_2(w)}{(z-w)^2} + \frac{U_2^\prime(w)}{z-w} \\
\nonumber
T_{1+\infty}(z) U_3(w) & \sim & \frac{(N+1)N(N-1)(N-2)\alpha_0^3}{(z-w)^5} + \frac{\frac{1}{2}(N+3)(N-1)(N-2)\alpha_0^2U_1(w)}{(z-w)^4} \\
& & + \frac{2(N-2)\alpha_0 U_2(w)}{(z-w)^3} + \frac{3U_3(w)}{(z-w)^2} + \frac{U_3^\prime(w)}{z-w} \\
\nonumber
T_{1+\infty}(z) U_4(w) & \sim & \frac{\frac{3}{2}(N+1)N(N-1)(N-2)(N-3)\alpha_0^4}{(z-w)^6} + \frac{(N+2)(N-1)(N-2)(N-3)\alpha_0^3 U_1(w)}{(z-w)^5} \\
& & + \frac{\frac{1}{2}(N+5)(N-2)(N-3) \alpha_0^2 U_2(w)}{(z-w)^4} + \frac{3(N-3)\alpha_0 U_3(w)}{(z-w)^3} + \frac{4U_4(w)}{(z-w)^2} + \frac{U_4^\prime(w)}{z-w}.
\end{eqnarray}}
For general $k$ we have
\begin{equation}
T_{1+\infty}(z) U_k(w) \sim \sum_{l=0}^{k-1} \frac{\frac{1}{2}\left[(k-1)(N+1)-l(N-1)\right] (N-k+1)_{k-l} \alpha_0^{k-l} U_l(w)}{(z-w)^{k+2-l}} + \frac{k U_k(w)}{(z-w)^2} + \frac{U_k^\prime(w)}{z-w}.
\end{equation}
Note that the right hand side is linear in $U_l(w)$ and the only term with derivative is the simple pole. The simplicity of this result is the reason computing $T_{1+\infty}$ first. Having computed OPE of $U_1(z)$ and $T_{1+\infty}(z)$ with $U_k(w)$, we can now compute OPE of $U_2(z)$ with $U_k(w)$. Again, for first few fields we find
\begin{eqnarray}
U_2(z) U_1(w) & \sim & \frac{-N(N-1)\alpha_0}{(z-w)^3} + \frac{(N-1)U_1(w)}{(z-w)^2} + \frac{(N-1)U_1^\prime(w)}{z-w} \\
\nonumber
U_2(z) U_2(w) & \sim & \frac{\frac{1}{2}N(N-1)(1+2(1-2N)\alpha_0^2)}{(z-w)^4} \\
& & + \frac{-2U_2(w)}{(z-w)^2} + \frac{(N-1)(U_1 U_1)(w)}{(z-w)^2} + \frac{N(N-1)\alpha_0 U_1^\prime(w)}{(z-w)^2} \\
\nonumber
& & + \frac{-U_2^\prime(w)}{z-w} + \frac{(N-1)(U_1^\prime U_1)(w)}{z-w} + \frac{\frac{1}{2} N(N-1)\alpha_0 U_1^{\prime\prime}(w)}{z-w} \\
\nonumber
U_2(z) U_3(w) & \sim & \frac{N(N-1)(N-2)(1+\alpha_0^2-3N\alpha_0^2)\alpha_0}{(z-w)^5} + \frac{\frac{1}{2}(N-1)(N-2)(1-4N\alpha_0^2)U_1(w)}{(z-w)^4} \\
\nonumber
& & + N(N-1)(N-2)\alpha_0^2 \left( \frac{U_1(w)}{(z-w)^4} + \frac{U_1^\prime(w)}{(z-w)^3} + \frac{U_1^{\prime\prime}(w)}{2(z-w)^2} + \frac{U_1^{\prime\prime\prime}(w)}{6(z-w)} \right) \\
\nonumber
& & + (N-1)(N-2)\alpha_0 \left( \frac{(U_1 U_1)(w)}{(z-w)^3} + \frac{(U_1^\prime U_1)(w)}{(z-w)^2} + \frac{(U_1^{\prime\prime}U_1)(w)}{2(z-w)}\right) \\
& & + \frac{-(N-2)(N+1)\alpha_0 U_2(w)}{(z-w)^3} + \frac{-2U_3(w)}{(z-w)^2} + \frac{-U_3(w)}{(z-w)^2} + \frac{-U_3^\prime(w)}{z-w} \\
\nonumber
& & + (N-2) \left( \frac{(U_1 U_2)(w)}{(z-w)^2} + \frac{(U_1^\prime U_2)(w)}{z-w} \right) \\
\end{eqnarray}
and
{\small \begin{eqnarray}
\nonumber
U_2(z) U_4(w) & \sim & \frac{\frac{1}{2}N(N-1)(N-2)(N-3)\alpha_0^2(3+2\alpha_0^2-8N\alpha_0^2)}{(z-w)^6} + \frac{(N-1)(N-2)(N-3)\alpha_0(1-3N\alpha_0^2) U_1(w)}{(z-w)^5} \\
\nonumber
& & + N(N-1)(N-2)(N-3)\alpha_0^3 \left( \frac{U_1(w)}{(z-w)^5} + \frac{U_1^\prime(w)}{(z-w)^4} + \frac{U_1^{\prime\prime}(w)}{2(z-w)^3} + \frac{U_1^{\prime\prime\prime}(w)}{6(z-w)^2} + \frac{U_1^{\prime\prime\prime\prime}(w)}{24(z-w)} \right) \\
\nonumber
& & + \frac{\frac{1}{2}(N-2)(N-3)(1-2\alpha_0^2-4N\alpha_0^2)U_2(w)}{(z-w)^4} + \frac{-(N-3)(N+2)\alpha_0 U_3(w)}{(z-w)^3} \\
\nonumber
& & + (N-1)(N-2)(N-3)\alpha_0^2 \left( \frac{(U_1 U_1)(w)}{(z-w)^4} + \frac{(U_1^\prime U_1)(w)}{(z-w)^3} + \frac{(U_1^{\prime\prime} U_1)(w)}{2(z-w)^2} + \frac{(U_1^{\prime\prime\prime} U_1)(w)}{6(z-w)} \right) \\
\nonumber
& & + (N-2)(N-3)\alpha_0 \left( \frac{(U_1 U_2)(w)}{(z-w)^3} + \frac{(U_1^\prime U_2)(w)}{(z-w)^2} + \frac{(U_1^{\prime\prime} U_2)(w)}{z-w} \right) \\
& & + (N-3) \left( \frac{(U_1 U_3)(w)}{(z-w)^2} + \frac{(U_1^\prime U_3)(w)}{z-w} \right) - \left( \frac{U_4(w)}{(z-w)^2} + \frac{U_4^\prime(w)}{z-w} \right) + \frac{-3U_4(w)}{(z-w)^2},
\end{eqnarray}}
which is quadratic and more complicated than OPE with $T_{1+\infty}(z)$. One property of OPE that can be noticed here is that the terms with derivatives can be absorbed into OPE if we allow for bilocal normal ordering (\ref{nobilocal}),
{\small \begin{eqnarray}
\nonumber
U_2(z) U_4(w) & \sim & \frac{\frac{1}{2}N(N-1)(N-2)(N-3)\alpha_0^2(3+2\alpha_0^2-8N\alpha_0^2)}{(z-w)^6} + \frac{(N-1)(N-2)(N-3)\alpha_0(1-3N\alpha_0^2) U_1(w)}{(z-w)^5} \\
\nonumber
& & + \frac{N(N-1)(N-2)(N-3)\alpha_0^3 U_1(z)}{(z-w)^5} + \frac{\frac{1}{2}(N-2)(N-3)(1-2\alpha_0^2-4N\alpha_0^2)U_2(w)}{(z-w)^4} \\
& & + \frac{-(N-3)(N+2)\alpha_0 U_3(w)}{(z-w)^3} + \frac{(N-1)(N-2)(N-3)\alpha_0^2(U_1(z) U_1(w))}{(z-w)^4} \\
\nonumber
& & + \frac{(N-2)(N-3)\alpha_0(U_1(z) U_2(w))}{(z-w)^3} + \frac{(N-3)(U_1(z) U_3(w))}{(z-w)^2} + \frac{-U_4(z)}{(z-w)^2} + \frac{-3U_4(w)}{(z-w)^2}
\end{eqnarray}}
We will see later that only a small generalization of this will allow us to understand all terms with derivatives in OPE of arbitrary $U_j(z)$ and $U_k(w)$. The general formula for OPE of $U_2(z)$ and $U_k(w)$ analogous to (\ref{opeu1uk}) is
\begin{eqnarray}
\label{opeu2uk}
\nonumber
U_2(z) U_k(w) & \sim & \sum_{m=0}^{k-1} \frac{[N-m]_{k-m} \alpha_0^{k-m-1} \big( U_1(z) U_m(w) \big)}{(z-w)^{k-m+1}} \\
& & + \sum_{m=0}^{k-1} \frac{[N-m]_{k-m} \alpha_0^{k-m-2} \left[ \frac{k-m-1}{2} + \alpha_0^2 \left( N(m-k) - m + 1 \right) \right] U_m(w)}{(z-w)^{k-m+2}} \\
\nonumber
& & + \frac{-m U_m(w)}{(z-w)^2} + \frac{-U_m^\prime(w)}{z-w}
\end{eqnarray}
It is easy to verify that all the previous formulas are special cases of this one. More importantly, the Jacobi identities $(U_1 U_2 U_j)$ and $(U_2 U_2 U_k)$ are satisfied as one can check using \texttt{OPEdefs}.

\subsection{Higher OPE using associativity conditions}
\label{quadraticopejacobi}
As discussed above, computation of OPE $U_j(z) U_k(w)$ using the free field representation is in principle a straightforward application of Wick theorem, but in practice the combinatorics of contractions is very complicated \footnote{But see \cite{Fateev:1987zh, Lukyanov:1987xg, Lukyanov1988} where the free-field manipulations are used.}. For this reason, in this section we will use the associativity conditions of the operator algebra as implemented in \texttt{OPEdefs} (bootstrap approach) to compute the operator product expansion of $U_j(z)$ fields.

Finding the most general algebra with given set of generators is more efficiently done in the basis of generators which are primaries with respect to Virasoro algebra \cite{Gaberdiel:2012ku} as was reviewed in the previous section. On the other hand, here we want to use as an input for bootstrap equations all that we already know from the free field computations. In particular, we will assume the following:
\begin{itemize}
  \item OPE $U_1(z) U_k(w)$ takes the form (\ref{opeu1uk})
	\item OPE $U_2(z) U_k(w)$ takes the form (\ref{opeu2uk})
	\item OPE $U_j(z) U_k(w)$ cannot have $U_l(w)$ in its singular part unless $l \leq j+k-2$
	\item we will solve the equations for generic values of $N$ and $\alpha_0$
	\item the operators appearing on the RHS of OPE are at most quadratic composites of $U_j$
\end{itemize}
The first two conditions fix the subalgebra generated by $U_1$ and $U_2$ as well as the normalization of higher $U_j$ fields and their transformation under $U_1$ and $U_2$ action. The third condition comes from the fact that the singular terms in OPE have at least one contraction of two free fields. The next assumption just expresses the fact that at this point we are interested in the full two-parametric family of solutions and not in solutions which could exist for special values of parameters. Finally, the last condition is nontrivial and seems to be very special for $\mathcal{W}_{\infty}$ and particular choice of the basis. It clearly is not true in the basis of primary fields, where the composite fields appearing in the OPE of two primaries have arbitrary nonlinearity, bounded only by the scaling dimensions of the fields (this was discussed in the previous section). But this quadraticity property holds classically \cite{FigueroaO'Farrill:1992cv, Khesin:1994ey} and follows also from the free-field computations of \cite{Fateev:1987zh, Lukyanov:1987xg, Lukyanov1988}.

Under these assumptions we can compute the OPE of $U_j(z)$ and $U_k(w)$ using the Jacobi identities. These identites can be solved order by order in $j+k$ using the Mathematica package \texttt{OPEdefs}. We start with an ansatz
\begin{equation}
\label{opequadansatz}
U_j(z) U_k(w) \sim \sum_{\substack{l+m+\alpha+\beta<j+k \\ (l,\alpha) \leq (m,\beta)}} \frac{C_{jk}^{lm\alpha\beta}(\alpha_0,N) (U_l^{(\alpha)} U_m^{(\beta)})(w)}{(z-w)^{j+k-l-m-\alpha-\beta}}
\end{equation}
which follows from the assumed quadraticity of OPE. The restriction $(l,\alpha)\leq(m,\beta)$ on possible operators on the right hand side expresses the fact that the operators $(U_j^{\alpha} U_k^{\beta})(w)$ and $(U_k^{\beta} U_j^{\alpha})(w)$ are not independent, but related via (\ref{opeasymmetry}).

Assuming (\ref{opequadansatz}) and restrictions summarized above, we can start solving the Jacobi identities. Jacobi identities for $(U_1 U_1 U_k)$, $(U_1 U_2 U_k)$ and $(U_2 U_2 U_k)$ are satisfied automatically given (\ref{opeu1uk}, \ref{opeu2uk}). Jacobi identities for triples $(U_1 U_3 U_3$) and $(U_2 U_3 U_3)$ determine uniquely OPE $U_3(z) U_3(w)$ except for the most singular term (coefficient of the identity) which we determine from the free field computation. To determine OPE $U_3(z) U_4(w)$ we use Jacobi identities for $(U_1 U_3 U_4)$, $(U_2 U_3 U_4)$ and $(U_3 U_3 U_3)$. We may proceed similarly at $j+k = 8$: Jacobi identities $(U_1 U_3 U_5)$ and $(U_2 U_3 U_5)$ determine $U_3(z) U_5(w)$ and Jacobi identities $(U_1 U_4 U_4)$, $(U_2 U_4 U_4)$ and $(U_3 U_3 U_4)$ determine $U_4(z) U_4(w)$. For illustration, some of these OPEs are given in appendix \ref{appendixquadratic}. We proceeded this way and determined OPE of $U_j(z) U_k(w)$ up to $j+k \leq 15$.

\subsection{Results of associativity constraints}
Following the procedure described in the preceeding section, we obtained a lot of data. The question is if any general patterns can be observed. We found that all the OPE coefficients that were computed are in fact consistent with the following: There exists a set of bilocal operators $U_{jk}(z,w)$ regular as $z \to w$ such that the full operator product expansion (not only the singular part) has the form
\begin{equation}
\label{opefullform}
\boxed{U_j(z) U_k(w) = \sum_{l+m \leq j+k} \frac{C_{jk}^{lm}(\alpha_0,N) U_{lm}(z,w)}{(z-w)^{j+k-l-m}}}
\end{equation}
Here $C_{jk}^{lm}$ are polynomials in two variables that we still have to determine. \footnote{This way of writing the OPE is similar to what was derived only using the free field computations in \cite{Fateev:1987zh, Lukyanov:1987xg, Lukyanov1988}.} Comparing this to (\ref{opequadansatz}) the difference is that the structure constants $C_{jk}^{lm\alpha\beta}$ have now been factorized into $C_{jk}^{lm}$ which does not depend on the order of derivatives $(\alpha,\beta)$ and $U_{jk}(z,w)$ which includes all the derivative terms.

Assuming the knowledge of $C_{jk}^{lm}$, we can use formula (\ref{opefullform}) to inductively compute $U_{lm}(z,w)$ as power series expansion in terms of usual normal ordered products of operators. For instance, since $U_0 = \mathbbm{1}$, and as we will see
\begin{equation}
C_{0j}^{lm} = \delta_0^l \delta_j^m,
\end{equation}
we have
\begin{equation}
U_j(w) = U_0(z) U_j(w) = U_{0j}(z,w)
\end{equation}
and similarly
\begin{equation}
U_j(z) = U_j(z) U_0(w) = U_{j0}(z,w).
\end{equation}
To determine $U_{11}(z,w)$, we have from (\ref{opefullform})
\begin{equation}
U_1(z) U_1(w) = \frac{C_{11}^{00} \mathbbm{1}}{(z-w)^2} + \frac{C_{11}^{10} U_{10}(z,w)}{z-w} + \frac{C_{11}^{01} U_{01}(z,w)}{z-w} + C_{11}^{11} U_{11}(z,w)
\end{equation}
Furthermore, the structure polynomials in this case are
\begin{eqnarray}
C_{11}^{00} & = & N \\
C_{11}^{10} & = & 0 \\
C_{11}^{01} & = & 0 \\
C_{11}^{11} & = & 1
\end{eqnarray}
so
\begin{equation}
U_{11}(z,w) = U_1(z) U_1(w) - \frac{N}{(z-w)^2} = \sum_{k=0}^{\infty} \frac{(z-w)^k}{k!} (U_1^{(\alpha)} U_1)(w) \equiv (U_1(z) U_1(w))
\end{equation}
where we used (\ref{nobilocal}). In this way we may determine inductively all $U_{jk}(z,w)$ assuming the form (\ref{opefullform}) of OPE and knowing the structure polynomials $C_{jk}^{lm}(\alpha_0,N)$. Furthermore, our explicit computation of OPE coefficients using the Jacobi identities lets us determine $C_{jk}^{lm}(\alpha_0,N)$ up to $j+k \leq 15$.

\subsubsection{Inverse formula}
There is a simple transformation of (\ref{opefullform}) which turns out to be quite useful. Considering indices $(j,k)$ as biindex and similarly for $(l,m)$, we see that we have a linear relation between $U_j(z) U_k(w)$ and $U_{lm}(z,w)$. The transformation matrix
\begin{equation}
\frac{C_{jk}^{lm}(\alpha_0,N)}{(z-w)^{j+k-l-m}}
\end{equation}
is lower triangular (consider the lower biindex to be the row index). Denoting $D_{jk}^{lm}(\alpha_0,N)$ the inverse matrix to $C_{jk}^{lm}(\alpha_0,N)$,
\begin{equation}
\label{cdinv}
C_{jk}^{lm} D_{lm}^{rs} = \delta_j^r \delta_k^s
\end{equation}
we can rewrite the OPE equivalently as
\begin{equation}
\label{opequadinv}
\boxed{U_{jk}(z,w) = \sum_{l+m \leq j+k} \frac{D_{jk}^{lm}(\alpha_0,N) U_l(z) U_m(w)}{(z-w)^{j+k-l-m}}}
\end{equation}
Since the left-hand side is an operator regular as $z \to w$, the main observation can also be stated in this form: there exist polynomials $D_{jk}^{lm}(\alpha_0,N)$ such that for each $(j,k)$ the following combination of OPE is regular:
\begin{equation}
\sum_{l+m \leq j+k} \frac{D_{jk}^{lm}(\alpha_0,N) U_l(z) U_m(w)}{(z-w)^{j+k-l-m}} \sim reg.
\end{equation}
If we denote this bilocal operator by $U_{jk}(z,w)$, reversing the steps above we can derive again (\ref{opefullform}).

\subsubsection{Connection to usual normal-ordered products}
If one wants to translate between the usual normal-ordered products and bilocal operators $U_{jk}(z,w)$, we can just take all the regular terms from the expression (\ref{opequadinv}) using (\ref{opebracket}). In this way we find
\begin{eqnarray}
U_{jk}(z,w) & = & \sum_{s=0}^{\infty} \sum_{l+m \leq j+k} D_{jk}^{lm}(\alpha_0,N) \left\{U_l U_m\right\}_{l+m-j-k-s}(w) (z-w)^s \\
& = & \sum_{s=0}^{\infty} \sum_{l+m \leq j+k} \frac{D_{jk}^{lm}(\alpha_0,N) \big(U_l^{(j+k-l-m+s)} U_m\big)(w)}{(j+k-l-m+s)!} (z-w)^s
\end{eqnarray}
Note that the sum over $s$ starts at $s=0$ since all the singular terms cancel by the definition of $D_{jk}^{lm}$ - otherwise $U_{jk}(z,w)$ would not be regular at $z=w$.

The inverse formula can be derived from (\ref{opefullform}) by Taylor expanding $U_{lm}(z,w)$:
\begin{equation}
\left\{U_j U_k\right\}_{r}(w) = \sum_{l+m \leq j+k} \frac{C_{jk}^{lm}(\alpha_0,N) \big(\partial_z^{j+k-l-m-r} U_{lm}\big)(z,w) \big|_{z\to w}}{(j+k-l-m-r)!}
\end{equation}

\subsection{Commutation relations of modes}
The formula (\ref{opequadinv}) together with contour deformation (\ref{contourpic}) is very useful for deriving the commutation relations between modes of $U_j(w)$ operators. The resulting commutation relations are
\begin{multline}
\label{quadcommutator}
\left[U_{j,a}, U_{k,b}\right] = - \sum_{l+m \leq j+k-1} \sum_{\alpha=0}^{\infty} D_{jk}^{lm}(\alpha_0,N) {j+k-l-m-1+\alpha \choose \alpha} \\
\Big( U_{l,m-k+a-\alpha} U_{m, k-m+b+\alpha} - (-1)^{j+k-l-m} U_{m, l-j+b-\alpha} U_{l, j-l+a+\alpha} \Big).
\end{multline}
To obtain this formula, we used the formula for the commutator of modes (\ref{modecommutator}) and applied it to (\ref{opequadinv}). The regular term on the left-hand side gives no contribution since it has no singularities as $z \to w$. The leading term $j+k = l+m$ gives the commutator on the left-hand side of (\ref{quadcommutator}) while the other terms after expanding the denominators and expressing the integrals in terms of modes give the right-hand side of (\ref{quadcommutator}). The infinite number of terms on the right-hand side of (\ref{quadcommutator}) seems to be an unavoidable consequence of having non-linear OPE of the $U_j(z)$ fields. It is analogous to the infinite sum in the expression (\ref{normordmodes}) for mode expansion of the normal ordered product. It is important to remember that as long as we work with the highest-weight representations of the chiral algebra where the highest-weight state is annihilated by positive modes, there will only be a finite number of non-zero terms in (\ref{quadcommutator}) when applied to a state at finite level.

\subsection{Formula for structure constants}
We still haven't determined the value of structure constants $C_{jk}^{lm}(\alpha_0,N)$. From the way they enter the OPE we see that they satisfy the symmetry relation
\begin{equation}
C_{jk}^{lm}(\alpha_0,N) = (-1)^{j+k-l-m} C_{kj}^{ml}(\alpha_0,N)
\end{equation}
Furthermore, from the OPE computed with \texttt{OPEdefs} we find a `translation symmetry'
\begin{equation}
\label{symtrans}
C_{j+n,k+n}^{l+n,m+n}(\alpha_0,N+n) = C_{jk}^{lm}(\alpha_0,N)
\end{equation}
with $n$ a positive integer. The origin of this symmetry is not clear to us. Using these two symmetries, we can reduce the computation of the structure constants to computation of
\begin{equation}
C_{0k}^{lm}(\alpha_0,N) \quad\quad\text{and}\quad\quad C_{jk}^{l0}(\alpha_0,N)
\end{equation}
The first one is easy, since the OPE of any field with identity field does not introduce any singularities. Hence we have
\begin{equation}
C_{0k}^{lm}(\alpha_0,N) = \delta_0^l \delta_m^k
\end{equation}
Determination of $C_{jk}^{l0}(\alpha_0,N)$ is more difficult. Focusing first on the two-point function $C_{jk}^{00}(\alpha_0,N)$ coefficient, one first observes from the computed OPE data that there exist universal polynomials $P_l(\alpha_0,N)$ such that for $j, k \geq 1$
\begin{equation}
C_{jk}^{00}(\alpha_0,N) = (-1)^{j+1} {N \choose j} {N \choose k} j! k! \sum_{l=0}^{\infty} {j-1 \choose l} {k-1 \choose l} \frac{(-1)^l \alpha_0^{j+k-2l-2} P_l(\alpha_0,N)}{{N \choose l+1} (l+1)!^2}
\end{equation}
which looks like a generalization of the binomial transform to two variables (the only terms that couple the $l$ dependence to $j$ and $k$ dependence are the binomial coefficients). The first few of these polynomials are
\begin{eqnarray}
\nonumber
P_0(\alpha_0,N) & = & 1 \\
\nonumber
P_1(\alpha_0,N) & = & 1-2N\alpha_0^2 \\
P_2(\alpha_0,N) & = & 1+4\alpha_0^2-6N\alpha_0^2-6N\alpha_0^4+6N^2\alpha_0^4 \\
\nonumber
P_3(\alpha_0,N) & = & 1+16\alpha_0^2-12N\alpha_0^2+36\alpha_0^4-84N\alpha_0^4\\
\nonumber
& & +36N^2\alpha_0^4-48N\alpha_0^6+72N^2\alpha_0^6-24N^3\alpha_0^6
\end{eqnarray}
One can use the OPE data to compute these polynomials to for rather high values of $l$, but we would like to find the formula for general $l$. Another important observation is that these polynomials factorize at $N=l$:
\begin{eqnarray}
\nonumber
P_0(\alpha_0,N=0) & = & 1 \\
\nonumber
P_1(\alpha_0,N=1) & = & \big(1-2\alpha_0^2\big) \\
P_2(\alpha_0,N=2) & = & \big(1-2\alpha_0^2\big)\big(1-6\alpha_0^2\big) \\
\nonumber
P_3(\alpha_0,N=3) & = & \big(1-2\alpha_0^2\big)\big(1-6\alpha_0^2\big)\big(1-12\alpha_0^2\big) \\
\nonumber
P_4(\alpha_0,N=4) & = & \big(1-2\alpha_0^2\big)\big(1-6\alpha_0^2\big)\big(1-12\alpha_0^2\big)\big(1-20\alpha_0^2\big)
\end{eqnarray}
or
\begin{equation}
\label{ppolrec}
P_l(\alpha_0,N=l) = \prod_{m=1}^l \big( 1 - m(m+1)\alpha_0^2 \big)
\end{equation}
Using the coproduct in $\mathcal{W}_{1+\infty}$ discussed in section \ref{seccoproduct} one finds a recurrence formula for $P_l(\alpha_0,N)$,
\begin{equation}
P_l(\alpha_0,N+1) - P_l(\alpha_0,N) = -\alpha_0^2 l(l+1)P_{l-1}(N)
\end{equation}
which together with the boundary condition (\ref{ppolrec}) and $P_0(\alpha_0,N)=1$ can be solved \footnote{Thanks to Masaki Murata for bringing optimism at this point.} in the region $N \geq l$ and one finds
\begin{equation}
P_l(\alpha_0,N) = \sum_{j=0}^l (-1)^j \alpha_0^{2j} \frac{l!(l+1)!}{(l-j)!(l-j+1)!} {N-l-1+j \choose j} \prod_{k=1}^{l-j} \big( 1 - k(k+1) \alpha_0^2 \big).
\end{equation}
Combining this with formula for $C_{jk}^{00}(\alpha_0,N)$ one finds finally that
\begin{equation}
\label{cjk00prod}
C_{jk}^{00}(\alpha_0,N) = \frac{(-1)^{j+1} N! \alpha_0^{j+k-2}}{(N-j)!(N-k)!} \sum_{l=0}^{\infty} \frac{(N-l-1)!(j+k-l-2)!}{(j-l-1)!(k-l-1)!} \prod_{m=1}^l \left( 1 - \frac{1}{m(m+1)\alpha_0^2} \right)
\end{equation}
which holds unless both $j$ and $k$ are zero. The sums and products can be resummed using hypergeometric summation identities and one arrives at
\begin{multline}
\label{cjk00hyper}
C_{jk}^{00}(\alpha_0,N) = \frac{(-1)^{j+1} N!(N-1)! (j+k-2)! \alpha_0^{j+k-2}}{(j-1)!(k-1)!(N-j)!(N-k)!} \times \\
\times {}_4 F_3 \left( \begin{array}{cc} 1-j, 1-k, \frac{3}{2} + \frac{1}{2\alpha_0} \sqrt{4+\alpha_0^2}, \frac{3}{2} - \frac{1}{2\alpha_0} \sqrt{4+\alpha_0^2} \\ 2, 2-j-k, 1-N \end{array}, 1 \right)
\end{multline}
The origin of square root factors in (\ref{cjk00hyper}) is the quadratic dependence of product (\ref{cjk00prod}) on $m$ and it is one of reasons why it was difficult to find form of $C_{jk}^{00}(\alpha_0,N)$. One may ask what do the values of $\alpha_0^2$ for which the product vanishes for large enough $l$:
\begin{equation}
\alpha_0^2 = \frac{1}{l(l+1)}
\end{equation}
For $\mathcal{W}_N$ minimal models the value of $\alpha_0^2$ is
\begin{equation}
\alpha_0^2 = \frac{(p^\prime-p)^2}{p^\prime p}
\end{equation}
where $p^\prime$ and $p$ are two coprime integers \cite{Bouwknegt:1992wg}. In particular, the unitary minimal models we can choose have $p^\prime = p+1$ and the product vanishes for
\begin{equation}
l = \frac{p}{p^\prime-p} = p.
\end{equation}

Having understood the $C_{jk}^{lm}(\alpha_0,N)$ for $l = m = 0$, we may try to express the results of OPE computations in terms of similar functions. One way of writing the result is
{\small
\begin{eqnarray}
C_{jk}^{l0} & = & \delta_j^l \delta_k^0 + \frac{(-1)^{j-1} (N-l)! N! \alpha_0^{j+k-l-2}}{(N-j)!(N-k)!} \times \\
\nonumber
& & \Bigg[ \sum_{a=0}^{\frac{l+1}{2}} \sum_{b=a-1}^{l-a-1} \frac{(-1)^{a+b+1}}{(N-a)!} \left( {l-a-1 \choose a}{l-2a-1 \choose b-a} + {l-a-1 \choose a-1}{l-2a \choose b-a+1} \right) \phi^{j+k-l-1, N-a}_{j-b-1, k-l+b+1} \\
\nonumber
& & + \sum_{a=0}^{\frac{l+1}{2}} \sum_{b=a-1}^{l-a} \frac{(-1)^{a+b}}{(N-a)!} \left( {l-a-1 \choose a}{l-2a-1 \choose b-a} + {l-a \choose a-1}{l-2a+1 \choose b-a+1} \right) \phi^{j+k-l-2, N-a}_{j-b-1, k-l+b} \Bigg]
\end{eqnarray}}
which is true for all $l>0$ and where we defined
\begin{eqnarray}
\phi^{ab}_{cd} (\alpha_0,N) & = & \sum_{\alpha=0}^{\min(c-1,d-1)} \frac{\Gamma(a-\alpha)\Gamma(b-\alpha)}{\Gamma(c-\alpha)\Gamma(d-\alpha)} \prod_{\beta=1}^{\alpha} \left(1 - \frac{1}{\beta(\beta+1)\alpha_0^2} \right) \\
& = & \frac{\Gamma(a)\Gamma(b)}{\Gamma(c)\Gamma(d)} {}_4 F_3 \left( \begin{array}{cc} 1-c, 1-d, \frac{3}{2} + \frac{1}{2\alpha_0} \sqrt{4+\alpha_0^2}, \frac{3}{2} - \frac{1}{2\alpha_0} \sqrt{4+\alpha_0^2} \\ 2, 1-a, 1-b \end{array}; 1 \right)
\end{eqnarray}
This is our main result for the structure coefficients. It would be nice if this formula could be rewritten in terms of functions that appear as structure constants of $\mathfrak{hs}(\lambda)$ algebra like the Wigner $3j$ or $6j$ symbols, but so far it has resisted our attempts to simplify it. But it is a closed form formula which together with symmetry relations gives the correct result for approximately $9000$ structure constants that we computed using \texttt{OPEdefs} - not including all the derivative terms which were already taken into account by (\ref{opefullform}).

\subsection{Correlation functions}
Having found explicit way of rewriting OPE (\ref{opequadinv}), we can easily compute the correlation functions of $U_j(z)$ operators on the sphere. The one-point functions are clearly (say because of the scaling or rotation symmetry)
\begin{equation}
\langle U_j(z) \rangle = \delta_{j0}.
\end{equation}
The two-point functions are given by the structure constant $C_{jk}^{00}$,
\begin{equation}
\label{twoptcor}
\langle U_j(z) U_k(w) \rangle = \frac{C_{jk}^{00}}{(z-w)^{j+k}}
\end{equation}
for which we have an explicit expression (\ref{cjk00prod}) when both $j$ and $k$ are nonzero. If any one of these is zero, the two-point function reduces to one-point function. It is easy to understand (\ref{twoptcor}). We have
\begin{equation}
\langle U_{jk}(z,w) \rangle = \delta_j^0 \delta_k^0
\end{equation}
because the expectation value of bilocal operators $U_{jk}(z,w)$ must be regular as $z \to w$ but at the same time proportional to $(z-w)^{-j-k}$ because of the scaling invariance. This is consistent only for $j = 0 = k$. Now (\ref{twoptcor}) follows immediatelly using (\ref{cdinv}).

To determine the three-point and higher correlation functions, we can use the Cauchy integral formula, since from the formula (\ref{opequadinv}) we understand the two-point singularities. Consider the three-point function
\begin{equation}
\langle U_j(x) U_k(y) U_l(z) \rangle
\end{equation}
and assume that $j \neq 0$ (otherwise the three-point function reduces to two-point function which we already know). This function at fixed values of $y$ and $z$ is a meromorphic function of $x$ on the Riemann sphere with possibly poles only at $y$ and $z$. From the Cauchy formula and the absence of poles at $x \to \infty$ \footnote{For this step we needed $j \neq 0$; the identity operator clearly does not have sufficient fall off at infinity.} we have
\begin{eqnarray}
\nonumber
\langle U_j(x) U_k(y) U_l(z) \rangle & = & \oint_x \frac{dw}{2\pi i} \frac{\langle U_j(w) U_k(y) U_l(z) \rangle}{w-x} \\
& = & -\oint_y \frac{dw}{2\pi i} \frac{\langle U_j(w) U_k(y) U_l(z) \rangle}{w-x} -\oint_z \frac{dw}{2\pi i} \frac{\langle U_j(w) U_k(y) U_l(z) \rangle}{w-x} \\
\nonumber
& = & \oint_y \frac{dw}{2\pi i} \sum_{a+b < j+k} \frac{D_{jk}^{ab} \langle U_a(w) U_b(y) U_l(z) \rangle}{(w-x)(w-y)^{j+k-a-b}} \\
& & + \oint_z \frac{dw}{2\pi i} \sum_{a+b < j+l} \frac{D_{jl}^{ab} \langle U_a(w) U_k(y) U_b(z) \rangle}{(w-x)(w-z)^{j+l-a-b}}
\end{eqnarray}
We see that the computation of our three-point function reduces to computation of three-point functions with lower values of $j+k+l$, so inductively we can compute all the three point functions. Four-point and higher-point correlation functions can be obviously computed using the same method.

The computation that we used singles out field $U_j(x)$. But of course we could do the same computation using $U_k(y)$ or $U_l(z)$. Consistency of these computations must follow from the associativity of OPE and it would be interesting to see what kind of algebraic constraints this computation of three-point functions gives us for structure constants $D_{jk}^{lm}$. If $U_j(x)$ were quasiprimary, the $x$-dependence of the correlation function would be fixed by global conformal invariance. Unfortunatelly, $U_j(x)$ do not transform under special conformal transformations as simply as quasiprimary fields and dilation, rotation and translation symmetries does not fix the functional dependence of three-point function. If the dependence of three-point function of $x$, $y$ and $z$ was fixed in terms of dimensions of fields, it would be easy to evaluate the integrals and we would obtain algebraic equations for the structure constants as compatibility equations for the three-point functions.

The reason we used the Cauchy formula was that the singular parts of the three-point functions as two points approach each other are not independent. Replacing the $U_j(x) U_k(y)$ by terms coming from (\ref{opequadinv}) determines the singularity as $x \to y$ but includes also some terms which are singular as $x \to z$ (in fact those that are singular both as $x \to y$ and $x \to z$). The same is true for singularity of $U_j(x) U_l(z)$ as $x \to z$. So if we just added the singular terms as determined by (\ref{opequadinv}), we would be overcounting - including terms which have both singularities twice.

It would be interesting to see what are the constraints on $D_{jk}^{lm}$ coming from consistency of $n$-point functions, what is the generating set of these equations and what kind of $\mathcal{W}$-algebras share this quadratic property with $\mathcal{W}_{1+\infty}$. Clearly the affine Lie algebras or the Virasoro algebra are of this type. For example, for Virasoro algebra we can rewrite the OPE as
\begin{equation}
T(z) T(w) - \frac{T(z)}{(z-w)^2} - \frac{T(w)}{(z-w)^2} - \frac{c/2}{(z-w)^4} \sim reg
\end{equation}
Note that the derivative terms disappear if we use this symmetric form of writing the OPE.

\subsection{Coproduct in $\mathcal{W}_{1+\infty}$}
\label{seccoproduct}
Although we succeeded in finding a closed-form formula for the structure constants $C_{jk}^{lm}(\alpha_0,N)$ it is useful as consistency check to derive some constraints on these structure constants and verify that these are satisfiied. One set of relations comes from the free field representation (\ref{miurar}). What we can do is to simply split $N$ free bosons we started with into two groups of $N_1$ and $N_2$ free bosons such that $N = N_1 + N_2$. Denoting the $U_j(z)$ fields constructed in this way by $U_{(1)j}(z)$ and $U_{(2)k}(w)$, we have
\begin{equation}
\label{coproductr}
R(z) = R_1(z) R_2(z) = \sum_{k=0}^{N_1} \sum_{l=0}^{N_2} U_{(1)k}(z) (\alpha_0 \partial)^{N_1-k} U_{(2)l}(z) (\alpha_0 \partial)^{N_2-l}
\end{equation}
We left out the free field normal orderings since the two sets of free fields commute with each other. Passing now the derivatives to the right \footnote{This construction is not symmetric in $U_{(1)}$ and $U_{(2)}$. There is an analogous version of this construction if we move the derivatives to the left, but we have not found any symmetric variant of the coproduct.}, we find
\begin{equation}
\label{coproductu}
U_j(z) = \sum_{k=0}^j \sum_{l=0}^{j-k} {N_1-k \choose j-k-l} \alpha_0^{j-k-l} U_{(1)k}(z) U_{(2)l}^{(j-k-l)}(z).
\end{equation}
The first set of fields generate the $\mathcal{W}_{1+\infty}$ with parameters $(\alpha_0, N_1)$ and the other with parameters $(\alpha_0, N_2)$. This formula tells us how to find the $\mathcal{W}_{1+\infty}$ algebra with parameters $(\alpha_0, N_1+N_2)$ in the product of the two theories. Using the explicit form of OPE $U_j(z) U_k(w)$ for $j+k \leq 15$ computed before, one can verify that (\ref{coproductu}) is consistent for arbitrary (not only positive integer) values of $N_1$ and $N_2$, which is however not very surprising since the structure constants $C_{jk}^{lm}(\alpha_0,N)$ are polynomials in $\alpha_0$ and $N$.

The consistency of coproduct together with quadratic form of the OPE (\ref{opequadinv}) allows us to derive consistency conditions for the structure constants. Plugging (\ref{coproductu}) in (\ref{opequadinv}) we find that
\begin{multline}
\sum_{l+m \leq j+k} \; \sum_{a+c+\rho \leq l} \; \sum_{b+d+\sigma \leq m} \; \sum_{\alpha+\beta \leq a+b} \; \sum_{\gamma+\delta \leq c+d} (-1)^{l-a-c-\rho} \alpha_0^{l+m-a-b-c-d} \\
\times {N_1-a \choose l-a-c} {N_1-b \choose m-b-d} {l-a-c \choose \rho} {m-b-d \choose \sigma} (c+d-\gamma-\delta)_{l+m-a-b-c-d-\rho-\sigma} \\
\times D_{jk}^{lm}(\alpha_0,N_1+N_2) C_{ab}^{\alpha\beta}(\alpha_0,N_1) C_{cd}^{\gamma\delta}(\alpha_0,N_2) \frac{U_{(1)\alpha\beta}(z,w) U_{(2)\gamma\delta}^{(\rho\sigma)}(z,w)}{(z-w)^{j+k-\alpha-\beta-\gamma-\delta-\rho-\sigma}} \sim reg.
\end{multline}
must be regular. Operators $U_{(1)\alpha\beta}(z,w)$ in the first $\mathcal{W}_{1+\infty}$ are independent so we can extract from this formula the coefficient of each of them. On the other hand, operators $U_{(2)\gamma\delta}^{(\rho\sigma)}(z,w)$ are not independent because of the derivatives and symmetry in $(\gamma,\delta)$. What we can do is to take the Taylor expansion of these operators at $z = w$, but the resulting formulas are not very illuminating. Specializing for simplicity to $\rho = \sigma = 0$ and fixed values of $\alpha, \beta, \gamma$ and $\delta$, which satisfy
\begin{equation}
\alpha+\beta+\gamma+\delta < j+k
\end{equation}
we have
\begin{multline}
0 = \sum_{l+m \leq j+k} \; \sum_{a+c \leq l} \; \sum_{b+d \leq m} \; (-1)^{l-a-c} \alpha_0^{l+m-a-b-c-d} {N_1-a \choose l-a-c} {N_1-b \choose m-b-d} \\
\times (c+d-\gamma-\delta)_{l+m-a-b-c-d} D_{jk}^{lm}(\alpha_0,N_1+N_2) C_{ab}^{\alpha\beta}(\alpha_0,N_1) \left( C_{cd}^{\gamma\delta}(\alpha_0,N_2) + C_{cd}^{\delta\gamma}(\alpha_0,N_2) \right)
\end{multline}
This equation holds trivially also for $\alpha+\beta+\gamma+\delta > j + k$, while the simple modification valid for any values of indices is
\begin{multline}
\delta_{j}^{\alpha+\gamma} \delta_{k}^{\beta+\delta} + \delta_{j}^{\alpha+\delta} \delta_{k}^{\beta+\gamma} = \sum_{l+m \leq j+k} \; \sum_{a+c \leq l} \; \sum_{b+d \leq m} \; (-1)^{l-a-c} \alpha_0^{l+m-a-b-c-d} {N_1-a \choose l-a-c} {N_1-b \choose m-b-d} \\
\times (c+d-\gamma-\delta)_{l+m-a-b-c-d} D_{jk}^{lm}(\alpha_0,N_1+N_2) C_{ab}^{\alpha\beta}(\alpha_0,N_1) \left( C_{cd}^{\gamma\delta}(\alpha_0,N_2) + C_{cd}^{\delta\gamma}(\alpha_0,N_2) \right)
\end{multline}
Inverting the $D_{jk}^{lm}$ matrix, we can write this also as
\begin{multline}
C_{lm}^{\alpha+\gamma,\beta+\delta}(\alpha_0,N_1+N_2) + C_{lm}^{\alpha+\delta,\beta+\gamma}(\alpha_0,N_1+N_2) = \\
= \sum_{a+c \leq l} \; \sum_{b+d \leq m} \; (-1)^{l-a-c} \alpha_0^{l+m-a-b-c-d} {N_1-a \choose l-a-c} {N_1-b \choose m-b-d} \\
\times (c+d-\gamma-\delta)_{l+m-a-b-c-d} C_{ab}^{\alpha\beta}(\alpha_0,N_1) \left( C_{cd}^{\gamma\delta}(\alpha_0,N_2) + C_{cd}^{\delta\gamma}(\alpha_0,N_2) \right)
\end{multline}
Specializing to $\gamma = 0 = \delta$,
\begin{multline}
C_{lm}^{\alpha\beta}(\alpha_0,N_1+N_2) = \sum_{a+c \leq l} \; \sum_{b+d \leq m} \; (-1)^{l-a-c} \alpha_0^{l+m-a-b-c-d} {N_1-a \choose l-a-c} {N_1-b \choose m-b-d} \\
\times (c+d)_{l+m-a-b-c-d} C_{ab}^{\alpha\beta}(\alpha_0,N_1) C_{cd}^{00}(\alpha_0,N_2)
\end{multline}
Choosing $N_2 = 1$ and noticing that $C_{cd}^{00}(\alpha_0,1)$ is nonzero only for $(c,d)=(0,0)$ or $(c,d)=(1,1)$, we arrive at recurrence relation
\begin{multline}
C_{lm}^{\alpha\beta}(\alpha_0,N+1) - C_{lm}^{\alpha\beta}(\alpha_0,N) = \sum_{a<l} \sum_{b<m} (-1)^{l-a-1} \alpha_0^{l+m-a-b-2} \\
\times {N-a \choose l-a-1} {N-b \choose m-b-1} (l+m-a-b-1)! \; C_{ab}^{\alpha\beta}(\alpha_0,N)
\end{multline}
These equations are sufficient to determine $C_{lm}^{\alpha\beta}(\alpha_0,N)$ assuming that we know these for any value of $N$. For example, we know that at $N=0$ and $N=1$ we $\mathcal{W}_{1+\infty}$ has a basis with known linear structure constants \cite{Pope:1989ew, Pope:1989sr, Pope:1990kc}, so one would only need to understand the transformation between the linear basis and quadratic basis of the algebra. If this is understood and $C_{jk}^{lm}(\alpha_0,0)$ or $C_{jk}^{lm}(\alpha_0,1)$ determined, we can use the identities following from the existence of the coproduct to determine $C_{jk}^{lm}(\alpha_0,N)$ for any $N$. Alternatively, we could also use the `translation symmetry' \ref{symtrans} which is clearly simpler, but its origin is not clear.

Let us now have a look at how the coproduct acts in the parameter space of $\mathcal{W}_{1+\infty}$. In the free field construction of the coproduct (\ref{coproductr}) it was important that both of the theories that we are composing have the same value of $\alpha_0^2$ but can possibly differ by value of $N_j$. The resulting theory has the same value of $\alpha_0^2$ and $N = N_1 + N_2$. Looking at (\ref{alphatolambda}) and (\ref{lambdaquadratic}) we see that both of these equations are homogeneous in $\lambda_j$. This means that the coproduct allows us to additively compose the triples $(\lambda_1, \lambda_2, \lambda_3)$ and $(\lambda_1^\prime,\lambda_2^\prime,\lambda_3^\prime)$ if they are proportional. We will see later that in fact this addition restricted to the set of minimal models produces again $\mathcal{W}_{1+\infty}$ with parameters of a minimal model, and that the minimal models are on various lines through the origin characterized by the value of $\alpha_0^2$.

\subsection{Virasoro subalgebras and quasiprimary fields}
Having understood the commutation modes in the quadratic $U$-basis, we can now try to compare it with the results of the primary basis computation.

\subsubsection{Quasiprimary fields}
One can check that with respect to $T_{1+\infty}(z)$, the generating fields $U_k(z)$ are not primary nor quasiprimary, but they do have definite scaling dimensions. Recall that a local field $A(z)$ is quasiprimary with respect to stress-energy tensor $T(z)$ if we have OPE
\begin{equation}
\label{opequasiprimary}
T(z) A(w) \sim \cdots + \frac{0}{(z-w)^3} + \frac{h_A A(w)}{(z-w)^2} + \frac{A^\prime(w)}{z-w}
\end{equation}
The quadratic pole determines the scaling dimension of $A(z)$ while the cubic pole must vanish is the field is quasiprimary with respect to $T(z)$. If there are no higher order poles than quadratic, the field $A(z)$ is primary. Computing OPE of $T_{1+\infty}(z)$ with $U_k(z)$ we find
\begin{equation}
T_{1+\infty}(z) U_k(w) \sim \frac{1}{2} \sum_{l=0}^{k-1} \frac{\big((k-l-1)N+(k+l-1)\big) (N-k+1)_{k-l} \alpha_0^{k-l} U_l(w)}{(z-w)^{k-l+2}} + \frac{k U_k(w)}{(z-w)^2} + \frac{U_k^\prime(w)}{z-w}
\end{equation}
so that $U_k(w)$ do have definite scaling dimension $k$ with respect to $T_{1+\infty}$ but they are not quasiprimary. But we can make linear combinations of $U_k(w)$ and their derivatives that are quasiprimary:
\begin{equation}
Q_j(z) = \frac{(j-1)!}{(2j-2)!} \sum_{k=0}^{j-1} (-1)^k \frac{(2j-k-2)!(N-j+1)_k \alpha_0^k}{k!(j-k-1)!}  U_{j-k}^{(k)}(z)
\end{equation}
The inverse transformation looks similar
\begin{equation}
U_j(z) = (j-1)! \sum_{k=0}^{j-1} \frac{(2j-2k-1)! (N-j+1)_k \alpha_0^k}{k!(j-k-1)!(2j-k-1)!} Q_{j-k}^{(k)}(z).
\end{equation}
These two identities generalize the binomial transform between two sequences. Since the transformation between $U$-basis and quasiprimary basis is linear, it preserves the quadratic form of the OPE. Furthermore, quasiprimary fields have simple transformation properties under the inversion, so they behave under BPZ conjugation better than $U$-fields. On the other hand, the structure constants of $\mathcal{W}_{1+\infty}$ are polynomials in each index if we fix the other indices. This is not true in quasiprimary basis, for example the coefficient of identity in OPE of $Q_j(z) Q_k(w)$ can be nonzero only if $j=k$ (because of the restrictions coming from the global conformal group on two-point functions).

\subsubsection{Primary fields}
\label{primarytou}
Having discussed $\mathcal{W}_{1+\infty}$ in primary basis as well as in quadratic basis, we should try to understand the transformation between these two bases. Apart from checking that we are actually dealing with the same chiral algebra, this comparison is also useful to understand the triality symmetry. Recall that although in quadratic basis we have simpler form of OPE and in fact we have a closed-form formula for it, the triality symmetry is hidden. On the other hand, the structure constants of $\mathcal{W}_{1+\infty}$ are manifestly triality invariant when we work with the primary basis.

As already discussed in section \ref{secprimary}, there is no canonical choice of primary fields. If we worked with $\mathcal{W}_{\infty}$ algebra by considering only fields which have regular OPE with $U_1(z)$, the non-uniqueness of primary fields would first appear at dimension $6$ where the composite primary field $(W_3 W_3) + \cdots$ appears. But working with $\mathcal{W}_{1+\infty}$ we saw in section \ref{secu1virasoro} that there are various choices of Virasoro subalgebras and fixing one of these, already at dimension $2$ there is a Virasoro-primary field. For example, with respect to stress-energy tensor (\ref{t1inf}), the following field is dimension $2$ primary:
\begin{equation}
-U_2(z) + \frac{(N-1)(N+1)\alpha_0^2}{2} (U_1 U_1)(z) + \frac{(N-1)\alpha_0}{2} U_1^\prime(z)
\end{equation}
Similarly to previous derivation in $\mathcal{W}_{\infty}$, we can derive the primary field counting function in $\mathcal{W}_{1+\infty}$ and we find
\begin{equation}
\sum_{h=0}^{\infty} P_h q^h = q + \prod_{s=2}^{\infty} \prod_{j=s}^{\infty} \frac{1}{1-q^j} \simeq 1 + q + q^2 + 2q^3 + 4q^4 + 6q^5 + 12q^6 + 18q^7 + 33q^8 + \cdots
\end{equation}

\paragraph{Orthogonality in two-point functions}
One possible choice of primary fields uses the two-point function (\ref{cjk00hyper}). Up to an overall normalization, we can find at each dimension a unique linear combination of fields of the same dimension, which has zero two-point function with all fields of lower dimension and with composite fields of the same dimension. For example at dimension $2$ we have only $3$ fields. The field
\begin{equation}
T_{\infty}(z) = -U_2(z) + \frac{N-1}{2N} (U_1 U_1)(z) + \frac{(N-1)\alpha_0}{2} U_1^\prime(z)
\end{equation}
is the only linear combination at dimension $2$ with coefficient of $U_2(z)$ equal to $-1$ and having zero two-point function with $U_1(z)$, $U_1^\prime(z)$ and $(U_1 U_1)(z)$. At dimension $3$ we have in total $6$ fields. The combination that has vanishing two-point function with all dimension $1$ and $2$ fields and dimension $3$ composite fields is
\begin{eqnarray}
\nonumber
W_3(z) & \sim & -U_3(z) + \frac{(N-2)}{N} (U_1 U_2)(z) - \frac{(N-1)(N-2)}{3N^2} (U_1(U_1 U_1))(z) + \frac{(N-2)\alpha_0}{2}U_2^\prime(z) \\
& & - \frac{(N-1)(N-2)\alpha_0^2}{12} U_1^{\prime\prime}(z) - \frac{(N-1)(N-2)\alpha_0}{2N}(U_1^\prime U_1)(z)
\end{eqnarray}
Similarly at dimension $4$ we have
{\footnotesize
\begin{eqnarray}
\nonumber
W_4(z) & \sim & \frac{\alpha_0(N-3)(N-2)(N-1)(5N+6)(\alpha_0^2N^2-\alpha_0^2N-1)(U_1^\prime(U_1 U_1))(z)}{2N^2(5\alpha_0^2 N^3-5\alpha_0^2 N-5N-17)} \\
\nonumber
& & + \frac{(N-3)(N-2)(N-1)(\alpha_0^2 N^2-\alpha_0^2 N-1) (2\alpha_0^2 N^2+3\alpha_0^2 N-3) (U_1^\prime U_1^\prime)(z)}{4 N^2 (5\alpha_0^2 N^3-5\alpha_0^2 N-5N-17)} \\
\nonumber
& & - \frac{\alpha_0(N-3)(N-2)(N-1) (5\alpha_0^2 N^2+7\alpha_0^2 N-5) (U_1^\prime U_2)(z)}{2N(5\alpha_0^2 N^3-5\alpha_0^2 N-5N-17)} -\frac{\alpha_0(N-3)(N-2)(U_1 U_2^\prime)(z)}{2N} \\
\nonumber
& & +\frac{(N-3)(N-2)(N-1)(5N+6)(\alpha_0^2 N^2-\alpha_0^2 N-1) (U_1 (U_1 (U_1 U_1)))(z)}{4N^3(5\alpha_0^2 N^3-5\alpha_0^2 N-5N-17)} \\
& & -\frac{(N-3)(N-2)(5N+6)(\alpha_0^2 N^2-\alpha_0^2 N-1) (U_1 (U_1 U_2))(z)}{N^2(5\alpha_0^2 N^3-5\alpha_0^2 N-5N-17)} \\
\nonumber
& & +\frac{(N-3)(N-2)(5\alpha_0^2 N^2+7\alpha_0^2 N-5) (U_2 U_2)(z)}{2N(5\alpha_0^2 N^3-5\alpha_0^2 N-5N-17)} +\frac{\alpha_0 (N-3)U_3^\prime(z)}{2} +\frac{(N-3)(U_1 U_3)(z)}{N} \\
\nonumber
& & +\frac{(N-3)(N-2)(N-1)(2\alpha_0^4 N^4-5\alpha_0^2 N^3-2\alpha_0^4 N^2-7\alpha_0^2 N^2-4\alpha_0^2 N+5 N-2) (U_1^{\prime\prime} U_1)(z)}{4N^2(5\alpha_0^2 N^3-5\alpha_0^2 N-5N-17)} \\
\nonumber
& & +\frac{\alpha_0 (N-3)(N-2)(N-1)(\alpha_0^4 N^4-10\alpha_0^2 N^3-\alpha_0^4 N^2-14\alpha_0^2 N^2-2\alpha_0^2 N+10N-1)U_1^{\prime\prime\prime}(z)}{24N(5\alpha_0^2 N^3-5\alpha_0^2 N-5N-17)} \\
\nonumber
& & -\frac{(N-3)(N-2)(2\alpha_0^4 N^4-2\alpha_0^4 N^2-5\alpha_0^2 N^2-11\alpha_0^2 N+3) U_2^{\prime\prime}(z)}{4N(5\alpha_0^2 N^3-5\alpha_0^2 N-5N-17)} -U_4(z)
\end{eqnarray}}
We see that already at dimension $4$ the result is quite complicated. The normal ordering prescription is not canonical so if we used a different normal ordering prescription or different nesting of higher non-linear terms, we would get different coefficients.

The procedure for obtaining primary fields as described above works also for fields of higher dimension. By making the two-point function of $W_j(z)$ with Virasoro descendants of lower dimension primaries vanish we find a combination of fields which is primary. By further restricting to a combination which has zero two-point function with other primary fields of the same dimension but constructed from lower spin fields we arrive at primary field of given dimension which is uniquely determined up to an overall normalization. Ultimately we want to have a triality-invariant combination, which can be checked by computing any OPE coefficient of our field with any other triality-invariant field.

As for the normalization, if we choose the coefficient of $U_j(z)$ in $W_j(z)$ to be $-1$, the coefficient of identity of OPE of $W_j(z) W_j(w)$ for first few fields is as follows:
\begin{eqnarray}
W_1(z) W_1(w) & \sim & \frac{N}{(z-w)^2} + \cdots \\
W_2(z) W_2(w) & \sim & \frac{S_1}{2(z-w)^4} + \cdots \\
W_3(z) W_3(w) & \sim & \frac{S_1 S_2}{6N(z-w)^6} + \cdots \\
W_4(z) W_4(w) & \sim & \frac{S_1 S_2 S_3 S_{-1}}{4N^2(5c+22)(z-w)^8} + \cdots \\
W_5(z) W_5(w) & \sim & \frac{S_1 S_2 S_3 S_4 S_{-1}}{10N^3(7c+114)(z-w)^{10}} + \cdots
\end{eqnarray}
where we introduced the notation
\begin{equation}
S_j = (\lambda_1-j)(\lambda_2-j)(\lambda_3-j) = (N-j)(j^2 - N j \alpha_0^2 - N^2 \alpha_0^2)
\end{equation}
which is obviously triality invariant. We see that up to an overall factor of $N^{2-j}$ all the two-point functions of $W_j(z)$ fields constructed above are manifestly triality-invariant.

\paragraph{Comparison to $\mathcal{W}_{\infty}$}

At this point we can finally compare the OPEs of these fields and primary generators of $\mathcal{W}_{\infty}$ discussed in section \ref{secprimary} and verify that the structure constants computed in both ways match. In this way we verify once again the formula (\ref{opexton}) which determines the structure constants of $\mathcal{W}_{\infty}$ in terms of the central charge $c$ and the rank parameter $N$. We choose the Virasoro subalgebra with Virasoro field $T_{\infty}(z)$ because this field has vanishing OPE with $\hat{\mathfrak{u}}(1)$. Note that in section \ref{secprimary} we discussed only $\mathcal{W}_{\infty}$ algebra which has no spin $1$ field. The main reason was that including this field would produce many composite primary fields and the computations would be much more difficult. On the other hand, when discussing the quadratic basis, the algebra $\mathcal{W}_{1+\infty}$ seems to be the one that is more natural. We can always reduce from $\mathcal{W}_{1+\infty}$ to $\mathcal{W}_{\infty}$ analogously to reduction from $GL(n)$ to $SL(n)$ or to reduction of Toda chain to center-of-mass system. To do it, we just need to find combinations of fields which have vanishing OPE with $U_1(z)$.

Luckily, the fields constructed using the orthogonality of two-point functions turn out not to be only primary with respect to $T_{\infty}(z)$ but they have at the same time vanishing OPE with $\hat{\mathfrak{u}}(1)$ algebra. So these are precisely the fields which we can use to make a comparison with results of the section \ref{secprimary}. For example, we can compute the $x^2$ coefficient and we find
\begin{equation}
x^2 = \frac{(N-3)(N+1)(1+N\alpha_0^2-N^2\alpha_0^2)(9-3N\alpha_0^2-N^2\alpha_0^2)}{(N-2)(N-1)(1-N\alpha_0^2-N^2\alpha_0^2)(4-2N\alpha_0^2-N^2\alpha_0^2)(17+5N+5N\alpha_0^2-5N^3\alpha_0^2)}
\end{equation}
which using the identification (\ref{ctoalpha}) is the same as (\ref{opexton}). This implies in particular that the parameter $N$ that we introduced in (\ref{opexton}) and which determines when $\mathcal{W}_{\infty}$ can reduce to $\mathcal{W}_N$ is the same as parameter $N$ that is given by the number of free bosons in free field representation (\ref{miurar}).

\paragraph{Triality in quadratic basis}
We saw earlier that there is a natural choice of stress-energy tensor $T_{1+\infty}(z)$ (\ref{t1inf}) and with respect to this field all the fields in $\mathcal{W}_{1+\infty}$ have the canonical engineering dimensions. Furthermore, we can construct a sequence of primary fields $W_j(z)$ such that they have vanishing two-point functions with composite primaries constructed from fields of lower spin. As discussed in section \ref{secprimary}, this basis has the property that up to overall rescaling of the primary generators the structure constants are manifestly invariant under the triality transformations.

We can use this transformation between the primary basis and the quadratic $U$-basis to express the nonlinear action of triality symmetry on the fields in the quadratic basis. Since the resulting equations are quite complicated, we only describe in words how it works. First we invert the expressions for $W_j(z)$ fields in terms of $U_k(z)$. The resulting formula expresses $U_j(z)$ which is not invariant under the triality transformations as (normal-ordered) polynomials in $W_k(z)$ and their derivatives. The $W_k(z)$ are themselves invariant under triality, but the coefficients are not. This is the origin of the non-trivial transformation properties of $U$-fields. If we now permute $\lambda_j$ in coefficients, say exchange $\lambda_1 \leftrightarrow \lambda_2$, we obtain another set of $U$-fields which must again satisfy the quadratic OPE, with structure constants obtained from $C_{jk}^{lm}(\alpha_0,N)$ by exchange $\lambda_1 \leftrightarrow \lambda_2$. In this way, we obtain three different quadratic bases of $W_{1+\infty}$. The transformation from one $U$-basis to another one is non-linear and the fact that OPE between these non-linearly transformed fields are again quadratic when expressed in terms of the new fields seems to be quite non-trivial.

One might hope that understanding the triality action on $U$-bases could shed some light on the old problem of the null states in the Virasoro algebra. The famous formula of \cite{Benoit:1988aw} which was later put in nice algebraic form by \cite{Bauer:1991qm, Bauer:1991ai} looks formally very similar to the transformation between $W$-basis and $U$-basis of $\mathcal{W}_N$ \cite{DiFrancesco:1990qr}. The triality could give a clue for this connection, since for example for the fixed value of $N$ where $U_j$ with $j>N$ decouple their triality images $\tilde{U}_k$ can still become null for a specific value of the central charge and thus giving us operators generating the null states.

\section{Representation theory}
Finally we are ready to use the commutation relations in $\mathcal{W}_{1+\infty}$ that were found in the previous section to learn something about the representation theory of $\mathcal{W}_{1+\infty}$. One could possibly ask many different questions but we start with the simplest one - for which values of parameteres of $\mathcal{W}_{1+\infty}$ we have vacuum representation with `maximal' number of null states.

\subsection{Virasoro algebra}
Let us recall what we know about the Virasoro algebra. The representations that one usually considers are the highest weight representations. To construct the irreducible highest weight representations of Virasoro algebra, we proceed in two steps. First step is the construction of so-called Verma module. We split the Virasoro algebra $\mathfrak{Vir}$ as vector space into two parts, \footnote{Note that as is usually done we treat the central charge $c$ as a number, although to get Lie algebra it should be a central element. For $\mathcal{W}$-algebras this is not a problem since they are not Lie algebras anyway.}
\begin{equation}
\mathfrak{Vir} = \mathfrak{Vir}_{<0} \oplus \mathfrak{Vir}_{\geq 0}
\end{equation}
where $\mathfrak{Vir}_{<0}$ are linear combinations of negative modes $L_{k}$, $k < 0$, and $\mathfrak{Vir}_{\geq 0}$ are linear combinations of zero mode $L_0$ and positive modes $L_k$, $k>0$. Note that both $\mathfrak{Vir}_{<0}$ and $\mathfrak{Vir}_{\geq 0}$ are subalgebras of the Virasoro algebra (they are closed under taking commutators).

To construct the Verma module, we start with the highest weight vector $\ket{h}$ which is a one-dimensional representation $F_h$ of $\mathfrak{Vir}_{\geq 0}$,
\begin{equation}
\label{virhw}
L_k \ket{h} = \delta_{k,0} h \ket{h}
\end{equation}
and take
\begin{equation}
M_h = \mathcal{U}(\mathfrak{Vir}) \otimes_{\mathcal{U}(\mathfrak{Vir}_{\geq 0})} F_h
\end{equation}
Here $\mathcal{U}(\mathfrak{Vir})$ is the universal envelopping algebra of Virasoro algebra. This is the formal construction of Virasoro Verma module $M_h$ with highest weight $h$. In other words, the Verma module is obtained by acting on the highest weight vector $\ket{h}$ by arbitrary finite product of mode operators $L_k$ and using only the commutation relations of the Virasoro algebra and the highest weight relations (\ref{virhw}). The important point is that there are no other relations that we are allowed to use. It is easy to see using the Poincar\'{e}-Birkhoff-Witt theorem that as a vector space the Verma module is isomorphic to  $\mathcal{U}(\mathfrak{Vir}_{<0})$.

The representations we are mainly interested in are not the Verma modules but the \emph{irreducible} highest weight representations. The nice fact about the Verma modules is that any irreducible highest weight representation is a quotient of a Verma module. Furthermore, any highest weight subrepresentation of Verma module is again a Verma module (with a different highest weight). So to construct an irreducible highest weight representation we only need to take the quotient of the Verma module by sum of all of its proper Verma submodules. Stated differently, to get the irreducible highest weight representation we need to remove all the `null states'. In this second step we are imposing all other relations between the Virasoro generators that are not consequences of the commutation relations and which hold for the irreducible module.

For generic values of the central charge $c$ and parameter $h$ the Verma module is irreducible. But for special values of $c$ and $h$ there are indeed some null states in Verma module and we only get irreducible representation if we remove these. The result for the Virasoro algebra is as follows: first we parametrize the central charge in terms of parameter $t$
\begin{equation}
c = 13 - 6t - 6t^{-1}.
\end{equation}
The Verma module of Virasoro algebra with central charge $c(t)$ and highest weight $h$ is reducible if and only if $h$ takes one of values
\begin{equation}
\label{virdegreps}
h_{rs} = \frac{r^2-1}{4} t + \frac{s^2-1}{4} t^{-1} - \frac{rs-1}{2}
\end{equation}
where $r$ and $s$ are nonnegative integers \cite{DiFrancesco:1997nk}. For $h$ of this form there is always a Verma submodule with highest weight $h+rs$ (but there can also be other Verma submodules).

To study the singular vectors in Verma modules systematically we introduce the Gram matrix. We choose an arbitrary basis of Verma module, for Virasoro algebra the conventional choice is the basis given by vectors
\begin{equation}
L_{-n_r} \cdots L_{-n_3} L_{-n_2} L_{-n_1} \ket{h}
\end{equation}
with $n_1 \leq n_2 \leq \cdots \leq n_r$. With respect to the natural hermitian conjugation \cite{Belavin:1984vu, DiFrancesco:1997nk} \footnote{One could in fact use the involutive antiautomorphism of the Virasoro algebra mapping $L_k \to L_{-k}$ but not taking complex conjugate of coefficients. Since the matrix elements are real, we get the same matrix as if we use the hermitian conjugation. The Gram matrix is in this context often called the Shapovalov form.} we have $L_k^\dagger = L_{-k}$ and the dual basis
\begin{equation}
\bra{h} L_{m_1} L_{m_2} L_{m_3} \cdots L_{m_s},
\end{equation}
$m_1 < m_2 < \cdots < m_s$. The Gram matrix is the matrix of inner products between elements of this basis. The inner products between two vectors can be nonzero only if the level is the same, $\sum_j n_j = \sum_k m_k$. This means that the matrix of inner products decomposes into blocks, one for each level, and the size of each block is the number of partitions of the level. The usefullness of introducing the Gram matrix lies in the fact that the Verma module with highest weight $h$ has Verma submodule at level $l$ if and only if the corresponding level $l$ block of the Gram matrix is degenerate. So by computing determinants of these matrices of inner products, we can easily determine for given $c$ the values of $h$ such that there are null states in the corresponding Verma module. These determinants have been computed for Virasoro algebra by Kac and proved in \cite{Feigin:1988se}. The result which was already stated above is: for every value of $c$ the Verma module is reducible if and only if the highest weight $h$ is one of (\ref{virdegreps}).

We would like to apply this procedure to the Verma module with $h=0$. But the problem is that for $r=s=1$ we get $h_{11}=0$, so the $h=0$ Verma module is always reducible for any value of the central charge $c$. In fact, we always have a null state at level $rs = 1$,
\begin{equation}
\label{virvacl1}
L_{-1} \ket{0} = 0
\end{equation}
in the irreducible vacuum representation which just expresses the translation invariance of the vacuum. For generic values of the central charge there are no other Verma submodules in the vacuum representation - all the null states are descendants of (\ref{virvacl1}). But there is an interesting discrete set of values of $c$ for which there are additional Verma submodules in the $h=0$ Verma module. These are the \emph{Virasoro minimal models}.

To find this special discrete set of values of $c$, we modify the construction of Shapovalov form above. We split the Virasoro algebra in different way,
\begin{equation}
\mathfrak{Vir} = \mathfrak{Vir}_{<-1} \oplus \mathfrak{Vir}_{\geq -1}
\end{equation}
and build the \emph{vacuum Verma module} on state $\ket{0}$ such that
\begin{equation}
L_k \ket{0} = 0, \quad k \geq -1
\end{equation}
by acting on it with products of Virasoro modes $L_k, k \leq -2$. In this way we explicitly exclude the states which would be descendants of $L_{-1} \ket{0}$ in the $h=0$ Verma module. Just as before, we can introduce the corresponding Shapovalov form and compute its determinant. This is now only a function of $c$ and its zeros are precisely the values of central charges of the Virasoro minimal models,
\begin{equation}
c = 1-6 \frac{(p-p^\prime)^2}{p p^\prime}
\end{equation}
with $p, p^\prime \geq 2$ coprime.

\subsection{Null states in vacuum representation of $\mathcal{W}_{1+\infty}$}

Now we apply the procedure explained in the previous section to $\mathcal{W}_{1+\infty}$. We focus only on the vacuum representation. Generic representations can be much more wild than in the case of finitely generated $\mathcal{W}$-algebras. The reason being that the natural generalization of the highest weight representations (\ref{virhw}) would be representations built on highest weight vector $\ket{u_1, u_2, \ldots}$ such that
\begin{eqnarray}
U_{s,k} \ket{u_1, u_2, \ldots} & = & 0, \quad k>0 \\
U_{s,0} \ket{u_1, u_2, \ldots} & = & u_s \ket{u_1, u_2, \ldots}
\end{eqnarray}
where $U_{s,k}$ is the $k$-th mode of field $U_s(z)$. The states in this Verma module would be generated by acting on this vector with products of negative mode operators and using the commutation relations (\ref{quadcommutator}). But the problem is that already at level $1$ there are infinite number of states (corresponding to the infinite set of generators). One cannot for example write the formal character of the Verma module. Although one can think of some ways around this problem (like using some refined characters which would distinguish various states and remove the infinite degeneracy), here we will only focus on the vacuum character which does not suffer from these difficulties.

The key simplification is that for each dimension $s$ field $U_s(z)$ the vacuum is annihilated by all modes operators $U_{s,k}$ with $k>-h$ (\ref{vacmode}). This is needed to have well-defined operator-state correspondence. So what we consider are the vacuum Verma modules built on the highest weight state $\ket{0}$ which satisfies
\begin{equation}
U_{s,k} \ket{0} = 0, \quad k>-s
\end{equation}
This removes the infinite degeneracy and we can proceed as in the case of the Virasoro algebra. Level by level, we can compute the Shapovalov form and study its zeros. There are few differences from the Virasoro case: first, we want to work in $U$-basis, where the generating fields are not quasiprimary - they do not transform as simply under special conformal transformations or the inversion as the quasiprimary fields. But the definition of conjugation depends on transformation under the inversion. The consequence of this will be that the Shapovalov form that we will compute will not be a symmetric matrix.

The second difference is that for Virasoro algebra we had just the central charge as parameter. The zeros of Shapovalov form were solutions to algebraic equations in $c$ so the Virasoro minimal models formed a discrete set of central charges. This would as well apply to any $\mathcal{W}_N$. For for $\mathcal{W}_{1+\infty}$ we have two parameters describing the algebra, so the zeros of Shapovalov form will be curves in the space of parameters. So what we will consider to be the minimal models will be the discrete intersections of pairs of these curves.

Let us now present the results of the calculation. We implemented the computation of determinant of Shapovalov form in Mathematica, using the commutation relations (\ref{quadcommutator}). Since we are studying zeros of quadratic form, the overall normalization is basis dependent. Our choice of basis of level $l$ states will be
\begin{equation}
U_{s_1,m_1} U_{s_2,m_2} \cdots U_{s_k,m_k} \ket{0}
\end{equation}
with $s_j \leq s_{j+1}$ and $m_j \leq m_{j+1}$ if $s_j = s_{j+1}$ \footnote{The ordering here is the reverse of the one that is conventionally used in Virasoro algebra.}. Furthermore, we always have $m_j \leq -s_j$. By the generalized Poincar\'{e}-Birkhoff-Witt theorem this set of states is a basis of the vacuum Verma module. We will choose the dual basis analogously,
\begin{equation}
\bra{0} U_{s_1,m_1} U_{s_2,m_2} \cdots U_{s_l,m_l}
\end{equation}
with $s_j \geq s_{j+1}$ and $m_j \leq m_{j+1}$ if $s_j = s_{j+1}$. Again, we have $m_j \geq s_j$. As explained above, we cannot anymore expect the matrix to be symmetric. The construction could be modified to make the matrix symmetric, but the zeros would anyway not depend on this (basically because of the triangularity of the $\mathcal{W}_{1+\infty}$ algebra). So computing the determinant of Shapovalov form with respect to this basis level by level, we find up to level $8$ polynomials as given in the following table:
\begin{center}
\small
\begin{tabular}{|c|c|c|c|c|c|c|c|c|}
\hline
level & 1 & 2 & 3 & 4 & 5 & 6 & 7 & 8 \\
\hline
& & & & & & & & \\[-2.2ex]
prefactor & $N$ & $2N^3$ & $6N^6$ & $384 N^{13}$ & $\sim N^{24}$ & $\sim N^{45}$ & $\sim N^{77}$ & $\sim N^{128}$ \\
$(N-1)M(0,1)$ & $0$ & $1$ & $3$ & $8$ & $17$ & $37$ & $71$ & $138$ \\
$(N-2)M(0,2)$ & $0$ & $0$ & $1$ & $3$ & $8$ & $19$ & $41$ & $85$ \\
$(N+1)M(0,-1)$ & $0$ & $0$ & $0$ & $1$ & $3$ & $10$ & $23$ & $54$ \\
$(N-3)M(0,3)$ & $0$ & $0$ & $0$ & $1$ & $3$ & $8$ & $19$ & $43$\\
$(N-4)M(0,4)$ & $0$ & $0$ & $0$ & $0$ & $1$ & $3$ & $8$ & $19$ \\
$N^3 M(1,2) M(1,-1) M(-1,-2) $ & $0$ & $0$ & $0$ & $0$ & $0$ & $1$ & $3$ & $10$ \\
$(N-5)M(0,5)$ & $0$ & $0$ & $0$ & $0$ & $0$ & $1$ & $3$ & $8$ \\
$(N-6)M(0,6)$ & $0$ & $0$ & $0$ & $0$ & $0$ & $0$ & $1$ & $3$ \\
$(N-7)M(0,7)$ & $0$ & $0$ & $0$ & $0$ & $0$ & $0$ & $0$ & $1$ \\
$N^3 M(1,3)M(2,-1)M(-2,-3)$ & $0$ & $0$ & $0$ & $0$ & $0$ & $0$ & $0$ & $1$ \\
\hline
\end{tabular}
\end{center}
Here we introduced
\begin{equation}
M(j,k) \equiv (j-k)^2-jk\alpha_0^2-(j+k)N\alpha_0^2-N^2\alpha_0^2 = (j-k)^2-\alpha_0^2(N+j)(N+k)
\end{equation}
For example, at level $4$ we have the polynomial
\begin{multline}
384N^{13}(N-3)(N-2)^3(N-1)^8(N+1)(1-N\alpha_0^2-N^2\alpha_0^2)^8 \times \\
\times (4-2N\alpha_0^2-N^2\alpha_0^2)^3(9-3N\alpha_0^2-N^2\alpha_0^2)(1+N\alpha_0^2-N^2\alpha_0^2)
\end{multline}
The polynomials $M_{jk}$ themselves are not triality invariant, but we can form triality-invariant products
\begin{equation}
N^3 M_{j,k} M_{j-k,-k} M_{k-j,-j}
\end{equation}
and these in fact are the factors that we found up to level $8$.

The power of $N$ in the prefactor given in the first line of the table can be explained as follows: the fields $U_j$ as defined in (\ref{miurar}) are not triality invariant. We can add to them combinations of lower dimension fields to make them primary as explained in section \ref{primarytou}. If we want $W_j(z)$ to be triality invariant, the coefficient of $U_j(z)$ should be chosen proportional to $N^{\frac{j-2}{2}}$. Since in the computation of the Kac determinant we are using the $U$-fields and not the properly normalized $W$-fields, we get from each mode operator $U_{s,j}$ an extra factor of $N^{\frac{2-j}{2}}$. Taking this into account, we can write down the function that counts these extra factors of $N$:
\begin{equation}
z \frac{d}{dz} \left( \prod_{s=1}^{\infty} \prod_{j=s}^{\infty} \frac{1}{1-z^{2-s} q^j} \right) \sim q + 3q^2 + 6q^3 + 13q^4 + 24q^5 + 45 q^6 + 77q^7 + 128q^8 + 201q^9 + \cdots
\end{equation}
The details are not important but the nice thing is that the origin of the powers of $N$ up to level $8$ is understood.

It would be nice to understand the powers of various factors as a function of level. For zeros of the form $(N-j)M(0,j)$ with $j>0$ these powers are the number of plane partitions of the level of height more than $j$. For example, the numbers of plane partitions of height more than $2$ are
\begin{equation}
0, 0, 1, 3, 8, 19, 41, 85, 167, 319, 588, 1066, \ldots
\end{equation}
which nicely matches the third row of the table above. The reason for appearance of these exponents is clear. For $N$ integer our $\mathcal{W}_{1+\infty}$ consistently truncates to $\mathcal{W}_{1+N}$. The number of states in the generic vacuum representation of this algebra is the number of plane partitions with height less than or equal to $N$. The null states of $\mathcal{W}_{1+\infty}$ at $N$ integer are exactly those states that are factored out when restricting to $\mathcal{W}_{1+N}$.

\paragraph{Plots}
When making plots of $\mathcal{W}_{1+\infty}$ parameter space, which has complex dimension two, we can restrict only to subspace of real dimension two, since this is where all the intersection occur. This is like in the case of Virasoro algebra where we have minimal models at real values of the central charge. The space of parameters of $\mathcal{W}_{1+\infty}$ can be parametrized by triples $(\lambda_1, \lambda_2, \lambda_3)$ subject to a quadratic equation (\ref{lambdaquadratic}). This defines a quadratic surface of a mixed signature if we consider $\lambda_j$ to be real. Since the triality group $\mathcal{S}_3$ acts in the space of $\lambda$ parameters by permuting the coordinate axes, we will project the cone to two dimensional plane along the $(1,1,1)$ axis to respect the symmetry. One possible parametrization is
\begin{eqnarray}
x & = & \frac{1}{3} (2\lambda_1 - \lambda_2 - \lambda_3) \\
y & = & \frac{1}{\sqrt{3}} (\lambda_2 - \lambda_3)
\end{eqnarray}
In $(x,y)$ plane the triality acts by $120^\circ$ rotations and by reflections with respect to the $x$ axis. The inverse transformation is
\begin{equation}
(\lambda_1, \lambda_2, \lambda_3) = \left(x, -\frac{x}{2} + \frac{\sqrt{3} y}{2}, -\frac{x}{2} - \frac{\sqrt{3} y}{2} \right) \pm \frac{\sqrt{x^2+y^2}}{2} (1,1,1)
\end{equation}
(there are two preimages, one from each half of the cone).

\begin{figure}
\begin{center}
\fbox{\includegraphics[scale=0.5]{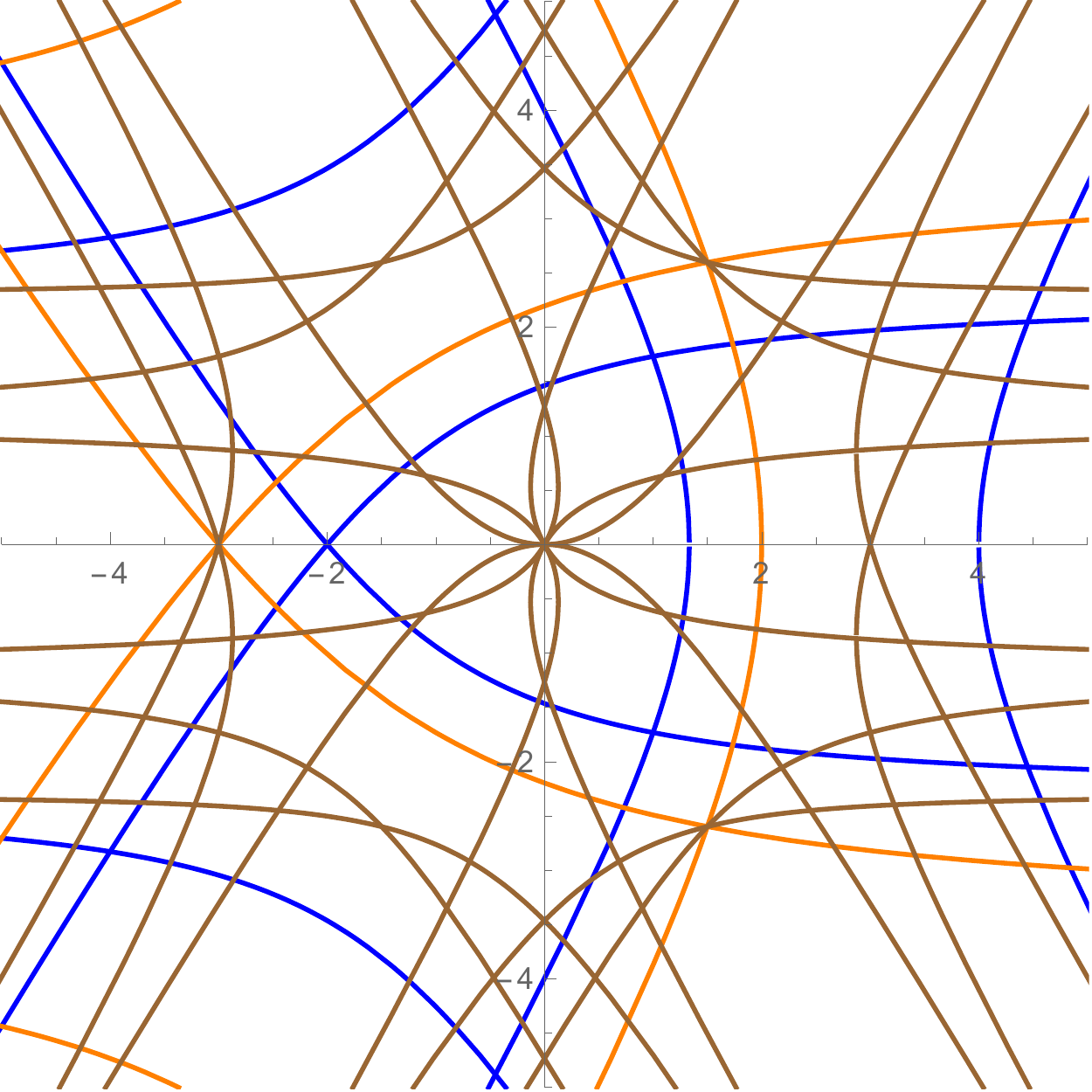}}
\end{center}
\caption{Vanishing curves of the Kac determinant for $\mathcal{W}_{1+\infty}$. The blue curves are the Virasoro $N=2$ curve and its triality images, the orange curves are the $\mathcal{W}_3$ $N=3$ curve and its triality images. The brown curves show the triality orbit of the first unitary minimal model curve.}
\label{figure1}
\end{figure}

\begin{figure}
\begin{center}
\fbox{\includegraphics[scale=0.5]{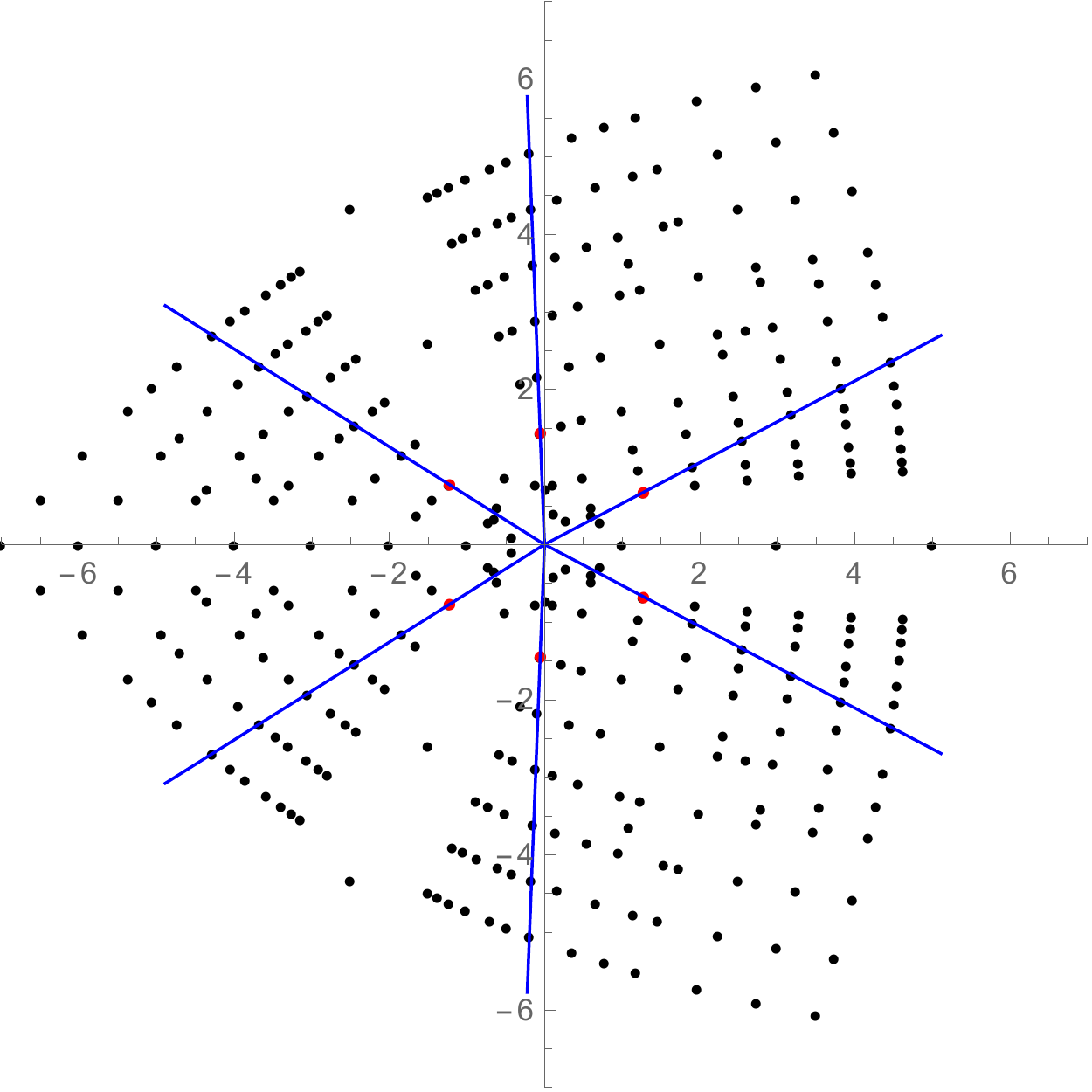}}
\end{center}
\caption{Intersection of vanishing curves are shown as black dots. The red dots represent the triality orbit of the Ising model. The blue lines are the rays connecting Ising model points to the origin. We can see other minimal models lying in these rays.}
\label{figure2}
\end{figure}

\paragraph{Minimal models}
Now as discussed above, zeros of these polynomials give us curves in two-dimensional parameter space of $\mathcal{W}_{1+\infty}$ where additional null states appear in the vacuum Verma module. We will call the discrete intersections of these curves the $\mathcal{W}_{1+\infty}$ minimal models. Let us look for example at the curve $N=2$ which corresponds to Virasoro plus $\hat{\mathfrak{u}}(1)$ truncation of $\mathcal{W}_{1+\infty}$. We find the following minimal models:

\begin{center}
\begin{tabular}{|c|c|c|c|c|}
\hline
$c$ & $\alpha_0^2$ & level & minimal model & $\lambda$ parameters \\
\hline
$0$ & $1/6$ & $2$ & $(2,3)$, trivial & $(1,2,-2/3)$ \\
$-2$ & $1/2$ & $3$ & $(1,2)$ & $(2,2,-1)$ \\
$-22/5$ & $9/10$ & $4$ & $(2,5)$, Lee-Yang, $\mathcal{W}_3$ & $(2,3,-6/5)$ \\
$-7$ & $4/3$ & $5$ & $(1,3)$ & $(2,4,-4/3)$ \\
$1/2$ & $1/12$ & $6$ & $(3,4)$, Ising & $(2,2/3,-1/2)$ \\
$-68/7$ & $25/14$ & $6$ & $(2,7)$ & $(2,5,-10/7)$ \\
$-25/2$ & $9/4$ & $7$ & $(1,4)$ & $(2,6,-3/2)$ \\
$-3/5$ & $4/15$ & $8$ & $(3,5)$ & $(2,4/3,-4/5)$ \\
$-46/3$ & $49/18$ & $8$ & $(2,9)$ & $(2,7,-14/9)$ \\
\hline
\end{tabular}
\end{center}

Note that apart from real Virasoro minimal models we also see here theories with $c=-2, -7, -25/2, \ldots$ for which the Virasoro Verma module is already irreducible. These models are not Virasoro minimal models in the strict sense, but they have extra null states which are descendants of $L_{-1} \ket{0}$, so it is not surprising that they appear in this table. Similarly, the intersections with $N=3$ are the $\mathcal{W}_3$ minimal models:

\begin{center}
\begin{tabular}{|c|c|c|c|c|}
\hline
$c$ & $\alpha_0^2$ & level & minimal model & $\lambda$ parameters\\
\hline
$0$ & $1/12$ & $2$ & $(3,4)$, trivial & $(1,3,-3/4)$ \\
$-22/5$ & $4/15$ & $3$ & $(3,5)$, Lee-Yang, Virasoro & $(2,3,-6/5)$ \\
$-2$ & $1/6$ & $4$ & $(2,3)$ & $(3,3/2,-1)$ \\
$-10$ & $1/2$ & $4$ & $(1,2)$ & $(3,3,-3/2)$ \\
$-114/7$ & $16/21$ & $5$ & $(3,7)$ & $(3,4,-12/7)$ \\
$4/5$ & $1/20$ & $6$ & $(4,5)$, first nontrivial unitary & $(3,3/4,-3/5)$ \\
$-23$ & $25/24$ & $6$ & $(3,8)$ & $(3,5,-15/8)$ \\
$-30$ & $4/3$ & $7$ & $(1,3)$ & $(3,6,-2)$ \\
$-186/5$ & $49/30$ & $8$ & $(3,10)$ & $(3,7,-21/10)$ \\
$-98/5$ & $9/10$ & $8$ & $(2,5)$ & $(3,9/2,-9/5)$ \\
\hline
\end{tabular}
\end{center}

In this way we should see $\mathcal{W}_N$ minimal models for every $N$ as special cases of $\mathcal{W}_{1+\infty}$. But there are also intersections of curves which don't have an integer value of $(\lambda_1, \lambda_2, \lambda_3)$. Up to level $8$ where we have computed the vacuum Kac determinant, there are up to triality $62$ different intersection points. Most of these are induced from $\mathcal{W}_N$ minimal models, but there are $5$ models,
\begin{equation}
(-1,-1,1/2), (-3,-3,3/2), (-5,-5,5/2), (-1,-1/2,1/3), (-1,-1/3,1/4), (1/2,1/3,-1/5)
\end{equation}
which do not come from any $\mathcal{W}_N$ minimal model. There are also $8$ $\mathcal{W}_1$ minimal models
\begin{equation}
\left(1,k,-\frac{k}{k+1}\right)
\end{equation}
with $k=1,\ldots,7$ and $(1,1/2,-1/3)$.

As explained in section (\ref{seccoproduct}) about coproduct, there is a way of realizing $\mathcal{W}_{1+\infty}$ representations in tensor product of two representations of $\mathcal{W}_{1+\infty}$ if the $\lambda$-parameters are proportional (lie on the same line through the origin in the $\lambda$-space). This follows from the condition of these theories to have the same value of $\alpha_0^2$. Although it is not very surprising, it is nice to see that composing in this way two minimal models produces again a minimal model. This is illustrated in figure \ref{figure2}. For example, the Ising model has $\lambda$-parameters
\begin{equation}
\left(2, \frac{2}{3}, -\frac{1}{2} \right)
\end{equation}
and there are $7$ other minimal models (with null states up to level $8$) with the same value of $\alpha_0^2$:
\begin{equation}
\left( 1, \frac{1}{3}, -\frac{1}{4} \right) \times \left\{-1,0,2,3,4,5,6,7\right\}
\end{equation}
The reason that say $(1,1/3,-1/4)$ is not in this list is probably because we are looking only at null states at low level (otherwise it could be generated by taking combinations of other models on this line).

To understand $\mathcal{W}_{1+\infty}$ models better, one would either need a formula for zeros of vacuum Kac determinant at all levels or combine information about $\mathcal{W}_N$ minimal models, which is something that we know much better. For instance, we know that the unitary minimal models of $\mathcal{W}_N$ have central charge
\begin{equation}
c_{N,k} = (N-1) \left(1 - \frac{N(N+1)}{(N+k)(N+k+1)} \right)
\end{equation}
which is exactly the value of the central charge at the intersection of $\mathcal{W}_N$ curve with curve $M(k,k+1)=0$. Figure \ref{figure1} shows these curves. The brown curves are zeros of $M(1,2)M(1,-1)M(-1,-2)$. They form a nice triality-invariant shape in the two-dimensional projection of $\lambda$-parameter space. The blue curve shows zeros of $(N-2)M(0,2)$ which are the curves on which all $\mathfrak{u}(1)$ plus Virasoro truncations of $\mathcal{W}_{1+\infty}$ lie. The orange curves are similarly $\mathcal{W}_{1+3}$ truncations. The Ising model lies on the intersection of the Virasoro curve and the first unitary minimal model curve, together with non-unitary minimal model with central charge $c = -68/7$.

\paragraph{Summary} To summarize, we verified that the commutation relations (\ref{quadcommutator}) in $U$-basis can be used to compute the vacuum Kac determinant for $\mathcal{W}_{1+\infty}$. We expected the result to be triality invariant, which is not a priori obvious since the triality symmetry is not manifest in $U$-basis computations. Luckily, the results were triality invariant (see for example figure \ref{figure1}) and furthermore consistent with everything we know about the properties of $\mathcal{W}_N$ vacuum representation.

As side-product, we found new minimal models of $\mathcal{W}_{1+\infty}$ which do not come from $\mathcal{W}_N$ minimal models and we also found indications of an interesting additive structure of $\mathcal{W}_{1+\infty}$ minimal models compatible with coproduct constructed using the free field representation of $\mathcal{W}_N$. Of course there are many other things to understand, starting from the character formula for vacuum representation to generalizing the Kac determinant to representations which are not vacuum. This is not as simple as in $\mathcal{W}_N$ because already at level $1$ the $\mathcal{W}_{1+\infty}$ Verma module has an infinite number of states, while the interesting $\mathcal{W}_N$ minimal models have of course only a finite number of states at any level. Having infinite possible null state already at level $1$ can make the continuation from $\mathcal{W}_N$ to $\mathcal{W}_{1+\infty}$ non-trivial, but at the same time it forces one to consider interesting mathematical objects like the infinite permutation group in connection with the Weyl character formula.

\section{Directions for further study}

\paragraph{Linear bases}
It is known there for certain values of parameters there exist generating sets of fields which have linear singular part of the OPE among themselves (linear bases). In fact, the oldest constructions of $\mathcal{W}_{1+\infty}$ by \cite{Pope:1989ew, Pope:1989sr, Pope:1990kc} are precisely in these linear bases. Preliminary analysis shows that $N=0$ and $N=1$ are not the only cases where the $\mathcal{W}_{1+\infty}$ linearizes. Furthermore, for $\alpha_0 = 0$ (which is triality equivalent to $N = 0$) the transformation between the quadratic $U$-basis and the linear basis is analogous to transformation between different bases of the symmetric functions \cite{macdonald1998symmetric}. Understanding this could lead to better understanding of the triality in this special case and to simplification of the structure constants of $\mathcal{W}_{1+\infty}$.

\paragraph{Integrability}
There are many connections between two-dimensional conformal field theory and two-dimensional integrable field theories. One of these connections is the presence of the infinite number of commuting conserved charges that can be constructed from the Virasoro field \cite{Bazhanov:1994ft, Bazhanov:1996dr, Bazhanov:1998dq}. The procedure of finding commuting charges constructed from fields of higher scaling dimension works also for $\mathcal{W}_{1+\infty}$ so it would be interesting if we can learn something new from this.

\paragraph{More about representation theory}
There are many questions that one can ask about representation theory of $\mathcal{W}_{1+\infty}$. There are many results known for $\mathcal{W}_N$ but because there is an infinite number of generating fields in $\mathcal{W}_{1+\infty}$, the representation theory should be richer. This was noticed already in \cite{Gaberdiel:2012ku} where some simple representations of $\mathcal{W}_\infty$ were studied. The representation theory of $\mathcal{W}_{1+\infty}$ must include everything we know about $\mathcal{W}_N$ in a smooth way. Furthermore, the triality symmetry which is not visible when one considers only $\mathcal{W}_N$ should play an important role. Understanding of modular invariance of $\mathcal{W}_{\infty}$ representations would be very useful in the context of Gaberdiel-Gopakumar higher-spin-CFT duality.

\paragraph{OPE algebra and other $\mathcal{W}$-algebras}
One of the important conclusions of this article is that the usual approach of expanding operator products at one of two points need not be the most efficient way of describing the chiral algebra. We introduced the bilocal combinations of fields which were regular and this enabled us to write down a closed-form expression for the structure constants of $\mathcal{W}_{1+\infty}$. Furthermore, we haven't lost anything, the computations that could in principle be complicated in the standard approach were quite simple (like the computation of commutation relations of modes or all the correlation functions). We got rid of all derivatives in the expansion and instead of specifying coefficients of all these fields, it is enough just to specify the matrix $D_{jk}^{lm}$. One can ask how general is this description of $\mathcal{W}$-algebra. Clearly all $\mathcal{W}_N$ algebras and in particular the Virasoro algebra and the affine Lie algebras can be specified in this way. To see how general this property is, one can try to find other algebras having this quadratic basis. We don't know yet what are the consistency conditions on $D_{jk}^{lm}$. But the discussion of correlation functions shows that clearly already the consistency of the three-point functions gives us conditions on $D_{jk}^{lm}$.

This reformulation of the usual OPE could perhaps be useful for understanding the `landscape' of $\mathcal{W}$-algebras. So far not much is known about the space of the possible chiral algebras. As explained in the beginning of this article, there are some procedures which produce a $\mathcal{W}$-algebra, but generally given a field content (dimensions of generating fields) we don't know anything about the space of theories with that field content. Usually there is either a discrete set or a one-parameter family of algebras for algebras like Virasoro algebra, $\mathcal{W}_N$ or affine Lie algebras. But $\mathcal{W}_\infty$ is example of two-parametric family of algebras. Extending the field content, one could expect even more complicated parameter spaces.

\paragraph{Higher products of operators}
It would be very useful to find some more canonical normal ordering of fields. The interesting transformations in $\mathcal{W}_{1+\infty}$ like the transformation between the primary basis and $U$-basis or the action of triality on various $U$-bases are non-linear. If we want to write down a formula for such transformations, it would be nice to first have some canonical bases for the non-linear combinations of fields. The usual normal ordering prescription (\ref{normalorder}) suffers from various problems like the lack of commutativity and complicated expressions for associativity conditions. The OPE in $U$-basis is expressed more simply in terms of bilocal fields $U_{jk}(z,w)$ introduced in (\ref{opequadinv}) and the coincident limit of these $U_{jk}(z,w)$ as $z \to w$ gives us quadratic product of $U_j$ and $U_k$ which has many nice properties compared to $(U_j U_k)$. But it is not clear if and how this quadratic product can be generalized to higher powers.

\paragraph{Combinatorics of plane partitions, topological strings}
The interesting observation \cite{Gaberdiel:2010ar} that the vacuum character of $\mathcal{W}_{1+\infty}$ is the MacMahon function connects the theory to combinatorics of plane partitions. Also the triality symmetry $S_3$ is mirrored naturally by $S_3$ symmetry of plane partitions. But the similarity goes further. One can easily check that the character of $\mathcal{W}_{1+N}$ which is the irreducible character of $\mathcal{W}_{1+\infty}(\lambda,c)$ at $\lambda = N$ for generic $c$ is a function that counts the number of partitions which are bounded from one direction by $N$. It would be nice to see if any of the structures found in $\mathcal{W}_{1+\infty}$ could be naturally interpreted in the combinatorial language of plane partitions.

As emphasized in \cite{Gaberdiel:2010ar}, MacMahon function counting plane partition also appears in topological strings \cite{Gopakumar:1998ii, Gopakumar:1998jq, Okounkov:2003sp, Iqbal:2003ds}.

\paragraph{Four-dimensional $\mathcal{N}=2$ supersymmetric theories}
$\mathcal{W}$-algebras play an important role in $4d$-$2d$ correspondence between $4d$ $\mathcal{N}=2$ supersymmetric gauge theory and $2d$ conformal field theory \cite{Alday:2009aq, Gaiotto:2009we, Mironov:2009by, Wyllard:2009hg}. For example, the instanton partition functions for theory with gauge group $SU(N)$ correspond to conformal blocks of $\mathcal{W}_N$. Another construction is that of chiral algebra of BPS operators of \cite{Beem:2013sza, Beem:2014kka}. In these constructions, the value of $N$ is always taken to be an integer - it would be interesting to see if considering the universal algebras like $\mathcal{W}_{\infty}$ and the symmetries similar to triality (which require continuation in $N$) could teach us something new about the space of $\mathcal{N}=2$ theories or about geometry of the instanton moduli spaces.

\paragraph{Kac-Moody algebras at critical level}
Another appearance of $\mathcal{W}$-algebras is in the context of the representation theory of affine Lie algebras \cite{Feigin:1991wy, Frenkel:1991rp}. If we consider the affine Lie algebra $\hat{\mathfrak{sl}}(N)$, there are two places were the classical $\mathcal{W}_N$ (the Gelfand-Dickey algebra) appears. One is in the classical limit of the infinite level, while the other appears at the critical level (which for $\mathfrak{sl}(N)$ is at $k=-N$) \footnote{The critical level is level at which the Sugawara construction of stress-energy tensor fails - there is a factor of $k+N$ in the denominator of $T(z)$ if we follow the usual normalization.}. For general algebra these two algebras are related by the Langlands duality. From point of view of quantum $\mathcal{W}_{1+\infty}$, this duality is what triality degenerates to in the classical limit. So in some sense triality in $\mathcal{W}_{1+\infty}$ is the quantum analogue of the classical Langlands duality. Similar phenomenon was also observed in \cite{Candu:2012ne} in the context of $\mathcal{W}_{\infty}^e$ algebra with only even spin generators. There the symmetry connects two Langlands dual algebras, $B_n$ and $C_n$.

\paragraph{Vasiliev theory and higher spin gravities}
The algebra $\mathcal{W}_{\infty}$ is the symmetry algebra of Gaberdiel-Gopakumar holographic duality between the three dimensional AdS higher spin gravity and two dimensional conformal field theory. The question is what can we learn about three dimensional higher spin theories if we understand the two dimensional dual theory. There are two ways of approaching the classical limit where the both theories can be compared. One possible limit is the 't Hooft limit studied in \cite{Gaberdiel:2010ar}. In this limit the CFT side has certain class of `light states' whose bulk duals are not understood. In different, `semiclassical' limit \cite{Perlmutter:2012ds} there are no light states, but one is forced to consider the non-unitary limit of the CFT. What triality teaches us about these two limits is that the symmetry algebra is equivalent in both of these cases and what differs is the spectrum, the representation content of the theory. The representations of $\mathcal{W}_\infty$ that are light in the semiclassical limit are representations which do not come from $\mathcal{W}_N$ representations. The restrictions on spectra should come from the requirement of the modular invariance, but so far this is not understood.

Another interesting consequence of the triality is the duality between the bulk solutions and scalar perturbations in the semiclassical limit. In \cite{Perlmutter:2012ds} the highest weight representations of $\mathcal{W}_N$ labeled by two Young diagrams $(\Lambda_+,\Lambda_-)$ were identified with scalar perturbations parametrized by $\Lambda_-$ on top of the background parametrized by $\Lambda_+$. But the $\mathbbm{Z}_2$ subgroup of the triality symmetry exchanges these two labels. So one may imagine that the quantized version of the bulk theory has an interesting S-duality symmetry. Thinking of the bulk solutions as being sourced by some heavy particles, the duality might just exchange the two species of particles. Depending on the semiclassical limit chosen, one species of particles would become heavy and the other light. The 't Hooft limit treats both labels $\Lambda_+$ and $\Lambda_-$ symmetrically and one could interpret the light states in the 't Hooft limit as bound states of the form $(\Lambda,\Lambda)$ with large binding energy.

\section*{Acknowledgements}
I would like to thank to Martin Ammon, Andrea Campoleoni, Stefan Fredenhagen, Matthias Gaberdiel, Rajesh Gopakumar, Maximilian Kelm, Libor K\v{r}i\v{z}ka, Mat\v{e}j Kudrna, Masaki Murata, Vasily Pestun, Joris Raeymaekers, Martin Schnabl for useful discussions and especially to Chris Thielemans without whose \texttt{OPEdefs} most of what was done would be impossible.

This research was supported by the Grant Agency of the Czech Republic under the grant 14-31689S.

Computational resources were provided by the MetaCentrum under the program LM2010005 and the CERIT-SC under the program Centre CERIT Scientific Cloud, part of the Operational Program Research and Development for Innovations, Reg. no. CZ.1.05/3.2.00/08.0144.

\appendix

\section{Appendix}

\subsection{Structure constants in primary basis}
\label{appendixprimary}
For completeness, we list here all of the structure constants of $\mathcal{W}_\infty$ in primary basis for primaries $W_j(z) W_k(w)$ up to spin $j+k \leq 10$. Because of the field redefinitions, the exact values between the structure constants are very non-canonical. To simplify the formulas, in this appendix we rescale the fields such that
{\small\begin{eqnarray*}
C_{33}^4 & = & x \\	
C_{33}^0 & = & 1 \\
C_{44}^0 & = & 1 \\
C_{34}^5 & = & 1 \\
C_{35}^6 & = & 1 \\
C_{36}^7 & = & 1 \\
C_{78}^8 & = & 1
\end{eqnarray*}}
and furthermore shift the primary fields $W_6(w)$, $W_7(w)$ and $W_8(w)$ such that
{\small\begin{eqnarray*}
C_{35}^{[33]} & = & 0 \\
C_{36}^{[34]} & = & 0 \\
C_{37}^{[35]} & = & 0 \\
C_{37}^{[44]} & = & 0 \\
C_{37}^{[33]^{\prime\prime}} & = & 0 \\
C_{37}^{[35]^\prime} & = & 0
\end{eqnarray*}}
This fixes completely all possible redefinitions of these fields (although in very non-canonical way). By shifting the primary fields, we can reconstruct from these data the structure constants with arbitrary other choice of primaries. Assuming the choice above, the first group of structure constants is
{\small\begin{eqnarray*}
C_{44}^4 & = & \frac{3(c+3)x}{c+2}-\frac{288(c+10)}{c(5c+22)x} \\
C_{44}^6 & = & \frac{4}{5x} \\
C_{45}^5 & = & \frac{5(c+7)(17c+126)x}{2(c+2)(7c+114)}-\frac{720(c+10)}{c(5c+22)x} \\
C_{45}^7 & = & \frac{2}{3x} \\
C_{46}^8 & = & \frac{4}{7x} \\
C_{55}^0 & = & \frac{5(c+7)(5c+22)x^2}{(c+2)(7c+114)}-\frac{60}{c} \\
C_{55}^6 & = & \frac{5(37c^2+425c+2202)x}{3(c+2)(7c+114)}-\frac{60(19c+218)}{c(5c+22)x} \\
C_{55}^8 & = & \frac{10}{21x} \\
C_{45}^{[34]^\prime} & = & \frac{240(c+7)x}{(c+2)(7c+114)}-\frac{2880}{c(5c+22)x} \\
C_{55}^{[35]^\prime} & = & \frac{60(3797c^3+82090c^2+387832c-306880)x}{7(c+2)(c+24)(7c+114)(11c+350)}-\frac{46080(73c^2+1149c-850)}{7c(c+24)   (5c+22)(11c+350) x}
\end{eqnarray*}}
The structure constants in this group are independent of shift of primary fields. In the next group we have structure constants
{\small\begin{eqnarray*}
C_{34}^3 & = & x \\
C_{35}^4 & = & -\frac{60}{c} + \frac{5(c+7)(5c+22)x^2}{(c+2)(7c+114)} \\
C_{45}^3 & = & -\frac{60}{c} + \frac{5(c+7)(5c+22)x^2}{(c+2)(7c+114)} \\
C_{55}^4 & = & \frac{43200(c+10)}{c^2(c+22)x} + \frac{25(c+7)^2(5c+22)(17c+126)x^3}{2(c+2)^2(7c+114)^2} - \frac{150(c+7)(41c+366)x}{c(c+2)(7c+114)}
\end{eqnarray*}}
The structure constants in this group are related to the previous ones by relations like
\begin{equation*}
C_{55}^4 C_{44}^0 = C_{45}^5 C_{55}^0
\end{equation*}
The third group of structure constants that we list are those that have shift-independent left hand side and composite fields on the right-hand side
{\small\begin{eqnarray*}
C_{44}^{[33]} & = & \frac{30(5c+22)}{(c+2)(7c+114)} \\
C_{45}^{[34]} & = & \frac{560(c+7)(c+10)x}{(c+2)(c+24)(7c+114)}-\frac{6720(c+10)}{c(c+24)(5c+22)x} \\
C_{55}^{[33]} & = & \frac{75(c+7)(5c+22)(39c+178)x^2}{2(c+2)^2(7c+114)^2}-\frac{450(39c+178)}{c(c+2)(7c+114)} \\
C_{55}^{[35]} & = & \frac{25(3343c^3+92550c^2+614104c+2418752x}{7(c+2)(c+24)(7c+114)(11c+350)} - \frac{9600(c+10)(169c+3370)}{7c(c+24)(5c+22)(11c+350) x} \\
C_{55}^{[44]} & = & \frac{400(c+7)(11c+166)x^2}{7(c+2)(c+24)(7c+114)} - \frac{4800(11c+166)}{7c(c+24)(5c+22)} \\
C_{55}^{[33]^{\prime\prime}} & = & \frac{100(c+7)(5c+22)(193c^3-12430c^2-299960c+243744)x^2}{63(c+2)^2(c+24)(5c-4)(7c+114)^2} - \frac{400(193c^3-12430c^2-299960c+243744)}{21c(c+2)(c+24)(5c-4)(7c+114)}
\end{eqnarray*}}
Note that since these coefficients are non-zero, it is clear that one cannot eliminate the composite primary operators from RHS of the OPE of all primaries at once. The penultimate group are the structure constants
{\small\begin{eqnarray*}
C_{36}^3 & = & \frac{5(c+7)(5c+22)(7c-8)x^3}{6(c+2)(c+24)(7c+114)} - \frac{10(7c-8)x}{c(c+24)} \\
C_{36}^5 & = & \frac{35(c+7)(7c^2+122c+688)x^2}{4(c+2)(c+24)(7c+114)} - \frac{960(c+10)^2}{c(c+24)(5c+22)} \\
C_{46}^4 & = & \frac{5(c+7)(5c+22)(7c^2+82c+288)x^3}{(c+2)^2(c+24)(7c+114)} + \frac{46080(c+10)^2}{c^2(c+24)(5c+22)x} - \frac{60(113c^3+3100c^2+26724c+77632)x}{c(c+2)(c+24)(7c+114)} \\
C_{37}^4 & = & \frac{23040(c+10)^2(29c-60)}{c^2(c+24)(5c+22)(11c+350)} \\
& & + \frac{5(c+7)(5c+22)(6860c^5+233021c^4+2210045c^3+2684318c^2-18804472c+11668160)x^4}{12(c+2)^2(c+23)(c+24)(5c-4)(7c+114)(11c+350)} \\
& & - \frac{5(103700c^6+5037443c^5+82080149c^4+484133372c^3+156571028c^2-4060675888c+2813946240)x^2}{c(c+2)(c+23)(c+24)(5c-4)(7c+114)(11c+350)} \\
C_{37}^6 & = & \frac{(1372c^4+57159c^3+985274c^2+8331408c+27861120)x^2}{(c+2)(c+24)(7c+114)(11c+350)} - \frac{18(1323c^3+52400c^2+759236c+3957600)}{c(c+24)(5c+22)(11c+350)} \\
C_{46}^6 & = & \frac{(147c^3+4237c^2+46786c+181360)x}{2(c+2)(c+24)(7c+114)} - \frac{192(7c^2+195c+1628)}{c(c+24)(5c+22)x}
\end{eqnarray*}}
If we chose the spin $6$ and spin $7$ primaries to have diagonal two-point functions, the structure constants $C_{36}^3$ and $C_{37}^4$ would vanish. But this would make other formulas more complicated. Furthermore, to compute the two-point functions of primaries directly we need to go to higher level. Finally, the remaining structure constants are
{\small\begin{eqnarray*}
C_{36}^{[34]^\prime} & = & \frac{1920(2c-1)}{c(c+24)(5c+22)} - \frac{160(c+7)(2c-1)x^2}{(c+2)(c+24)(7c+114)} \\
C_{46}^{[33]} & = & \frac{120(c+7)(2c-1)(5c+22)(7c+68)x^2}{(c+2)^2(c+24)(7c+114)^2}-\frac{1440(2c-1)(7c+68)}{c(c+2)(c+24)(7c+114)} \\
C_{37}^{[33]} & = & \frac{15(c+7)(2c-1)(5c+22)(7c+68)(167c^2+2186c-1392)x^3}{4(c+2)^2(c+23)(c+24)(5c-4)(7c+114)^2} - \frac{45(2c-1)(7c+68)(167c^2+2186c-1392)x}{c(c+2)(c+23)(c+24)(5c-4)(7c+114)} \\
C_{46}^{[35]} & = & \frac{15(8611c^3+301020c^2+3170988c+11305504)x}{7(c+2)(c+24)(7c+114)(11c+350)} - \frac{11520(c+10)(169c+3370)}{7c(c+24)(5c+22)
  (11c+350)x} \\
C_{46}^{[44]} & = & \frac{15360(2c-1)}{7c(c+24)(5c+22)} - \frac{1280(c+7)(2c-1)x^2}{7(c+2)(c+24)(7c+114)} \\
C_{46}^{[33]^{\prime\prime}} & = & \frac{640(c+7)(2c-1)(5c+22)(29c^2+533c-870)x^2}{21(c+2)^2(c+24)(5c-4)(7c+114)^2} - \frac{2560(2c-1)(29c^2+533
  c-870)}{7c(c+2)(c+24)(5c-4)(7c+114)} \\
C_{46}^{[35]^\prime} & = & \frac{2304(499c^2-1404c-184900)}{7c(c+24)(5c+22)(11c+350)x} - \frac{3(28961c^3+239956c^2-7359452c-62692000)x}{7(c+2)
  (c+24)(7c+114)(11c+350)} \\
C_{37}^{[35]^\prime} & = & \frac{2304(73c^2+1149c-850)}{c(c+24)(5c+22)(11c+350)} - \frac{3(3797c^3+82090c^2+387832c-306880)x^2}{(c+2)(c+24)(7
  c+114)(11c+350)}
\end{eqnarray*}}
The main purpose of giving these structure constants explicitly is to show how complicated the OPEs are if we use the primary fields as generating fields of $\mathcal{W}_{\infty}$ but also to illustrate the manifest triality invariance of these expressions.

\subsubsection{Composite primary fields}
\label{appendixcomposite}
In our computation of OPE in the primary basis there were four composite primary fields which involved derivatives of primaries and were obtained from the higher order regular terms in the OPE. These were denoted by $[W_3 W_4]^{(1)}$, $[W_3 W_3]^{(2)}$ and $[W_3 W_5]^{(1)}$ and $[W_3 W_4]^{(2)}$. Let us show the explicit expressions for two of these: $[W_3 W_4]^{(1)}$ can be obtained using \texttt{OPEconf} command \texttt{OPEPPole[-1][W3,W4]} with result
\begin{eqnarray*}
[W_3 W_4]^{(1)} & = & \frac{(323c^2+1578c-608) C_{34}^3 (T^{(3)} W_3)}{42(c+2)(5c-4)(7c+114)} + \frac{(151c^2+336c-796) C_{34}^3 (T^{\prime\prime} W_3)}{7(c+2)(5c-4)7c+114)} \\
& & + \frac{(245c^2 + 396c+244) C_{34}^3 (T^\prime W_3^{\prime\prime})}{14(c+2)(5c-4)(7c+114)} + \frac{5(43c^2-261c-34) C_{34}^3 (T W_3^{(3)})}{42(c+2)(5c-4)(7c+114)} \\
& & - \frac{6(127c+18) C_{34}^3 (T^\prime (T W_3))}{7(c+2)(5c-4)(7c+114)} + \frac{4(127c+18) C_{34}^3 (T (T W_3^\prime))}{7(c+2)(5c-4)(7c+114)} - \frac{3C_{34}^5 (T^\prime W_5)}{7(c+7)} \\
& & + \frac{6C_{34}^5 (T W_5^\prime)}{35(c+7)} + \frac{4}{7}(W_3^\prime W_4) - \frac{3}{7} (W_3 W_4^\prime) + \frac{(5c+32) C_{34}^5 W_5^{(3)}}{210(c+7)} \\
& & + \frac{(5c^3-245c^2+616c+92) (C_{34}^3)^2 W_3^{(5)}}{280(c+2)(5c-4)(7c+114)}
\end{eqnarray*}
and $[W_3 W_3]^{(2)}$ is obtained similarly using \texttt{OPEPPole[-2][W3,W3]} with result
\begin{eqnarray*}
[W_3 W_3]^{(2)} & = & -\frac{108(T(W_3 W_3))}{13c+516} +\frac{3(c+48) (W_3^{\prime\prime} W_3)}{13c+516} -\frac{(7c+228) (W_3^\prime W_3^\prime)}{2(13c+516)} \\
& & -\frac{18(4c^2+211c-4083) C_{33}^4 (T^{\prime\prime} W_4)}{(c+31)(13c+516)(55c-6)} +\frac{18(5c^2-218c-4218) C_{33}^4 (T^\prime W_4^\prime)}{(c+31)(13c+516)(55c-6)} \\
& & +\frac{(805c^2+18649c+28254) C_{33}^4 (T W_4^{\prime\prime})}{(c+31)(13c+516)(55c-6)} +\frac{12(1927c-3543) C_{33}^4 (T(TW_4))}{(c+31)(13c+516)(55c-6)} \\
& & +\frac{(35c^3+1883c^2+31434c-36504) C_{33}^4 W_4^{(4)}}{16(c+31)(13c+516)(55c-6)} \\
& & + \frac{24 (1861c^2+14814c+50184) C_{33}^0 (T^{\prime\prime}(TT))}{c(3c+46)(5c+3)(5c+22)(13c+516)} \\
& & +\frac{6 (6895c^2+80424c-67212) C_{33}^0 (T^\prime(T^\prime T))}{c(3c+46)(5c+3)(5c+22)(13c+516)} \\
& & +\frac{(805c^3+29516c^2+197676c+169488) C_{33}^0 (T^{(4)} T)}{2c(3c+46)(5c+3)(5c+22)(13c+516)} \\
& & +\frac{9(149c^3+6116c^2+77580c-85392) C_{33}^0 (T^{\prime\prime} T^{\prime\prime})}{4c(3c+46)(5c+3)(5c+22)(13c+516)} \\
& & +\frac{(935c^3+61940c^2+793908c+767376) C_{33}^0 (T^{(3)} T^\prime)}{2c(3c+46)(5c+3)(5c+22)(13c+516)} \\
& & +\frac{144(1919c-642) C_{33}^0 (T (T (TT)))}{c(3c+46)(5c+3)(5c+22)(13c+516)} \\
& & +\frac{(175c^4+15990c^3+178120c^2-721656c-19152) C_{33}^0 T^{(6)}}{240c(3c+46)(5c+3)(5c+22)(13c+516)}.
\end{eqnarray*}

\subsection{Some OPE in quadratic basis}
\label{appendixquadratic}
Although this article is trying to convince the reader that the quadratic basis of $\mathcal{W}_{1+\infty}$ has its advantages over the primary basis, still the OPEs are not very simple. This appendix shows some of these for reader to get an idea how they look like. For the OPEs of $\mathcal{W}_3$ algebra we need (\ref{opeu1uk}, \ref{opeu2uk}) and
\begin{eqnarray*}
U_3(z) U_3(w) & \sim & \frac{\frac{1}{6}N(N-1)(N-2)(1+28\alpha_0^2-18N\alpha_0^2+12\alpha_0^4-72N\alpha_0^4+36N^2\alpha_0^4}{(z-w)^6} \\
& & + \frac{-(N-2)(1+N\alpha_0^2-N^2\alpha_0^2) U_2(w)}{(z-w)^4} + \frac{-(N-2)(1+N\alpha_0^2-N^2\alpha_0^2) U_2(z)}{(z-w)^4} \\
& & + \frac{-(N-1)(N-2)\alpha_0(1-N-5N\alpha_0^2-3N^2\alpha_0^2)U_1(z)}{(z-w)^5} \\
& & + \frac{(N-1)(N-2)\alpha_0(1-N-5N\alpha_0^2-3N^2\alpha_0^2)U_1(w)}{(z-w)^5} \\
& & + \frac{\frac{1}{2}(N-1)(N-2)(1+6\alpha_0^2-4N\alpha_0^2)(U_1(z) U_1(w))(w)}{(z-w)^4} \\
& & + \frac{2\alpha_0 U_3(w)}{(z-w)^3} + \frac{-2\alpha_0 U_3(z)}{(z-w)^3} + \frac{-2 U_4(w)}{(z-w)^2} + \frac{-2 U_4(z)}{(z-w)^2} \\
& & + \frac{-(N-1)(N-2)\alpha_0 (U_1(z) U_2(w))}{(z-w)^3} + \frac{(N-1)(N-2)\alpha_0 (U_2(z) U_1(w))}{(z-w)^3} \\
& & -\frac{\frac{1}{2}(N-1)^2(N-2)\alpha_0 U_1^{\prime\prime}}{(z-w)^3} + \frac{-\frac{3}{2}(N+1)N^2(N-1)(N-2)\alpha_0^3 U_1^{\prime\prime}(w)}{(z-w)^3} \\
& & + \frac{-\frac{1}{6}(N-1)^2(N-2)\alpha_0 U_1^{\prime\prime\prime}(w)}{(z-w)^2} + \frac{-\frac{1}{6}N(N-1)(N-2)\alpha_0(1+6N\alpha_0^2) U_1^{\prime\prime\prime}(w)}{(z-w)^2} \\
& & + \frac{-\frac{1}{24}(N-1)^2(N-2)\alpha_0 U_1^{(4)}(w)}{z-w} + \frac{-\frac{1}{24}N(N-1)(N-2)\alpha_0(1+6N\alpha_0^2) U_1^{(4)}(w)}{z-w} \\
& & + \frac{-(U_1(z) U_3(w))}{(z-w)^2} + \frac{-(U_3(z) U_1(w))}{(z-w)^2} + \frac{\frac{1}{6}(N-1)(N-2)\alpha_0 U_1^{\prime\prime\prime}(w)}{(z-w)^2} \\
& & + \frac{\frac{1}{24}(N-1)(N-2)\alpha_0 U_1^{(4)}(w)}{z-w} + \frac{(N-2)}{(z-w)^2} \left( (U_2 U_2)(w) -\frac{1}{2}(N-1)(U_1^{\prime\prime} U_1)(w) \right) \\
& & + \frac{(N-2)}{z-w} \left( (U_2^\prime U_2)(w) -\frac{1}{6}(N-1)(U_1^{\prime\prime\prime} U_1)(w) \right) \\
\end{eqnarray*}

\subsubsection{Bilocal fields}
To write down $\mathcal{W}_4$, we must know the OPE of $U_3(z) U_4(w)$ and $U_4(z) U_4(w)$. We can compute these from (\ref{opefullform}), but to write the singular part, we need the bilocal fields up to dimension $7$.
{\small
\begin{eqnarray*}
U_{0j}(z,w) & = & U_j(w) \\
U_{1j}(z,w) & = & (U_1 U_j)(w) + (z-w) (U_1^\prime U_j)(w) + \frac{(z-w)^2}{2} (U_1^{\prime\prime} U_j)(w) + \cdots \\
& = & (U_1(z) U_j(w)) \\
U_{22}(z,w) & = & (U_2 U_2)(w) - \frac{N-1}{2} (U_1^{\prime\prime} U_1)(w) + \frac{1}{2} U_2^{\prime\prime}(w) - \frac{N(N-1)\alpha_0}{6} U_1^{(3)}(w) \\
& & + (z-w) \Big( (U_2^\prime U_2)(w) - \frac{N-1}{6} (U_1^{(3)} U_1)(w) + \frac{1}{6} U_2^{(3)}(w) - \frac{N(N-1)\alpha_0}{24} U_1^{(4)}(w) \Big) \\
& & + (z-w)^2 \Big( \frac{1}{2} (U_2^{\prime\prime} U_2)(w) - \frac{N-1}{24} (U_1^{(4)} U_1)(w) + \frac{1}{24} U_2^{(4)}(w) - \frac{N(N-1)\alpha_0}{120} U_1^{(5)}(w) \Big) \\
& & + (z-w)^3 \Big( \frac{1}{6} (U_2^{(3)} U_w)(w) - \frac{N-1}{120} (U_1^{(5)} U_1)(w) + \frac{1}{120} U_2^{(5)}(w) - \frac{N(N-1)\alpha_0}{720} U_1^{(6)}(w) \Big) + \cdots \\
U_{23}(z,w) & = & (U_2 U_3)(w) - \frac{N-2}{2} (U_1^{\prime\prime} U_2)(w) - \frac{(N-1)(N-2)\alpha_0}{6} (U_1^{(3)} U_1)(w) + \frac{1}{2} U_3^{\prime\prime}(w) \\
& & - \frac{N(N-1)(N-2)\alpha_0^2}{24} U_1^{(4)}(w) + (z-w) \Big( (U_2^{\prime} U_3)(w) - \frac{N-2}{6} (U_1^{(3)} U_2)(w) \\
& & - \frac{(N-1)(N-2)\alpha_0}{24} (U_1^{(4)} U_1)(w) + \frac{1}{6} U_3^{(3)}(w) - \frac{N(N-1)(N-2)\alpha_0^2}{120} U_1^{(5)}(w) \Big) \\
& & + (z-w)^2 \Big( \frac{1}{2} (U_2^{\prime\prime} U_3)(w) - \frac{N-2}{24} (U_1^{(4)} U_2)(w) - \frac{(N-1)(N-2)\alpha_0}{120} (U_1^{(5)} U_1)(w) \\
& & + \frac{1}{24} U_3^{(4)}(w) - \frac{N(N-1)(N-2)\alpha_0^2}{720} U_1^{(6)}(w) \Big) + \cdots \\
U_{24}(z,w) & = & (U_2 U_4)(w) - \frac{N-3}{2} (U_1^{\prime\prime} U_3)(w) - \frac{(N-2)(N-3)\alpha_0}{6} (U_1^{(3)} U_2)(w) \\
& & - \frac{(N-1)(N-2)(N-3)\alpha_0^2}{24} (U_1^{(4)} U_1)(w) + \frac{1}{2} U_4^{\prime\prime}(w) - \frac{N(N-1)(N-2)(N-3)\alpha_0^3}{120} U_1^{(5)}(w) \\
& & + (z-w) \Big( (U_2^\prime U_4)(w) - \frac{N-3}{6} (U_1^{(3)} U_3)(w) - \frac{(N-2)(N-3)\alpha_0}{24} (U_1^{(4)} U_2)(w) \\
& & - \frac{(N-1)(N-2)(N-3)\alpha_0^2}{120} (U_1^{(5)} U_1)(w) + \frac{1}{6} U_4^{(3)}(w) - \frac{N(N-1)(N-2)(N-3)\alpha_0^3}{720} U_1^{(6)}(w) \Big) + \cdots \\
U_{25}(z,w) & = & (U_2 U_5)(w) - \frac{N-4}{2} (U_1^{\prime\prime} U_4)(w) - \frac{(N-3)(N-4)\alpha_0}{6} (U_1^{(3)} U_3)(w) \\
& & - \frac{(N-2)(N-3)(N-4) \alpha_0^2}{24} (U_1^{(4)} U_2)(w) - \frac{(N-1)(N-2)(N-3)(N-4)\alpha_0^3}{120} (U_1^{(5)} U_1)(w) \\
& & + \frac{1}{2} U_5^{\prime\prime}(w) - \frac{N(N-1)(N-2)(N-3)(N-4)\alpha_0^4}{720} U_1^{(6)}(w) + \cdots \\
U_{33}(z,w) & = & (U_3 U_3)(w) + \frac{1}{2} (U_1^{\prime\prime} U_3)(w) - \frac{N-2}{2} (U_2^{\prime\prime} U_2)(w) + \frac{(N-1)(N-2)\alpha_0}{6} (U_1^{(3)} U_2)(w) \\
& & + \frac{(N-1)(N-2)(1-6\alpha_0^2+4N\alpha_0^2)}{48}(U_1^{(4)} U_1)(w) + U_4^{\prime\prime}(w) + \frac{\alpha_0}{3} U_3^{(3)}(w) \\
& & - \frac{N(N-1)(N-2)\alpha_0^2}{24} U_2^{(4)}(w) + \frac{(N-1)(N-2)\alpha_0 (1-5N\alpha_0^2+3N^2\alpha_0^2)}{120} U_1^{(5)}(w) \\
& & - \frac{(N-1)(N-2)\alpha_0}{6} (U_1 U_2^{(3)})(w) + \frac{1}{2} (U_1 U_3^{\prime\prime})(w) \\
& & + (z-w) \Big( (U_3^\prime U_3)(w) + \frac{1}{6} (U_1^{(3)} U_3)(w) - \frac{N-2}{6} (U_2^{(3)} U_2)(w) + \frac{(N-1)(N-2)\alpha_0}{24} (U_1^{(4)} U_2)(w) \\
& & + \frac{(N-1)(N-2)(1-6\alpha_0^2+4N\alpha_0^2)}{240} (U_1^{(5)} U_1)(w) + \frac{1}{3} U_4^{(3)}(w) \\
& & + \frac{\alpha_0}{12} U_3^{(4)}(w) - \frac{N(N-1)(N-2)\alpha_0^2}{120} U_2^{(5)}(w) + \frac{(N-1)(N-2)\alpha_0(1-5N\alpha_0^2+3N^2\alpha_0^2)}{720} U_1^{(6)}(w) \\
& & - \frac{(N-1)(N-2)\alpha_0}{24} (U_1 U_2^{(4)})(w) + \frac{1}{6} (U_1 U_3^{(3)})(w) \Big) + \cdots \\
U_{34}(z,w) & = & (U_3 U_4)(w) + (U_1^{\prime\prime} U_4)(w) - \frac{N-3}{2} (U_2^{\prime\prime} U_3)(w) + \frac{N(N-3)\alpha_0}{6} (U_1^{(3)} U_3)(w) \\
& & - \frac{(N-2)(N-3)\alpha_0}{6} (U_2^{(3)} U_2)(w) + \frac{(N-2)(N-3)(1-4\alpha_0^2+4N\alpha_0^2)}{48} (U_1^{(4)} U_2)(w) \\
& & + \frac{(N-1)(N-2)(N-3)\alpha_0(1-4\alpha_0^2+3N\alpha_0^2)}{120} (U_1^{(5)} U_1)(w) + U_5^{\prime\prime}(w) + \frac{\alpha_0}{3} U_4^{(3)}(w) \\
& & - \frac{N(N-1)(N-2)(N-3)\alpha_0^3}{120} U_2^{(5)}(w) - \frac{(N-1)(N-2)(N-3)\alpha_0^2}{24} (U_1 U_2^{(4)})(w) \\
& & + \frac{(N-1)(N-2)(N-3)\alpha_0^2 (N+2-12N\alpha_0^2+8N^2\alpha_0^2)}{1400} U_1^{(6)}(w) + \frac{1}{2} (U_1 U_4^{\prime\prime})(w) + \cdots
\end{eqnarray*}}

\subsubsection{Structure constants}
Here we give some structure constants of $\mathcal{W}_{1+\infty}$ in quadratic basis. We leave out those structure constants $C_{jk}^{lm}$ that are trivial due to
\begin{equation}
C_{jk}^{lm} = \delta_j^l \delta_k^m, \quad \quad j+k=l+m
\end{equation}
or
\begin{equation}
C_{jk}^{lm} = 0, \quad \quad j+k=l+m+1
\end{equation}
and also those related to others by symmetry
\begin{equation}
C_{jk}^{lm} = (-1)^{j+k-l-m} C_{kj}^{ml}
\end{equation}
and translation symmetry
\begin{equation}
C_{jk}^{lm}(\alpha_0,N) = C_{j+1,k+1}^{l+1,m+1}(\alpha_0,N+1)
\end{equation}
The remaining independent structure constants $C_{jk}^{lm}$ up to $j+k < 6$ and $j = k = 3$ are
\begin{eqnarray*}
C_{11}^{00} & = & N \\
C_{21}^{00} & = & -N(N-1)\alpha_0 \\
C_{21}^{10} & = & N-1 \\
C_{21}^{01} & = & 0 \\
C_{31}^{00} & = & N(N-1)(N-2)\alpha_0^2 \\
C_{31}^{10} & = & -(N-1)(N-2)\alpha_0 \\
C_{31}^{20} & = & N-2 \\
C_{31}^{01} & = & 0 \\
C_{31}^{02} & = & 0 \\
C_{22}^{00} & = & \frac{1}{2}N(N-1)(1+2\alpha_0^2-4N\alpha_0^2) \\
C_{22}^{10} & = & N(N-1)\alpha_0 \\
C_{22}^{20} & = & -1 \\
C_{41}^{00} & = & -N(N-1)(N-2)(N-3)\alpha_0^3 \\
C_{41}^{10} & = & (N-1)(N-2)(N-3)\alpha_0^2 \\
C_{41}^{20} & = & -(N-2)(N-3)\alpha_0 \\
C_{41}^{30} & = & N-3 \\
C_{41}^{01} & = & 0 \\
C_{41}^{02} & = & 0 \\
C_{41}^{03} & = & 0 \\
C_{32}^{00} & = & -N(N-1)(N-2)\alpha_0 (1+\alpha_0^2-3N\alpha_0^2) \\
C_{32}^{10} & = & \frac{1}{2}(N-1)(N-2)(1-4N\alpha_0^2) \\
C_{32}^{20} & = & (N-2)(N+1)\alpha_0 \\
C_{32}^{30} & = & -2 \\
C_{32}^{01} & = & N(N-1)(N-2)\alpha_0^2 \\
C_{32}^{02} & = & 0 \\
C_{32}^{03} & = & -1 \\
C_{33}^{00} & = & \frac{1}{6}(N-2)(N-1)N\left(12\alpha_0^4+28\alpha_0^2+36\alpha_0^4N^2-72\alpha_0^4N-18\alpha_0^2N+1\right) \\
C_{33}^{10} & = & -(N-1)(N-2)\alpha_0 \left(3\alpha_0^2N^2-5\alpha_0^2N-N+1\right) \\
C_{33}^{20} & = & (N-2) \left(\alpha_0^2N^2-\alpha_0^2N-1\right) \\
C_{33}^{30} & = & -2\alpha_0 \\
C_{33}^{40} & = & -2.
\end{eqnarray*}

\bibliography{winfcom}

\end{document}